\documentclass[12pt]{article}
\usepackage{lineno}
\usepackage{graphicx}
\usepackage{amsmath}
\usepackage{hhline}
\usepackage{amssymb}
\usepackage{times}
\usepackage{cite}
\usepackage{array}
\usepackage{rotating}
\usepackage{multirow}
\usepackage{bigstrut}
\usepackage{color}
\definecolor{rltred}{rgb}{0.75,0,0}
\definecolor{rltgreen}{rgb}{0,0.5,0}
\definecolor{rltblue}{rgb}{0,0,0.75}
\usepackage{dcolumn}
\newcolumntype{d}{D{.}{.}{1}}
\newcolumntype{x}{D{<}{<}{10}}
\newcolumntype{e}{D{E}{E}{7}}
\newcolumntype{q}{D{Q}{Q}{7}}

\usepackage{thumbpdf}
\usepackage[pdftex,
        colorlinks=true,
        urlcolor=rltblue,       
        filecolor=rltgreen,     
        linkcolor=rltred,       
        pdftitle={Example File for pdflatex},
        pdfauthor={H1},
        pdfsubject={Template file},
        pdfkeywords={High-Energy Physics, Particle Physics},
        pdfpagemode=None,
        bookmarksopen=true]{hyperref}
 
\newlength{\dinwidth}
\newlength{\dinmargin}
\begin{document}  

\newcommand{\pom}{{I\!\!P}}
\newcommand{\reg}{{I\!\!R}}
\newcommand{\slowpi}{\pi_{\mathit{slow}}}
\newcommand{\fiidiii}{F_2^{D(3)}}
\newcommand{\fiidiiiarg}{\fiidiii\,(\beta,\,Q^2,\,x)}
\newcommand{\n}{1.19\pm 0.06 (stat.) \pm0.07 (syst.)}
\newcommand{\nz}{1.30\pm 0.08 (stat.)^{+0.08}_{-0.14} (syst.)}
\newcommand{\fiidiiiful}{F_2^{D(4)}\,(\beta,\,Q^2,\,x,\,t)}
\newcommand{\fiipom}{\tilde F_2^D}
\newcommand{\ALPHA}{1.10\pm0.03 (stat.) \pm0.04 (syst.)}
\newcommand{\ALPHAZ}{1.15\pm0.04 (stat.)^{+0.04}_{-0.07} (syst.)}
\newcommand{\fiipomarg}{\fiipom\,(\beta,\,Q^2)}
\newcommand{\pomflux}{f_{\pom / p}}
\newcommand{\nxpom}{1.19\pm 0.06 (stat.) \pm0.07 (syst.)}
\newcommand {\gapprox}
   {\raisebox{-0.7ex}{$\stackrel {\textstyle>}{\sim}$}}
\newcommand {\lapprox}
   {\raisebox{-0.7ex}{$\stackrel {\textstyle<}{\sim}$}}
\def\gsim{\,\lower.25ex\hbox{$\scriptstyle\sim$}\kern-1.30ex%
\raise 0.55ex\hbox{$\scriptstyle >$}\,}
\def\lsim{\,\lower.25ex\hbox{$\scriptstyle\sim$}\kern-1.30ex%
\raise 0.55ex\hbox{$\scriptstyle <$}\,}
\newcommand{\pomfluxarg}{f_{\pom / p}\,(x_\pom)}
\newcommand{\dsf}{\mbox{$F_2^{D(3)}$}}
\newcommand{\dsfva}{\mbox{$F_2^{D(3)}(\beta,Q^2,x_{I\!\!P})$}}
\newcommand{\dsfvb}{\mbox{$F_2^{D(3)}(\beta,Q^2,x)$}}
\newcommand{\dsfpom}{$F_2^{I\!\!P}$}
\newcommand{\gap}{\stackrel{>}{\sim}}
\newcommand{\lap}{\stackrel{<}{\sim}}
\newcommand{\fem}{$F_2^{em}$}
\newcommand{\tsnmp}{$\tilde{\sigma}_{NC}(e^{\mp})$}
\newcommand{\tsnm}{$\tilde{\sigma}_{NC}(e^-)$}
\newcommand{\tsnp}{$\tilde{\sigma}_{NC}(e^+)$}
\newcommand{\st}{$\star$}
\newcommand{\sst}{$\star \star$}
\newcommand{\ssst}{$\star \star \star$}
\newcommand{\sssst}{$\star \star \star \star$}
\newcommand{\tw}{\theta_W}
\newcommand{\sw}{\sin{\theta_W}}
\newcommand{\cw}{\cos{\theta_W}}
\newcommand{\sww}{\sin^2{\theta_W}}
\newcommand{\cww}{\cos^2{\theta_W}}
\newcommand{\trm}{m_{\perp}}
\newcommand{\trp}{p_{\perp}}
\newcommand{\trmm}{m_{\perp}^2}
\newcommand{\trpp}{p_{\perp}^2}
\newcommand{\alp}{\alpha_s}

\newcommand{\alps}{\alpha_s}
\newcommand{\sqrts}{$\sqrt{s}$}
\newcommand{\LO}{$O(\alpha_s^0)$}
\newcommand{\Oa}{$O(\alpha_s)$}
\newcommand{\Oaa}{$O(\alpha_s^2)$}
\newcommand{\PT}{p_{\perp}}
\newcommand{\JPSI}{J/\psi}
\newcommand{\sh}{\hat{s}}
\newcommand{\uh}{\hat{u}}
\newcommand{\MP}{m_{J/\psi}}
\newcommand{\PO}{I\!\!P}
\newcommand{\xbj}{x}
\newcommand{\xpom}{x_{\PO}}
\newcommand{\ttbs}{\char'134}
\newcommand{\xpomlo}{3\times10^{-4}}  
\newcommand{\xpomup}{0.05}  
\newcommand{\dgr}{^\circ}
\newcommand{\pbarnt}{\,\mbox{{\rm pb$^{-1}$}}}
\newcommand{\gev}{\,\mbox{GeV}}
\newcommand{\WBoson}{\mbox{$W$}}
\newcommand{\fbarn}{\,\mbox{{\rm fb}}}
\newcommand{\fbarnt}{\,\mbox{{\rm fb$^{-1}$}}}
%
%
\newcommand{\qsq}{\ensuremath{Q^2} }
\newcommand{\gevsq}{\ensuremath{\mathrm{GeV}^2} }
\newcommand{\et}{\ensuremath{E_t^*} }
\newcommand{\rap}{\ensuremath{\eta^*} }
\newcommand{\gp}{\ensuremath{\gamma^*}p }
\newcommand{\dsiget}{\ensuremath{{\rm d}\sigma_{ep}/{\rm d}E_t^*} }
\newcommand{\dsigrap}{\ensuremath{{\rm d}\sigma_{ep}/{\rm d}\eta^*} }

\def\Journal#1#2#3#4{{#1} {\bf #2} (#3) #4}
\def\NCA{\em Nuovo Cimento}
\def\NIM{\em Nucl. Instrum. Methods}
\def\NIMA{{\em Nucl. Instrum. Methods} {\bf A}}
\def\NPB{{\em Nucl. Phys.}   {\bf B}}
\def\PLB{{\em Phys. Lett.}   {\bf B}}
\def\PRL{\em Phys. Rev. Lett.}
\def\PRD{{\em Phys. Rev.}    {\bf D}}
\def\ZPC{{\em Z. Phys.}      {\bf C}}
\def\EJC{{\em Eur. Phys. J.} {\bf C}}
\def\CPC{\em Comp. Phys. Commun.}

\begin{titlepage}
\noindent

\begin{flushleft}

DESY 10-083 \hfill ISSN 0418-9833 \\
June 2010
\end{flushleft}

\vspace{2cm}

\begin{center}
\begin{Large}
  
   {\bf Measurement of Charm and 
 Beauty Jets in Deep Inelastic Scattering
     at HERA}

\vspace{2cm}

H1 Collaboration

\end{Large}
\end{center}

\vspace{2cm}

\begin{abstract}
  \noindent Measurements of cross sections for events with charm and
  beauty jets in deep inelastic scattering at HERA are presented.
  Events with jets of transverse
  energy $E_T^{\rm jet}>6~{\rm GeV}$ and pseudorapidity
  $-1.0<\eta^{\rm jet}<1.5$ in the laboratory frame 
  are selected in the kinematic region of photon virtuality $
  Q^2 > 6~{\rm GeV}^2$ and inelasticity variable $0.07 < y < 0.625$.
  Measurements are also made requiring a jet in the Breit frame with
  $E_T^{* {\rm jet}}>6~{\rm GeV}$. The data were collected with the H1
  detector in the years 2006 and 2007 corresponding to an integrated
  luminosity of $189~{\rm pb^{-1}}$.  The numbers of charm and beauty
  jets are determined using variables reconstructed using the H1
  vertex detector with which the impact parameters of the tracks to
  the primary vertex and the position of secondary vertices are
  measured.  The measurements are compared with QCD predictions and
  with previous measurements where heavy flavours are identified using
  muons.
\end{abstract}

\vspace{1.5cm}

\begin{center}
Submitted to Eur. Phys. J.\ C.
\end{center}

\end{titlepage}

\begin{flushleft}

F.D.~Aaron$^{5,49}$,           
C.~Alexa$^{5}$,                
V.~Andreev$^{25}$,             
S.~Backovic$^{30}$,            
A.~Baghdasaryan$^{38}$,        
E.~Barrelet$^{29}$,            
W.~Bartel$^{11}$,              
K.~Begzsuren$^{35}$,           
A.~Belousov$^{25}$,            
J.C.~Bizot$^{27}$
% ...

\bigskip{\it
% ...
$ ^{5}$ National Institute for Physics and Nuclear Engineering (NIPNE) ,
          Bucharest, Romania$^{ m}$ \\
% ...
$ ^{11}$ DESY, Hamburg, Germany \\
% ...
$ ^{25}$ Lebedev Physical Institute, Moscow, Russia$^{ e}$ \\
% ...
$ ^{27}$ LAL, Universit\'e Paris-Sud, CNRS/IN2P3, Orsay, France \\
% ...
$ ^{29}$ LPNHE, Universit\'e Pierre et Marie Curie Paris 6,
          Universit\'e Denis Diderot Paris 7, CNRS/IN2P3, Paris, France \\
 $ ^{30}$ Faculty of Science, University of Montenegro,
          Podgorica, Montenegro$^{ e}$ \\
% ...
$ ^{35}$ Institute of Physics and Technology of the Mongolian
          Academy of Sciences, Ulaanbaatar, Mongolia \\
% ...
$ ^{38}$ Yerevan Physics Institute, Yerevan, Armenia \\
% ...
$ ^{49}$ Also at Faculty of Physics, University of Bucharest,
          Bucharest, Romania \\
% ...
$ ^e$ Supported by the Deutsche Forschungsgemeinschaft \\
% ...
$ ^m$ Supported by the Romanian National Authority for Scientific Research
      under the contract PN 09370101 \\
}

\end{flushleft}

\newpage

\section{Introduction}

The production of heavy flavour quarks in deep inelastic scattering
(DIS) at the HERA electron--proton collider is of particular interest
for testing calculations in the framework of perturbative quantum
chromodynamics (QCD). The process has the special feature of involving
two hard scales: the square root of the photon virtuality $Q$ and the
heavy quark mass $m$. In the case of jet production the transverse
energy $E_T$ of the jet provides a further hard scale. In leading
order (LO) QCD, the photon-gluon fusion (PGF) processes $ep
\rightarrow ec\bar{c}X$ and $ep \rightarrow eb\bar{b}X$ are the
dominant production mechanisms for charm ($c$) and beauty ($b$) quarks
respectively. 

The inclusive $c$ and $b$ quark cross sections and the derived
structure functions have been measured in DIS at HERA using the
`inclusive lifetime' technique~\cite{f2bcall,Aktas:2005iw} and found
to be well described by next to leading order (NLO) QCD. Measurements
of the charm cross section using the technique of $D$ meson tagging
have also been made~\cite{H1ZEUSDstar, H1Dstar} and are found to be in
good agreement with those using inclusive lifetime
information. Measurements of the total charm and beauty cross sections
have been made by identifying their decays to
muons~\cite{ZEUSincmuons}.
In the charm case these measurements show good agreement with the data 
extracted using the inclusive lifetime technique, but are somewhat larger in
the case of beauty.

Measurements of beauty quark production using muon tagging have also
been made for DIS events containing a high $E_T$ jet in either the
Breit frame~\cite{ZEUSmuons,H1muons} or in the laboratory
frame~\cite{ZEUSmuonslab}. As in the muon inclusive
case~\cite{ZEUSincmuons} the results were found to be somewhat higher
than NLO QCD predictions, in particular at low values of $Q^2$. In
photoproduction, measurements of beauty have been made using various
lepton tagging techniques and have been found to be either somewhat
higher than~\cite{Highgammapleptons} or in agreement
with~\cite{Agreegammapleptons} NLO QCD.  A measurement in the Breit
frame of the production of $D^*$ mesons in association with high $E_T$
dijets~\cite{Aktas:2006py} was found to be in agreement with NLO QCD
predictions within the statistics of the measurement. A measurement of
$c$ and $b$ jets in photoproduction has been made~\cite{Aktas:2006vs},
which uses a similar method to distinguish heavy flavour jets as in
the present analysis. The results were found to be in good agreement
with NLO QCD.

This paper reports on measurements of the cross sections for events
with a $c$ or $b$ jet in DIS at HERA. The analysis uses an inclusive
lifetime technique following a similar procedure as used
in~\cite{f2bcall} to distinguish the jets that contain $c$ or $b$
flavoured hadrons from those containing light flavoured hadrons only.  The
data are analysed in the laboratory frame of reference to
match the acceptance of the H1 detector and a heavy flavour
jet with the highest transverse energy  $E_T^{\rm jet}>6~{\rm
GeV}$ is required.
The measurements in the laboratory frame are compared with 
$b$ quark production measurements obtained from muon tagging~\cite{ZEUSmuonslab}.
The analysis is extended to the Breit frame of reference requiring a jet with 
transverse energy of $E_T^{* {\rm jet}}>6~{\rm GeV}$. The
results are also compared with $b$ quark measurements obtained from muon 
tagging~\cite{H1muons}.
The cross section measurements in both frames of reference are compared
with an NLO QCD program employing mass factorisation~\cite{hvqdis}. 

The data for this analysis were recorded in the years 2006 and 2007
with integrated luminosities of $135$~${\rm pb}^{-1}$ taken in $e^+p$
mode and $54$~${\rm pb}^{-1}$ taken in $e^-p$ mode .  The $ep$ centre
of mass energy is $\sqrt{s} = 319~{\rm GeV}$, with a proton beam
energy of $920~{\rm GeV}$ and electron\footnote{In this paper the term
`electron'  also denotes `positron' unless explicitly stated.} beam
energy of $27.6~{\rm GeV}$. The measurements are made 
for the kinematic region of photon virtuality $ Q^2 >6~{\rm GeV}^2$ and 
inelasticity variable $0.07 < y < 0.625$. 

Jets containing heavy flavoured hadrons are distinguished from those
containing only light flavours using variables reconstructed using the
H1 vertex detector.  The most important of these inputs are the
transverse displacement of tracks from the primary vertex and the
reconstructed position of a secondary vertex in the transverse
plane. Hadrons from heavy quark decays typically have longer lifetimes
than light hadrons and thus produce tracks that have a significant
displacement from the primary vertex.  For jets with three or more
tracks in the vertex detector the reconstructed variables are used as
input to a neural network to discriminate beauty from charm jets.

\section{Monte Carlo Simulation}
\label{sec:MC}
Monte Carlo simulations are used to correct for the effects of the
finite detector resolution, acceptance and efficiency.  The Monte
Carlo program RAPGAP\cite{Jung:1993gf} is
used to generate DIS events for the processes
$ep \rightarrow eb\bar{b}X$, $ep \rightarrow ec\bar{c}X$ and
$ep \rightarrow eq X$ where $q$ is a light
quark of flavour $u$, $d$ or $s$.
RAPGAP combines $\cal{O}$($\alpha_s$) matrix elements with higher
order QCD effects modelled by parton showers. The
heavy flavour event samples are generated according to the massive
photon gluon fusion (PGF) matrix element~\cite{massive} with the mass of 
the $c$ and
$b$ quarks set to $m_c=1.5~{\rm GeV}$ and $m_b=4.75~{\rm GeV}$,
respectively.  The DIS cross section is calculated using the leading
order 3-flavour parton density function (PDF) 
set  MRST2004F3LO~\cite{Martin:2006qz}.

The partonic system for the generated events is fragmented according
to the Lund string model~\cite{Andersson:1983ia} implemented within the PYTHIA
program~\cite{Sjostrand:2001yu}.  
The $c$ and $b$ quarks are hadronised according to the Bowler fragmentation
function~\cite{bowler} using the parameters $a=0.4~{\rm GeV^{-2}}$, 
$b=1.03~{\rm GeV^{-2}}$ and $r_Q=1$.
The HERACLES
program\cite{Kwiatkowski:1990es} calculates single photon radiative
 emissions off the lepton line, virtual and electroweak corrections.

PYTHIA is used to simulate the background contribution from
photoproduction $\gamma p \rightarrow X$. 
The assumed heavy flavour cross sections are in agreement with the 
measurements made by H1~\cite{Aktas:2006vs}.

The samples of events generated for the $uds$, $c$, and $b$ processes
are passed through a detailed simulation of the detector response
based on the GEANT3 program\cite{Brun:1978fy}, and through the same
reconstruction software as is used for the data.

\section{QCD models}
\label{sec:results:models}

The jet cross section data in this paper are compared with two
approaches within QCD:

Firstly, the data are compared with the predictions of Monte Carlo 
programs based on
leading order matrix elements with the effect of higher orders
modelled by initial and final state parton showers.  The predictions
from the RAPGAP Monte Carlo program are calculated with the same
settings as described in section~\ref{sec:MC}.  The renormalisation and
factorisation scales are set to $\mu_r = \mu_f = Q$.  The Monte Carlo
program CASCADE~\cite{cascade} is also used to produce predictions for
the $b$ and $c$ jet cross sections.  CASCADE is based on
the CCFM\cite{ccfm2} evolution equation and uses off shell matrix
elements convoluted with $k_T$ unintegrated proton parton
distributions.  The CASCADE predictions use the A0 PDF set with
$m_c=1.5~{\rm GeV}$ and $m_b=4.75~{\rm GeV}$, and $\mu_r =
 \sqrt{Q^2+p_T^2+4m^2}$, where $p_T$ is the transverse momentum of the
heavy quark in the virtual photon-proton centre of
mass frame.  Due to the fact that the predictions are
based on leading order matrix elements the uncertainty on the
normalisation of the cross sections is large, and is not quantified here.

Secondly, the data are compared with the predictions of the NLO QCD
program \linebreak
HVQDIS~\cite{hvqdis}.  The program is based on the fixed flavour numbering scheme 
(FFNS) which uses the massive PGF
$\cal{O}$($\alpha_s^2$) matrix element \cite{massivenlo}
and provides weighted events
with two or three outgoing partons, i.e. a heavy quark pair and
possibly an additional light parton.  The calculations are made using
the same settings for the choice of the quark masses as for the Monte
Carlo programs above: $m_c=1.5~{\rm GeV}$, $m_b=4.75~{\rm GeV}$.  At
NLO the predictions of QCD depend on the choice of the scales $\mu_r$
and $\mu_f$.  To investigate the dependence of the predictions on the
scales two example choices are made. Firstly, the scale $\mu_r = \mu_f
= \sqrt{(Q^2+p_T^2+m^2)/2}$, where $p_T$ is the transverse momentum of
the heavy quark with the highest value of $p_T$
in the virtual photon-parton centre of
mass frame, is used.  This choice of scale is motivated by the
comparison of NLO QCD with recent measurements of inclusive jet data
by H1~\cite{h1jets}. Secondly, the scale $\mu_r =
\mu_f = \sqrt{Q^2+4m^2}$ is selected. This scale has been used in the
comparison of HVQDIS with H1 inclusive and dijet $D^*$ DIS
data~\cite{Aktas:2006py,H1Dstar}.  Since HVQDIS provides cross
sections at the parton level, corrections to the hadron level are needed
in order to compare to the data. These corrections are calculated
using the RAPGAP Monte Carlo event generator. In each kinematic bin of
the measurement, the ratio $C_{\rm had}$ of the RAPGAP hadron level to
parton level cross sections is calculated and applied as a correction
factor to the NLO calculation. The hadron level corrections generally
amount to a change in the prediction by $\le 6\%$ for charm and $ \le
15\%$ for beauty.

In QCD fits to global hard-scattering data the parton density
functions are usually extracted using the general mass variable
flavour number scheme (GM
VFNS)~\cite{VFNS1part1,VFNS1part2,cteqvfns,VFNS2,NNLO,VFNS3,Alekhin}
for heavy quarks.
This scheme, which interpolates from the massive approach at low
scale values to a `massless' approach at high scale values, 
provides a theoretically accurate description of heavy flavour
production.  
Recently, a set of PDFs~\cite{mstw08ff3} compatible with the FFNS 
 were generated, using the standard
GM VFNS PDFs~\cite{Martin:2009iq} to facilitate comparison of the heavy
flavour final state data with up-to-date PDFs.

Predictions are made using three different sets of PDFs: the
MSTW08FF3~\cite{mstw08ff3} set
extracted using the GM VFNS but
evolved using the FFNS in order to be compatible with HVQDIS; 
the CTEQ5F3~\cite{cteq5} set extracted using the FFNS;
and with the CTEQ6.6~\cite{cteqvfns} set extracted using the 
GM VFNS.  The CTEQ6.6 PDF set uses a variable flavour definition of the
running coupling $\alpha_s$ which is different to the fixed flavour 
definition assumed in HVQDIS. However, the inaccuracy introduced
by this incompatibility is likely to be compensated by using
an up-to-date PDF set~\cite{Thorne:2008xf}.

As an estimate of the uncertainty on each of the NLO QCD predictions
the scales $\mu_r$ and $\mu_f$  are varied simultaneously 
by factors of  $0.5$  and $2$, $m_c$ is changed by $\pm 0.2~{\rm GeV}$
and $m_b$ is changed by $\pm 0.25~{\rm GeV}$.  The uncertainty from
fragmentation is estimated by replacing the Bowler~\cite{bowler}
function  by the symmetric function in the
Lund model \cite{Andersson:1983jt}, corresponding to $r_Q=0$.

\section{H1 Detector}

Only a short description of the H1 detector is given here; a more
complete description may be found elsewhere \cite{Abt:1997xv,
Nicholls:1996di}. A right-handed coordinate system is employed at H1,
with its origin at the nominal interaction vertex, that has its
$Z$-axis pointing in the proton beam, or forward, direction and $X$
($Y$) pointing in the horizontal (vertical) direction. The
pseudorapidity is related to the polar angle $\theta$ by $\eta = - \ln
\tan(\theta/2)$.

Charged particles are measured in the central tracking detector (CTD).
This device consists of two cylindrical drift chambers interspersed
with orthogonal chambers to improve the $Z$-coordinate reconstruction and
multi-wire proportional chambers mainly used for triggering. The CTD
is operated in a uniform solenoidal $1.16\,{\rm T}$ magnetic field, enabling the
momentum measurement of charged particles over the polar angular
range\footnote{\noindent{The angular coverage of each detector
component is given for the interaction vertex in its nominal position
i.e. the position of the centre of the detector.}}  
$20^\circ< \theta<160^\circ$.

The CTD tracks are linked to hits in the vertex detector, 
the central silicon tracker (CST)~\cite{cst}, to provide precise spatial track
reconstruction. The CST consists of two layers of double-sided silicon
strip detectors surrounding the beam pipe, covering an angular range
of $30^\circ< \theta<150^\circ$ for tracks passing through both
layers.   The information on the $Z$-coordinate of the CST tracks 
is not used in the analysis presented in this paper. For CTD tracks
with CST hits in both layers the transverse distance of closest
approach (DCA) to the nominal vertex in $X$--$Y$,
averaged over the azimuthal angle, is measured to have
a resolution of $43\;\mu\mbox{m} \oplus 51 \;\mu\mbox{m} /(P_T
[\mbox{GeV}])$, where the first term represents the intrinsic
resolution (including alignment uncertainty) and the second term is
the contribution from multiple scattering in the beam pipe and the
CST; $P_T$ is the transverse momentum of the particle. The efficiency
for linking hits in both layers of the CST to a CTD track is 
around $84\%$.  The efficiency for finding tracks in the CTD is 
greater than $95\%$.

The track detectors are surrounded in the forward and central
directions ($4^\circ<\theta<155^\circ$) by a finely grained liquid argon
calorimeter (LAr) and in the backward region
($153^\circ<\theta<178^\circ$) by a lead-scintillating fibre
calorimeter (SPACAL)
with electromagnetic and
hadronic sections. These calorimeters provide energy and angular
reconstruction for final state particles from the hadronic system and
are also used in this analysis to measure and identify the scattered
electron.  

Electromagnetic calorimeters situated downstream in the electron beam
direction allow detection of photons and electrons scattered at very
low $Q^2$. The luminosity is measured with these calorimeters
from the rate of photons
produced in the Bethe-Heitler process $ep\rightarrow ep\gamma$.

\section{Experimental Method}

\subsection{DIS Event Selection}
\label{sec:disselect}
The events are triggered by  a compact, isolated
electromagnetic cluster in either the LAr or SPACAL calorimeters in
combination with a loose track requirement such that the overall
trigger efficiency is almost 100\%.  The electromagnetic cluster with
the highest transverse energy, which also passes stricter offline
criteria is taken as the scattered electron.  The $Z$-position of the
interaction vertex, reconstructed by one or more charged tracks in the
tracking detectors, must be within $\pm 20~{\rm cm}$ of the centre of
the detector to match the acceptance of the CST.  

Photoproduction
events and DIS events with a hard photon radiated from the initial
state electron are suppressed by requiring $\sum_{i} (E_i - p_{Z,i})
>35~{\rm GeV}$.  Here, $E_i$ and $p_{Z,i}$ denote the energy and
longitudinal momentum components of a particle and the sum is over all
final state particles including the scattered electron and the
hadronic final state (HFS). The HFS particles are reconstructed using
a combination of tracks and calorimeter deposits in an energy flow
algorithm that avoids double counting\cite{peezportheault}. 

The event
kinematics, $Q^2$ and  $y$, are reconstructed
with the `$e\Sigma$' method\cite{Bassler:1994uq}, which uses the
scattered electron and the HFS.  In order to have good acceptance 
for the scattered electron in the calorimeters the events are selected
in the range $ Q^2 > 6 \ {\rm GeV^2}$. The analysis is
restricted to $0.07<y<0.625$ in order to ensure there is a high
probability of at least one jet  within the acceptance of the CST and
to reduce the photoproduction background.
The position of the beam interaction region in $X$ and $Y$ (beam spot)
is derived from tracks with CST hits and updated
regularly to account for drifts during beam storage.

\subsection{Jet Reconstruction}
Jets are reconstructed using the inclusive longitudinally invariant
$k_T$ algorithm with the massless $P_T$ recombination scheme and the
distance parameter $R_0=1$ in the $\eta-\phi$ plane~\cite{KTJET}. The
algorithm is first run in the laboratory frame using all reconstructed HFS
particles and the resultant jets are required to have transverse
energy $E_T^{\rm jet} >1.5~{\rm GeV}$, in the angular range $-1.0 < \eta^{\rm jet}< 1.5$. The $\eta$ range is asymmetric since the $y$ range chosen
means few jets have $\eta <-1.0$. This cut also means that the
jets are not near the boundary between the LAr and SPACAL
calorimeters. 
Jets are reconstructed from the Monte Carlo simulation 
using an identical procedure to that of the data.

The Monte Carlo simulation is also used to define hadron and parton
level jets before they are processed by the simulation of the
detector response.  Hadron level jets are defined by running the same jet
algorithm as for reconstructed jets using all final state
particles, including neutrinos, but excluding the scattered electron.  A
Monte Carlo jet at the reconstructed or hadron level is defined as a
`$b$ jet' if there is at least one $b$ hadron within a cone of radius
$1$ about the jet axis in the $\eta-\phi$ plane.  A jet is defined as a
`$c$ jet' if there is at least one $c$ hadron within the same cone and
that $c$ hadron does not arise from the decay of a $b$ hadron. Jets
that have not been classified as $c$ or $b$ jets are called `light
jets'. Parton level jets are defined for the Monte Carlo samples and for the
NLO calculation by running the same jet algorithm on final state
partons.  A parton level jet is defined as a $b$ jet if there is
at least one $b$ quark within a cone of radius $1$ about the jet
axis in the $\eta-\phi$ plane.  
A parton level jet is defined as a
$c$ jet if there is at least one $c$ quark and no $b$ quark 
within the cone.

In order to compare with perturbative calculations a good correlation
between the parton level and hadron level jets is necessary. A jet
with high transverse energy is required in either the laboratory frame
of reference $E_T^{\rm jet} >6~{\rm GeV}$ or in the Breit frame
$E_T^{* \rm jet} >6~{\rm GeV}$.  For the analysis in the laboratory
frame the cross section is measured as a function of $E_T^{\rm jet}$,
$\eta^{\rm jet}$ , $Q^2$, the number of jets  $N_{\rm jet}$ with $E_T^{\rm
jet}>6~{\rm GeV}$  and also for the integrated sample.
For the analysis in the Breit frame the flavour of the jet is defined
in the laboratory, as described above, for jets in the range $E_T^{\rm
jet} >1.5~{\rm GeV}$ and $-1.0 < \eta^{\rm jet}< 1.5$. For events
satisfying this condition all the final state particles are then
boosted to the Breit frame using the four vector of the scattered
electron and the value of Bjorken $x$ obtained from $x = Q^2/sy$. The
jet finding algorithm is rerun on the boosted particles. The jets in
the Breit frame are required to have a transverse energy $E_T^{* {\rm
jet}} >6~{\rm GeV}$ and to have a pseudorapidity, when boosted back to
the laboratory frame, in the range $-1.0 < \eta^{\rm jet}< 1.5$.  The
cross section is measured as a function of $E_T^{* {\rm jet}}$ and
$Q^2$ for the selected events.  The data, measured as a function of
$Q^2$, in both reference frames are compared with $b$ jet data
obtained from muon tagging, after correcting those results for the
muon phase space and other, smaller, differences between the kinematic
ranges of the measurements.

\subsection{Jet Flavour Separation}
\label{sec:jetflavoursep}
In the analysis presented in this paper the flavour of the
event is defined as the flavour of the jet with the highest 
$E_T^{\rm jet}$ in the laboratory.  Therefore, the measured cross sections 
are proportional to the number of events with a jet rather than the number 
of jets in an event.

The separation of $b$, $c$ and light jets is only briefly described
here. The procedure closely follows that described
in~\cite{f2bcall}. The separation is performed using the properties of
those tracks which are within a cone of radius $1$ from the jet axis
in the $\eta-\phi$ plane. The tracks are reconstructed in the CTD and
must have at least $2$ CST hits and have transverse momentum greater
than $0.3~{\rm GeV}$.  The impact parameter $\delta$ of a track is the
transverse DCA of the track to the beam spot point. Tracks with
$\delta>0.1 \ {\rm cm}$ are rejected to suppress contributions from
the decays of long-lived strange particles.

The number of tracks in the jet after these selections is called $N_{\rm track}$. 
The track significance $S$ is defined as
$S = \delta/\sigma(\delta)$, where $\sigma(\delta)$ is the uncertainty on
$\delta$. If the angle $\alpha$ between the azimuthal angle of the jet
$\phi_{\rm jet}$ and the line joining the primary vertex to the point
of DCA is less than $90^\circ$, the significance is defined as
positive~\cite{f2bcall}. 
It is defined as negative otherwise. The significances
$S_1$, $S_2$ and $S_3$ are defined as the significance of the track
with the highest, second highest and third highest absolute
significance, respectively.  The selected tracks are also used to
reconstruct the position of the secondary vertex.

The jets are separated into three independent samples. For each sample
a different distribution is used to separate the light, $b$ and $c$
jets. The $S_1$ distribution is used for jets where $N_{\rm track}=1$
or $S_1$ and $S_2$ have opposite signs. The $S_2$ distribution is used
for the remaining jets with $N_{\rm track}=2$ or where $S_3$ has a
different sign to $S_1$ and $S_2$. Generally $S_2$ has a better
discrimination between light and heavy flavour jets than $S_1$, since
the chance of reconstructing $2$ high significance tracks is small for
jets where all the tracks arise at the primary vertex. For jets with
$N_{\rm track} \ge 3$ where $S_1$, $S_2$ and $S_3$ all have the same
sign an artificial neural network (NN) is used to produce a
distribution that combines several variables in order to provide an
optimal discrimination between $b$ and $c$ jets.  The
inputs to the NN are $S_1$, $S_2$, $S_3$, the significance of the
transverse distance between the secondary and primary vertex, the
transverse momenta of the tracks with the highest and second highest
transverse momentum, $N_{\rm track}$, and the number of reconstructed
tracks at the secondary vertex. The NN is trained using a sample of
inclusive heavy flavour DIS Monte Carlo events, with $b$ events as
`signal' and $c$ events as `background', as described
in~\cite{f2bcall}. The NN output is signed according to the sign of
$S_1$.

\begin{figure}[tb]
\begin{center}
\includegraphics[width=0.49\textwidth]{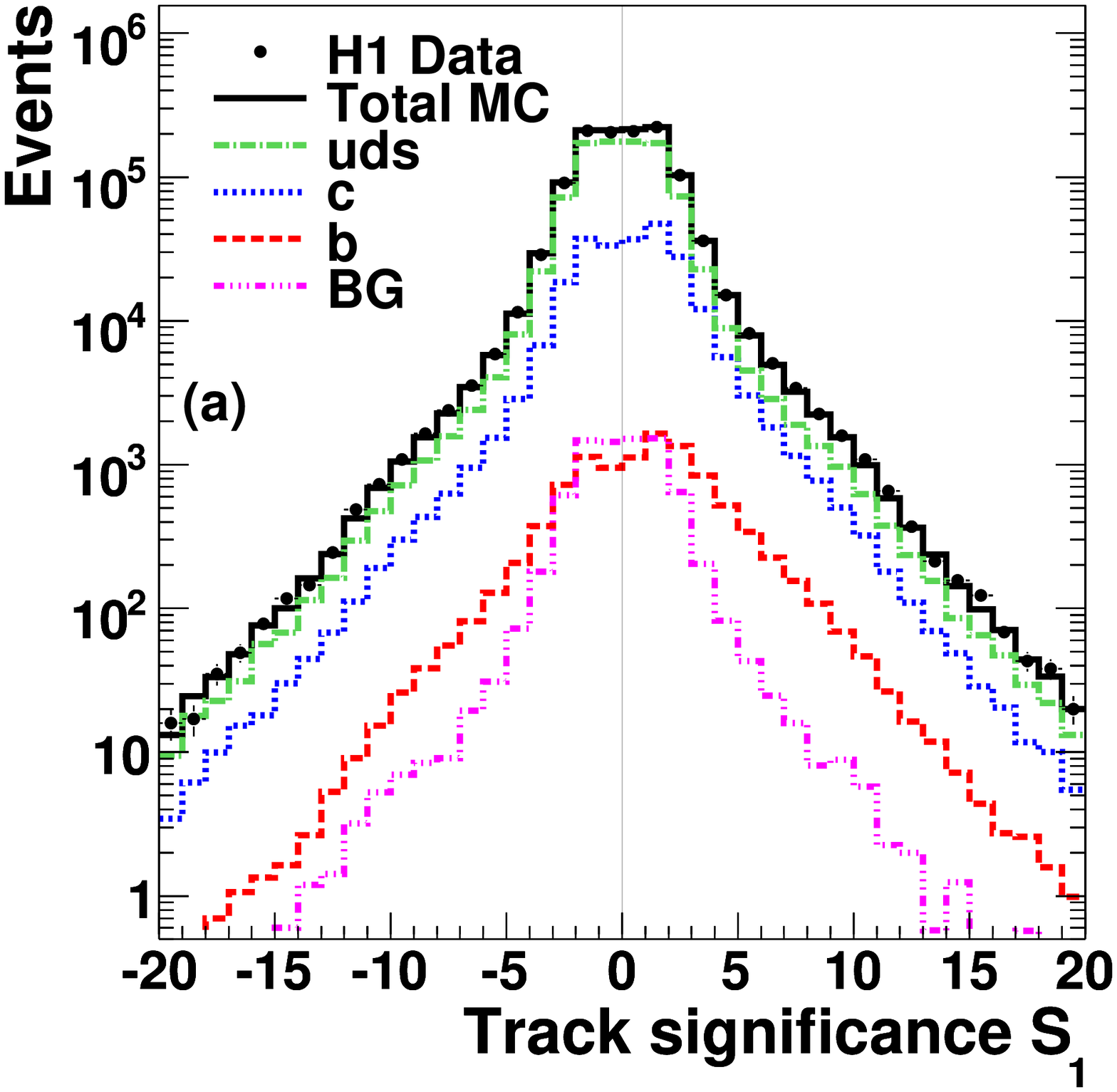}
 \includegraphics[width=0.49\textwidth]{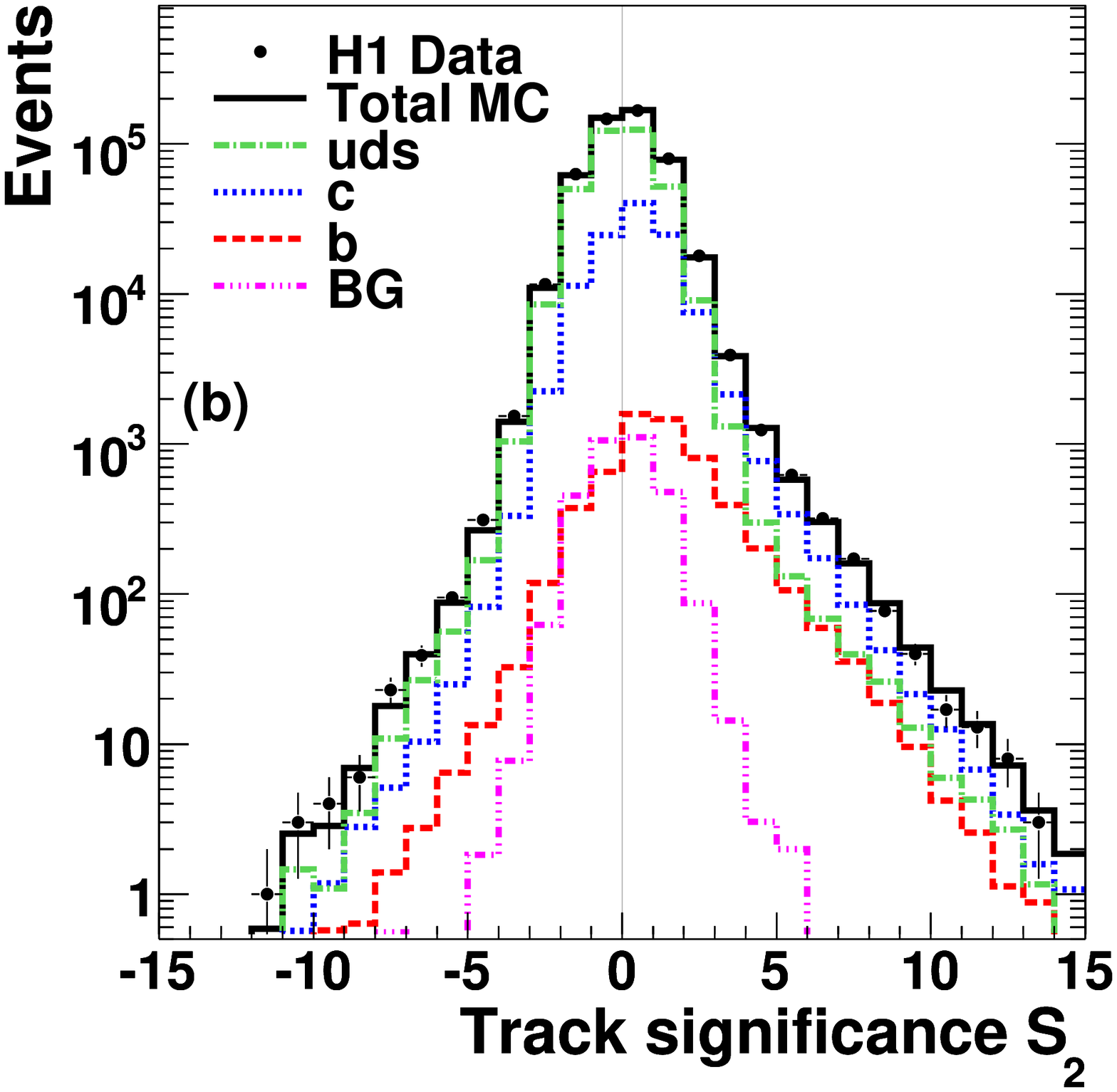}
 \includegraphics[width=0.49\textwidth]{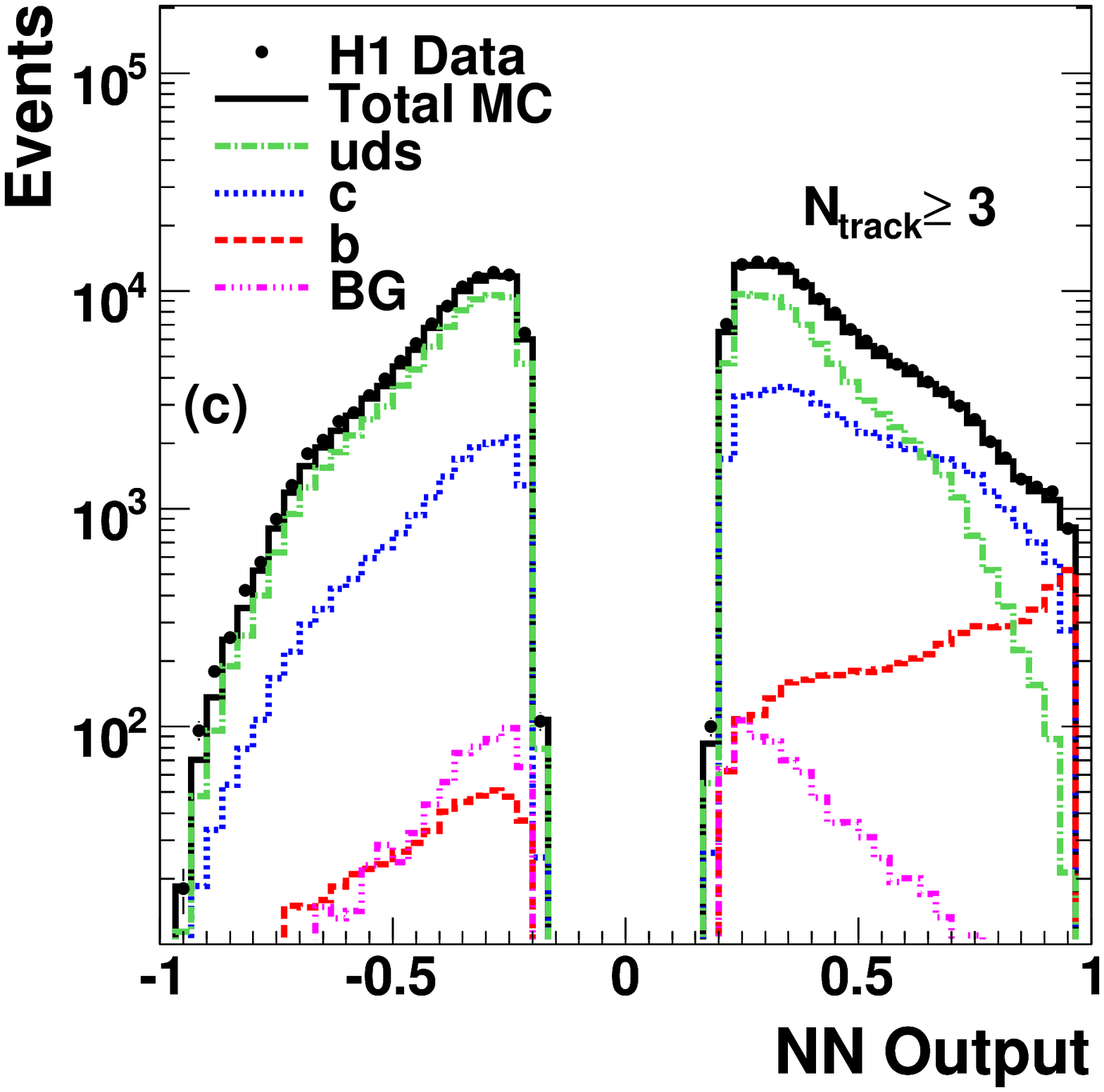}
\end{center}
 \caption{The
 significance distribution $S_1$ (a), $S_2$ (b) and the output of the
 neural network (NN Output) (c) for tracks of the highest transverse energy jet in the event.  Included in the
 figure is the expectation from the Monte Carlo simulation for $uds$ (light),
 $c$ and $b$ events. The contributions from the various quark flavours
 in the Monte Carlo simulation are shown after applying the scale factors
 $\rho_l$, $\rho_c$ and $\rho_b$, as described in the text.
The background (BG) contribution from a photoproduction Monte Carlo simulation
is also shown.}  \label{fig:s1s2nn}
\end{figure}

The three distributions that are used in the flavour separation are
shown in figure~\ref{fig:s1s2nn}.  It can be seen that the
distributions are asymmetric, mainly due to the tracks arising from
heavy flavour decays. The NN output gives absolute values in the range
from about $0.2$ to $0.95$.  The light jet distribution is approximately
symmetric and peaks towards low absolute values; the $c$ and $b$
distributions are asymmetric with more positive than negative entries;
the $b$ events are peaked towards $1$, whereas the $c$ events are
peaked towards $0$.  For the $S_1$, $S_2$ and NN output distributions the
data are well described by the Monte Carlo simulation and the
contribution from photoproduction is very small.

\begin{figure}[htb]

\begin{center}
\includegraphics[width=0.49\textwidth]{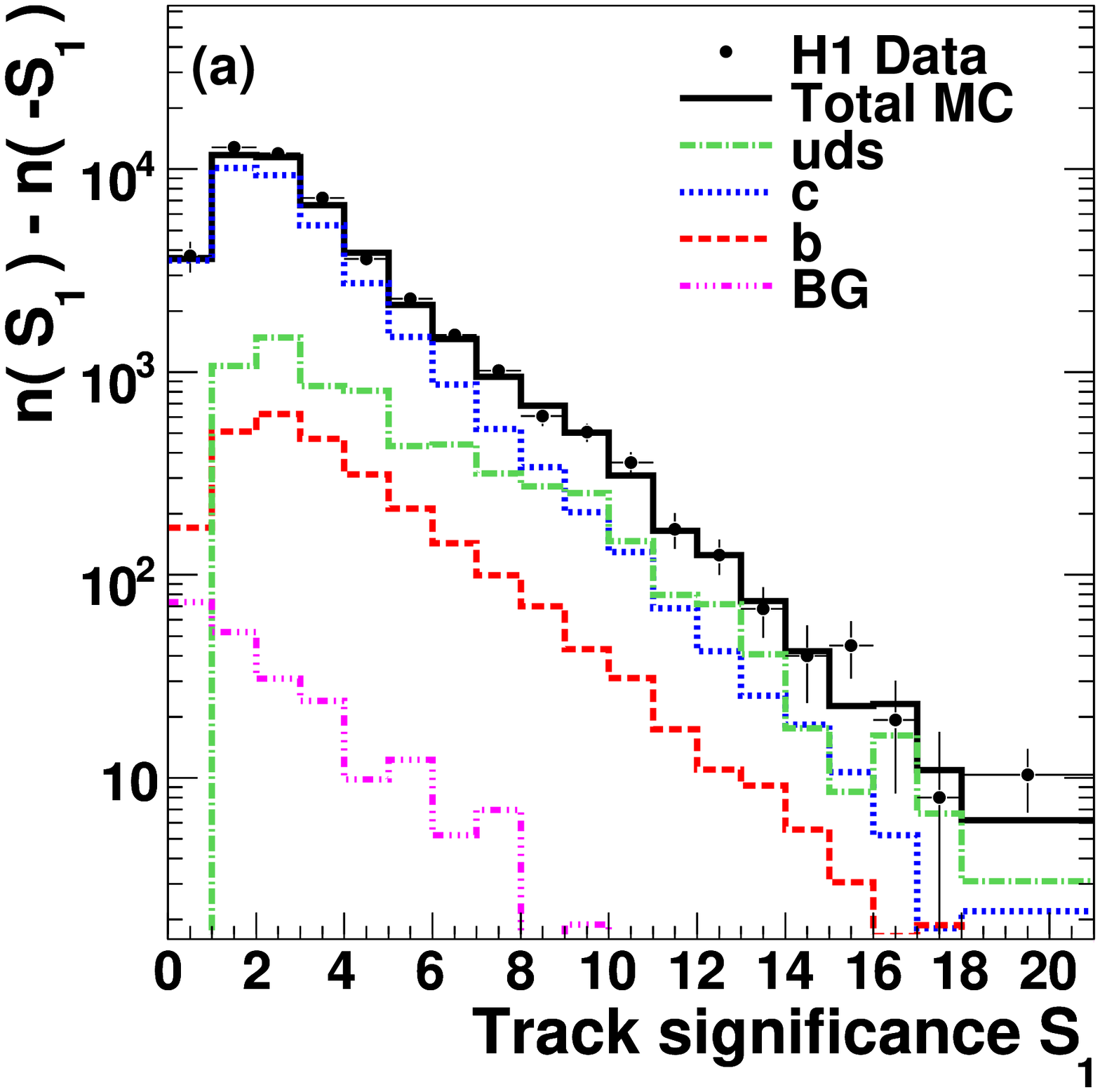}
 \includegraphics[width=0.49\textwidth]{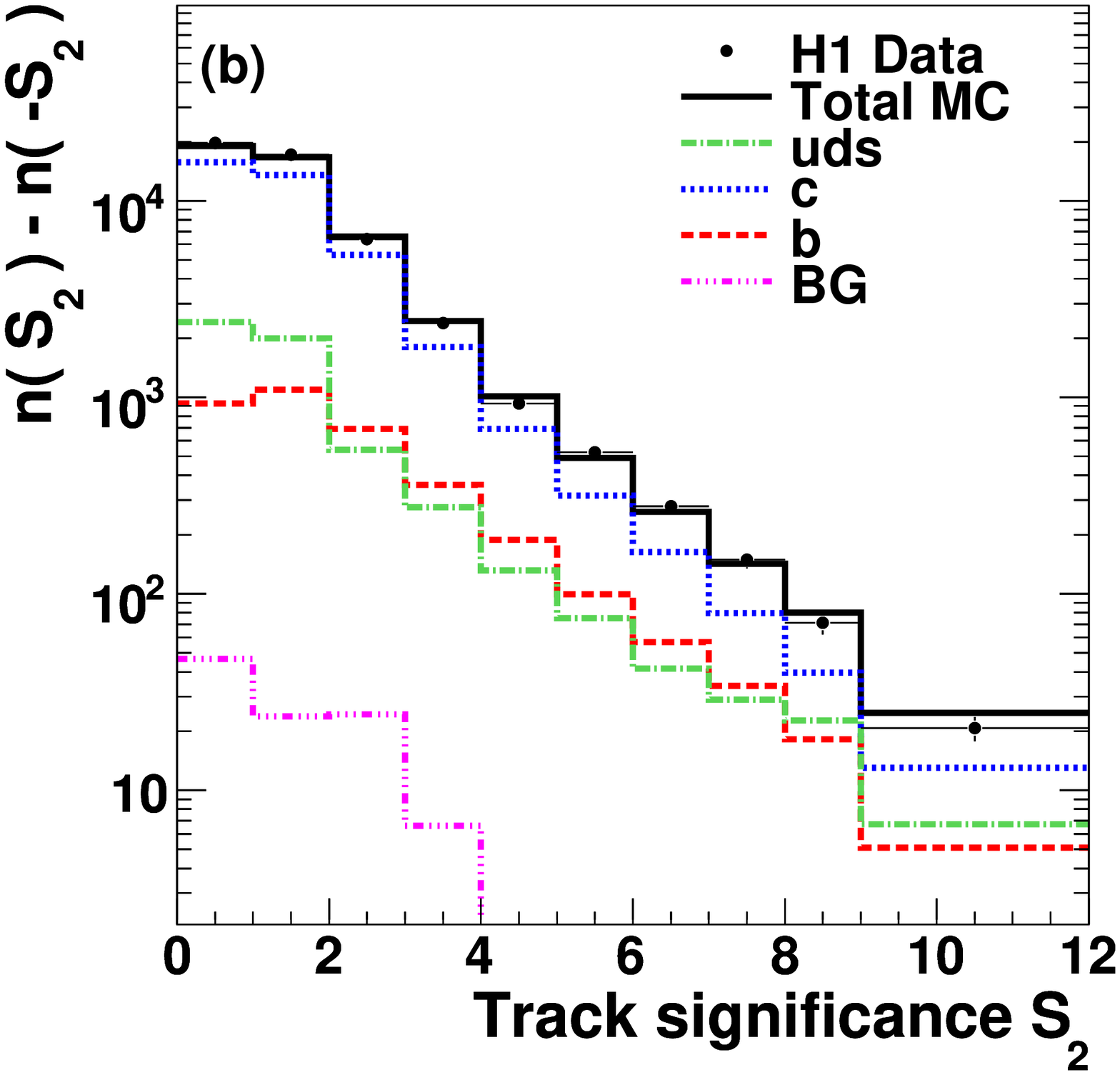}
 \includegraphics[width=0.49\textwidth]{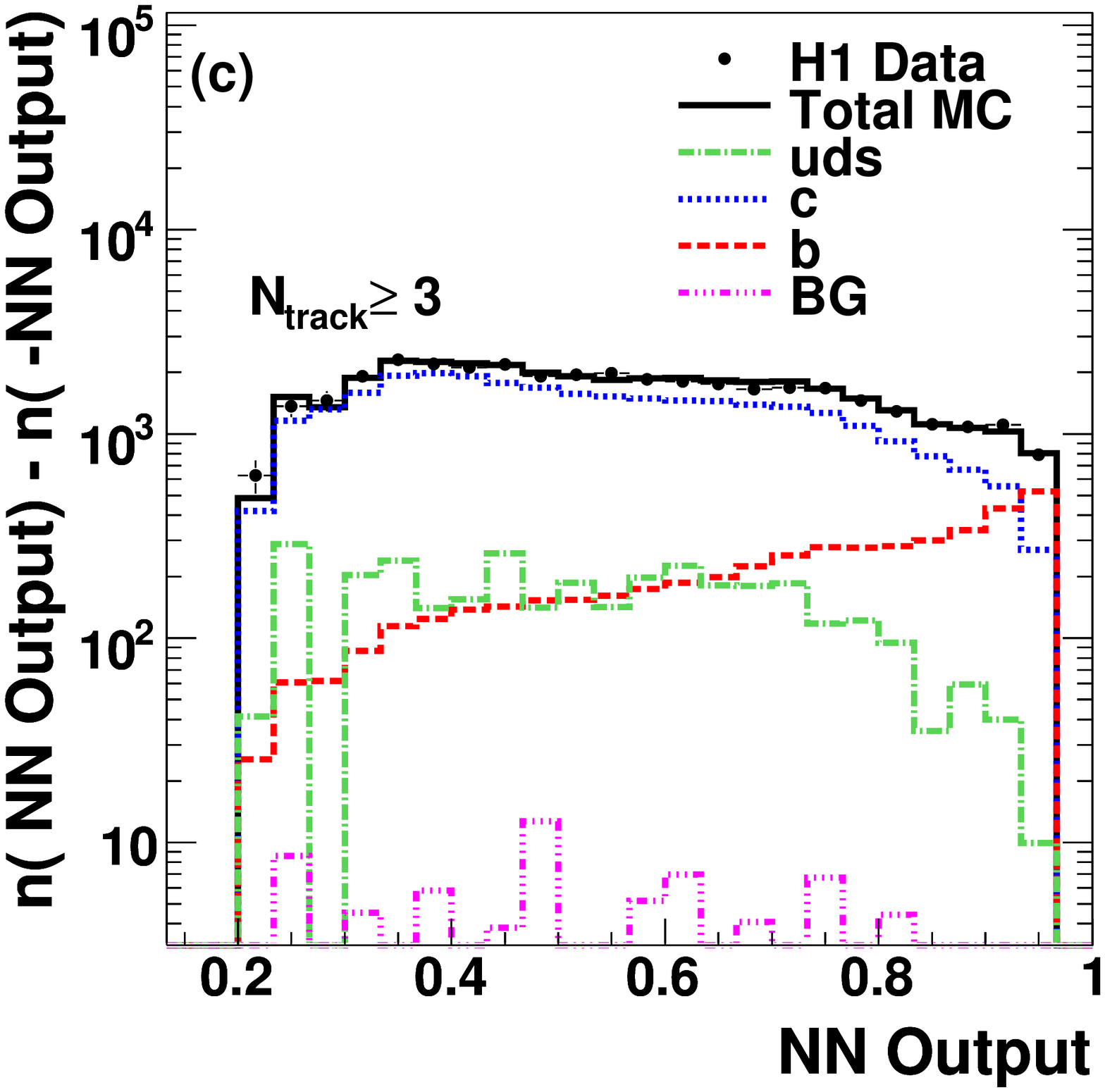}
\end{center}

 \caption{The subtracted
   distributions of $S_1$ (a), $S_2$ (b)
 and the neural network output (c) for the highest  transverse energy jet in 
the event.
  Included in the figure is
  the result from the fit to the data of the Monte Carlo simulation
  distributions 
of the $uds$ (light), $c$ and $b$ quark flavours to obtain the scale factors $\rho_l$, $\rho_c$ and $\rho_b$, as described in the text. 
The background (BG) contribution from a photoproduction 
Monte Carlo simulation is also shown.}  

  \label{fig:s1s2nnnegsub}

\end{figure}

Since the  $S_1$, $S_2$ and NN output distributions for light jets are nearly
symmetric around zero the sensitivity to the modelling of the light
jets can be reduced by subtracting the contents of the negative bins
from the contents of the corresponding positive bins. The subtracted
distributions are shown in figure~\ref{fig:s1s2nnnegsub}.  The resulting 
distributions are dominated by
$c$ jets, with a $b$ jet fraction increasing towards the
upper end of the distributions. Overall the light jets contribute only a
small fraction.

The fractions of events with $c$, $b$ and light jets in the data are extracted
using a least squares simultaneous fit to the subtracted $S_1$, $S_2$
and NN output distributions (as in figure~\ref{fig:s1s2nnnegsub}) and the
total number of events after DIS and jet selection. 
Only those bins in the significance
distributions which have at least $25$ events before subtraction are
considered in the fit, since Gaussian errors are assumed.  The last
fitted bin of the significance distributions, which usually has the
lowest statistics, is made $3$ times as wide as the other bins (see
figure~\ref{fig:s1s2nnnegsub}).  

The $uds$ (light), $c$ and $b$ RAPGAP Monte Carlo simulation samples are used
as templates.  The templates are scaled by factors $\rho_l$, $\rho_c$
and $\rho_b$, respectively, to give the best fit. The Monte Carlo samples
are weighted to the equivalent luminosity of the data sample so that the
$\rho$ scale factors are the ratio between the cross sections
of the Monte Carlo models and the data. The PYTHIA Monte
Carlo program is used to estimate photoproduction background
and found to be $0.8\%$ overall. The contributions of
light, $c$ and $b$ jets in photoproduction are fixed to the PYTHIA
prediction.  Only the statistical errors of the data and Monte Carlo
simulations are considered in the fit.

The fitted $\rho$ parameters for the whole kinematic range and 
for each of the differential distributions 
are listed in table~\ref{tab:rhotable}.  
The table includes the correlation coefficients of the fit parameters.
The fitted parameter $\rho_c$ is seen to be anti-correlated with both 
$\rho_l$ and $\rho_b$, due to $c$ jets being a significant contribution to the
total jet cross section.  The magnitude of the correlation 
$C_{lc}$ is greater than $C_{bc}$ 
reflecting the fact that the shapes of the Monte Carlo templates for $c$ jets are more 
similar to those for the light  jets than those for $b$ jets.
Also included in the table is the $\chi^2/$n.d.f. for each fit evaluated
using statistical errors only. Acceptable values are obtained for all
fits.

The fitted $\rho_c$ value for each bin is converted to a $c$ jet cross
section using
\begin{equation}
\sigma_c = \frac{\rho_c N^{\rm MC gen}_c}  { \mathcal{L} C_{\rm rad}},
\end{equation}
where $N^{\rm MC gen}_c$ is the number of generated events  that pass
the DIS kinematic selection of the bin and which contain a $c$ jet
passing the jet cuts of the bin at the hadron level, $\mathcal{L}$ is
the integrated luminosity of $189~{\rm pb^{-1}}$ and $C_{\rm rad}$ is
a radiative correction, calculated from the HERACLES Monte Carlo
program. The number of generated events $N^{\rm MC gen}_c$ is calculated
after normalising the luminosity of the Monte Carlo samples to that of the 
data as described above.
The $b$ cross sections are evaluated in a corresponding manner. 
The differential cross sections are obtained from the 
cross sections integrated over the bin interval by dividing by the
size of the bin interval, and no further bin centre correction is applied.  


\section{Systematic Uncertainties}
The following uncertainties are taken into account in order to evaluate
the systematic error.
\label{sec:systematics}
\begin{itemize}
\item The uncertainty in the $\delta$ resolution of the tracks is estimated
 by varying the resolution by an amount that encompasses any
 difference between the data and the simulation.  This was achieved by
 applying an additional Gaussian smearing in the Monte Carlo
 simulation of $200$~$\mu{\rm m}$ to $5\%$ of randomly selected tracks
 and $12$~$\mu{\rm m}$ to the rest.

\item The uncertainty due to the track efficiency uncertainty is estimated by 
varying the  efficiency of the CTD by $\pm 1\%$ and that of the CST by $\pm 2\%$.

\item The uncertainties on the various $D$ and $B$ meson lifetimes,
  decay branching fractions and mean charge multiplicities are
  estimated by varying the input values of the Monte Carlo simulation
  by the errors on the world average measurements.  For the branching
  fractions of $b$ quarks to hadrons and the lifetimes of the $D$ and
  $B$ mesons the central values and errors on the world averages are
  taken from\cite{pdg2006}. For the branching fractions of $c$
  quarks to hadrons the values and uncertainties are taken
  from the $e^+e^-$ average of \cite{Gladilin:1999pj}, which are consistent with measurements
  made in DIS at HERA\cite{Aktas:2004ka}. For the mean charged track multiplicities the
  values and uncertainties for $c$ and $b$ quarks are taken from
  MarkIII\cite{Coffman:1991ud} and LEP/SLD\cite{lepjetmulti}
  measurements, respectively. 

\item 
The uncertainty on the fragmentation function of the heavy quarks is
estimated by reweighting the events according to the longitudinal
string momentum fraction $z$ carried by the heavy hadron in the Lund
model using weights of $(1 \mp 0.7)\cdot(1-z) + z \cdot (1 \pm 0.7)$
for charm quarks and by $(1 \mp 0.5)\cdot(1-z) + z \cdot (1 \pm 0.5)$
for beauty quarks. The variations for the charm fragmentation 
are motivated by encompassing the differences between  
the Monte Carlo simulation and H1 $D^*$ data~\cite{h1frag}. The
size of the variations is reduced for beauty compared with charm 
since the fragmentation spectrum is harder.

\item The uncertainty on the QCD model of heavy quark production
 is estimated by reweighting the jet transverse momentum and
 pseudorapidity by $(E^{\rm jet}_T/(10 ~{\rm GeV}))^{\pm 0.2}$ and $(1 \pm
 \eta^{\rm jet})^{\pm 0.15}$ for charm jets and $(E^{\rm jet}_T/(10
 ~{\rm GeV}))^{\pm 0.3}$ and $(1 \pm \eta^{\rm jet})^{\pm 0.3}$ for beauty jets. These values are obtained by comparing these variations with the
 measured cross sections.

\item The uncertainty on the asymmetry of the light jet $\delta$
  distribution is estimated by repeating the fits with the subtracted
  light jet distributions (figure~\ref{fig:s1s2nnnegsub}) changed by
  $\pm30\%$. The light jet asymmetry was checked to be within this
  uncertainty by comparing the asymmetry of Monte Carlo simulation
  events to that of the data for $K^0$ candidates, in the region
  $0.1<|\delta|<0.5~{\rm cm}$, where the light jet asymmetry is
  enhanced.

\item The uncertainty on the reconstruction of  $\phi_{\rm jet}$ is estimated
by shifting its value by $\pm 2^\circ$.  The uncertainty was
evaluated by comparing the distribution of the difference between
$\phi_{\rm jet}$ and the track azimuthal angle in data and Monte
Carlo simulation.

\item The uncertainty arising from
 the hadronic energy scale is estimated by changing
the hadronic energy by $\pm 2\%$ for jets in the laboratory 
and $\pm 4\%$ for the jets in the Breit frame. 

\item The uncertainty arising from the electron energy scale and polar angle is estimated by changing the electron energy by $\pm 1\%$ and the polar angle by $\pm 1~{\rm mrad}$.

\item The uncertainty in the photoproduction background is estimated by
varying the expected number of events by $\pm 100\%$.  

\item The uncertainty on the luminosity is $4\%$.

\item The uncertainty on the radiative correction is $2\%$.

\end{itemize}

The above systematic uncertainties are evaluated by making the changes
described above to the Monte Carlo simulation and repeating the
procedure to evaluate the $c$ and $b$ cross sections, including the
fits.  The uncertainties are evaluated separately for each
measurement bin and are treated as correlated except for the radiative corrections.

The most important sources of systematic error for
the charm jets are the uncertainty on the light jet contribution, the
uncertainty of the impact parameter resolution and the contribution of
the uncorrelated errors. For the beauty jets, the systematic
uncertainties are considerably larger with the main sources of
uncertainty being those due to the multiplicity of $b$ quark decays,
the track efficiency, the hadronic energy scale and the impact
parameter resolution.

\section{Results}
\label{sec:results}
The cross sections for $c$ and $b$ jets are presented 
in the laboratory frame of reference (section \ref{sec:results:lab})
and in the Breit frame (section \ref{sec:results:breit}).
The $b$ jet data are also compared with measurements obtained
from muon tagging (section \ref{sec:results:muon}).
The cross sections for events with $c$ or $b$ jets are
shown together with theoretical predictions in table~\ref{tab:sigtheory}.
The cross section values for all the measurements are given in
table~\ref{tab:sig} with the contribution of
the systematic errors for each measurement listed in table~\ref{tab:sys}.

\subsection{Jet Cross Sections in the Laboratory Frame}
\label{sec:results:lab}

The  jet cross sections in the laboratory  frame 
are measured in the kinematic range
$Q^2>6~{\rm GeV}^2$ and $0.07<y<0.625$ for the heavy flavour jet 
with the highest $E_T^{\rm jet}$ with
$E_T^{\rm jet}>6~{\rm GeV}$ and $-1.0<\eta^{\rm jet}<1.5$. 
 The hadron level $c$ and $b$ cross sections with jets are 
\begin{equation*}
3290 \pm 50({\rm stat.})  \pm 260({\rm syst.}) ~{\rm pb}
\end{equation*}
and
\begin{equation*}
189 \pm 9({\rm stat.}) \pm 42({\rm syst.})~{\rm pb}, 
\end{equation*}
respectively. Here
the first error is statistical and the second is systematic.

These cross sections are compared in table~\ref{tab:sigtheory} to the
expectations of the Monte Carlo programs RAPGAP and CASCADE as well as
to the NLO predictions 
with HVQDIS including hadronisation corrections. The NLO predictions
are given for three different sets of PDFs and two different scale
choices, $\mu = \sqrt{(Q^2+p_T^2+m^2)/2}$ and $\mu = \sqrt{Q^2+4m^2}$.
Overall, RAPGAP agrees well with data for both charm and
beauty. CASCADE predicts a significantly larger $c$ cross section than
observed in data while for beauty the discrepancy is much
reduced. Within uncertainties the NLO predictions agree reasonably
well with the data both for charm and beauty. In general the NLO
expectations for beauty display a smaller dependence on scale than for charm.

\begin{figure}[htb]
  \begin{center} 
  \includegraphics[width=0.49\textwidth]{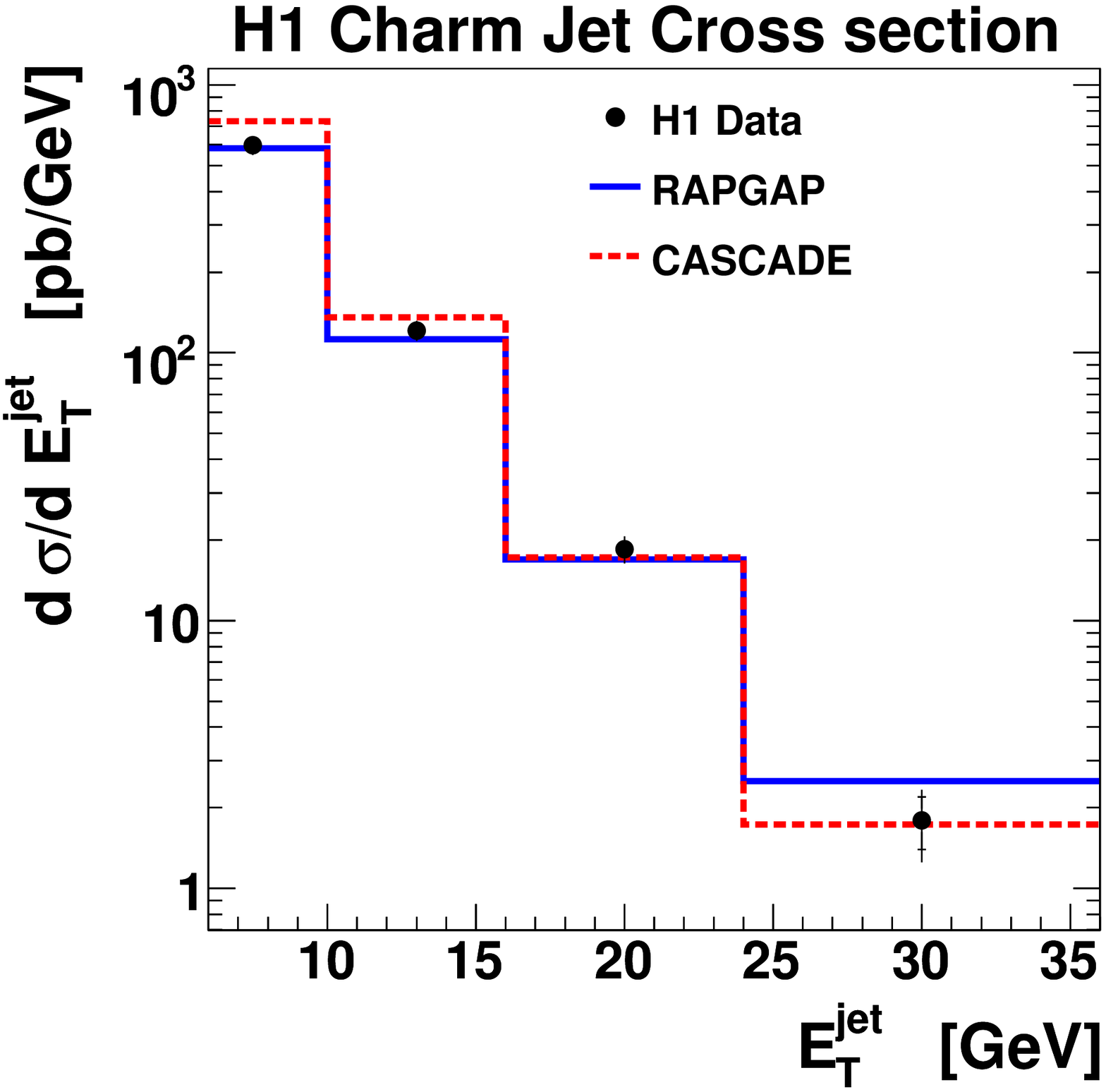}
  \includegraphics[width=0.49\textwidth]{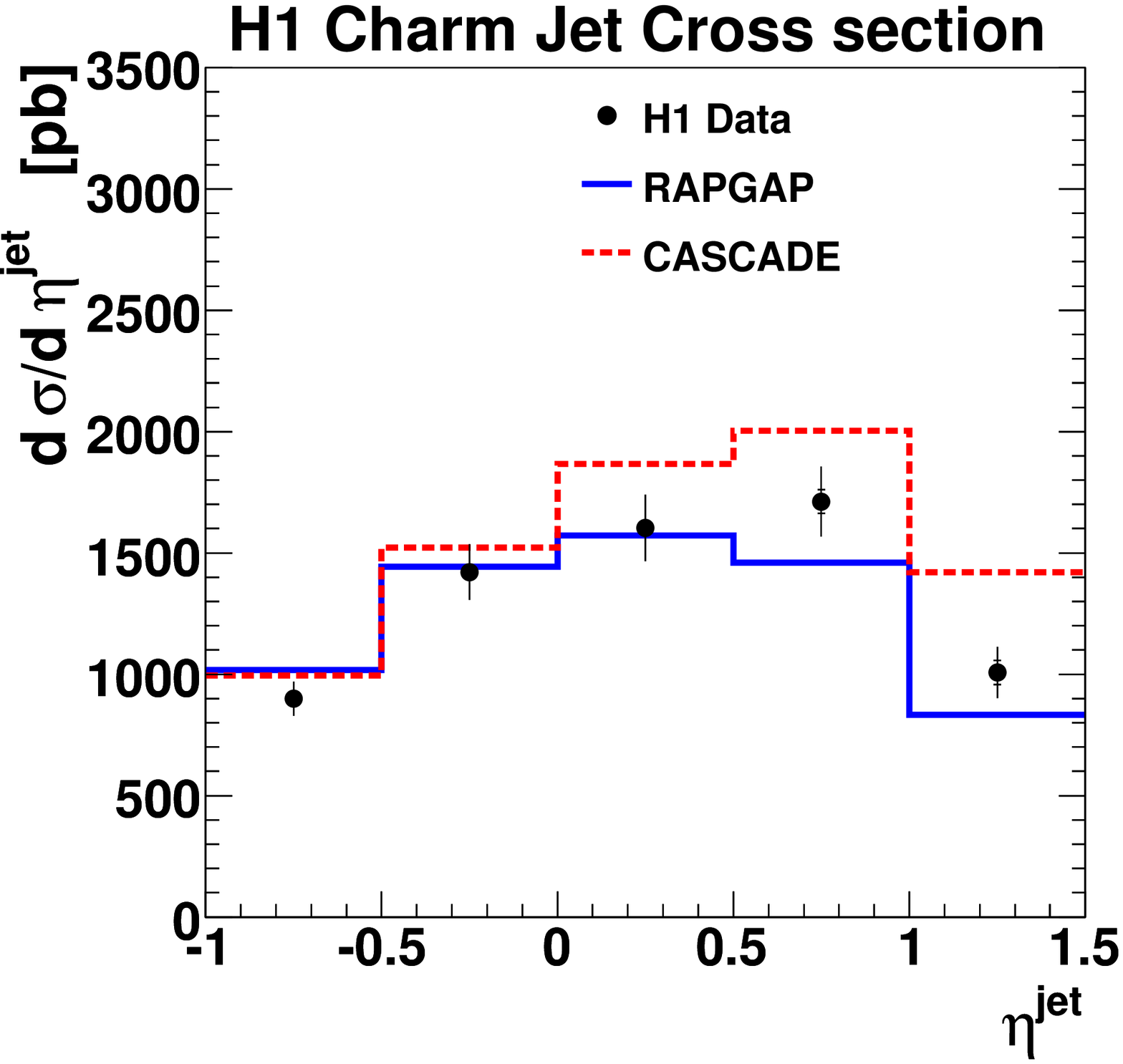}
  \includegraphics[width=0.49\textwidth]{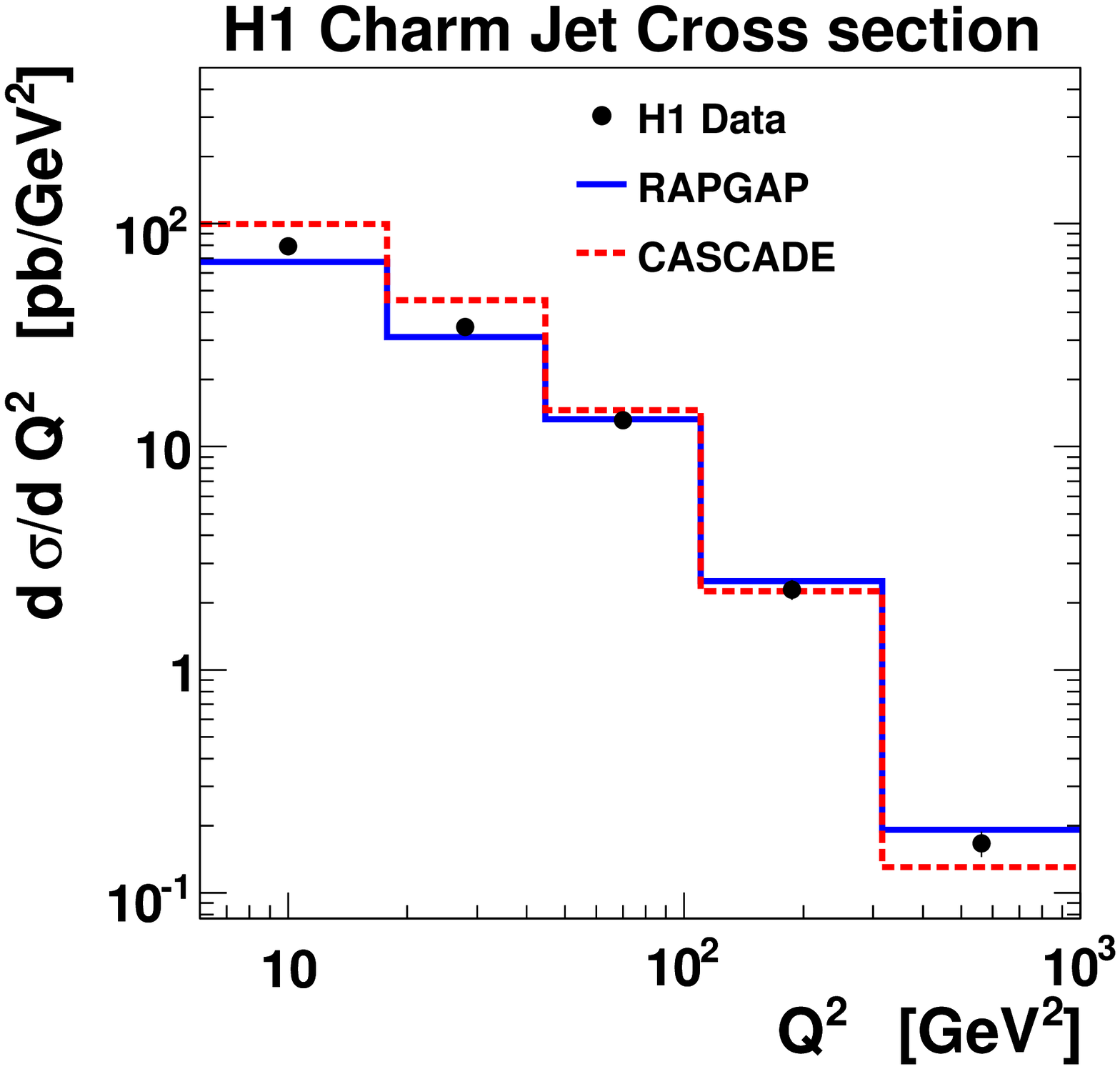}
  \includegraphics[width=0.49\textwidth]{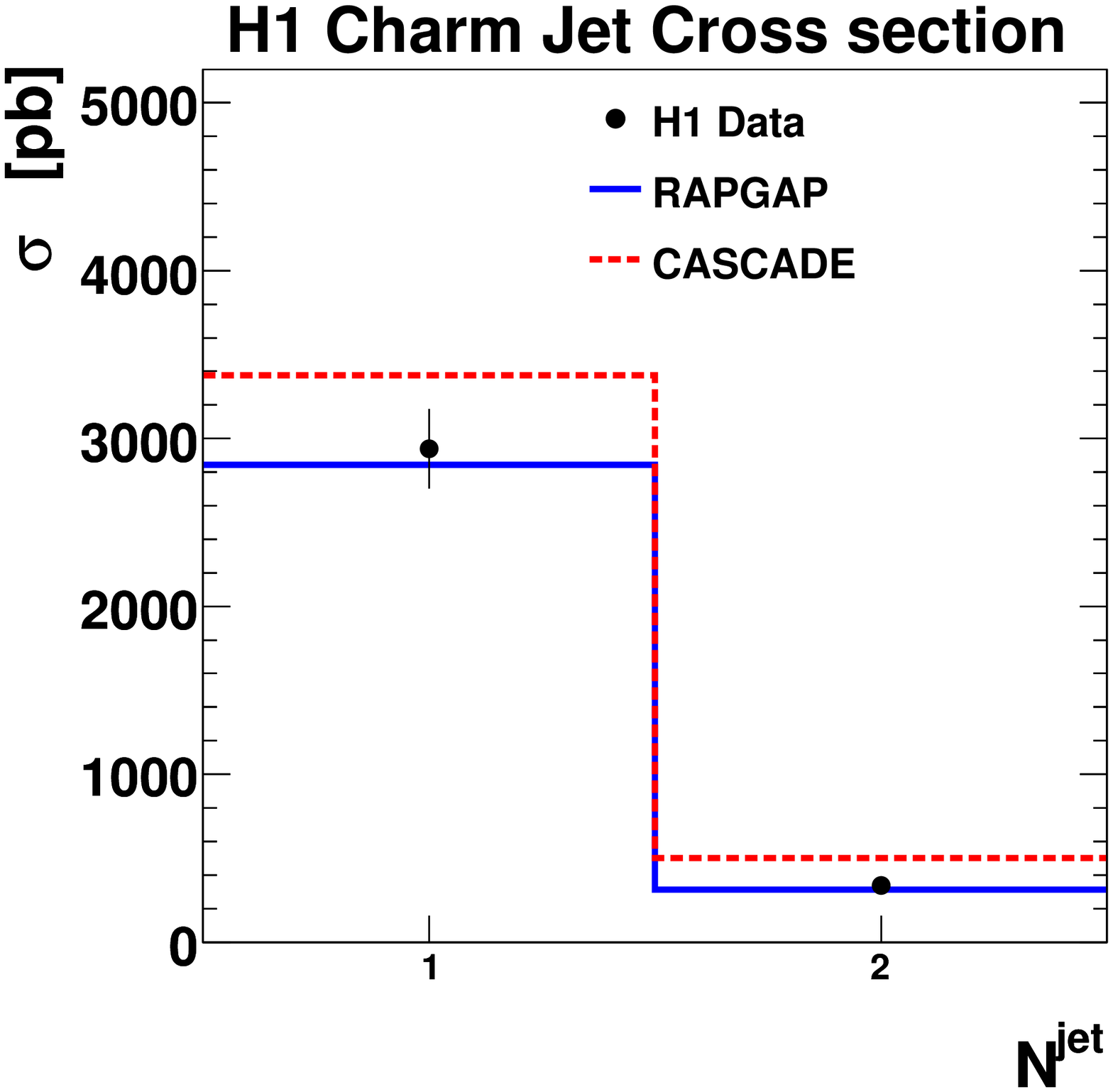}
\caption{
The differential cross sections for
the highest transverse energy charm jet in the laboratory frame 
as a function of $E_T^{\rm jet}$, $\eta^{\rm jet}$, $Q^2$ and
the number of laboratory frame jets in the event $N^{\rm jet}$. 
The measurements are made for the kinematic range
$E_T^{\rm jet}>6~{\rm GeV}$, $-1<\eta^{\rm jet}<1.5$, $Q^2>6~{\rm GeV}^2$ and $0.07 < y <0.625$.
The inner error bars show the statistical error, the
outer error bars represent the statistical and systematic errors
added in quadrature. 
The data are compared with the predictions from
the Monte Carlo models RAPGAP and CASCADE.}
\label{fig:xscmc} 
\end{center}
\end{figure}

Differential $c$ and $b$ jet cross sections are measured as a
function of $E_T^{\rm jet}$, $\eta^{\rm jet}$, $Q^2$, and the number of
jets $N^{\rm jet}$ with $E_T^{\rm jet}>6~{\rm GeV}$ (table \ref{tab:sig}).
The differential $c$ cross sections 
are shown in figure~\ref{fig:xscmc} in comparison to Monte Carlo
expectations.  The RAPGAP model describes all these distributions
reasonably well in shape and normalisation. CASCADE exceeds the data
at low $E_T^{\rm jet}$ as well as at low $Q^2$ but provides a good
description at high $Q^2$ and high $E_T^{\rm jet}$, respectively. The
excess of CASCADE is concentrated in the forward $\eta^{\rm jet}$
region.  As expected from the visible cross section given in
table~\ref{tab:sigtheory}, CASCADE lies above the data in the $N^{\rm
jet}$ distribution. The model does, however, give a reasonable description of
the $\eta^{\rm jet}$ distribution after accounting for the difference in normalisation.

\begin{figure}[htb]
  \begin{center} 
  \includegraphics[width=0.49\textwidth]{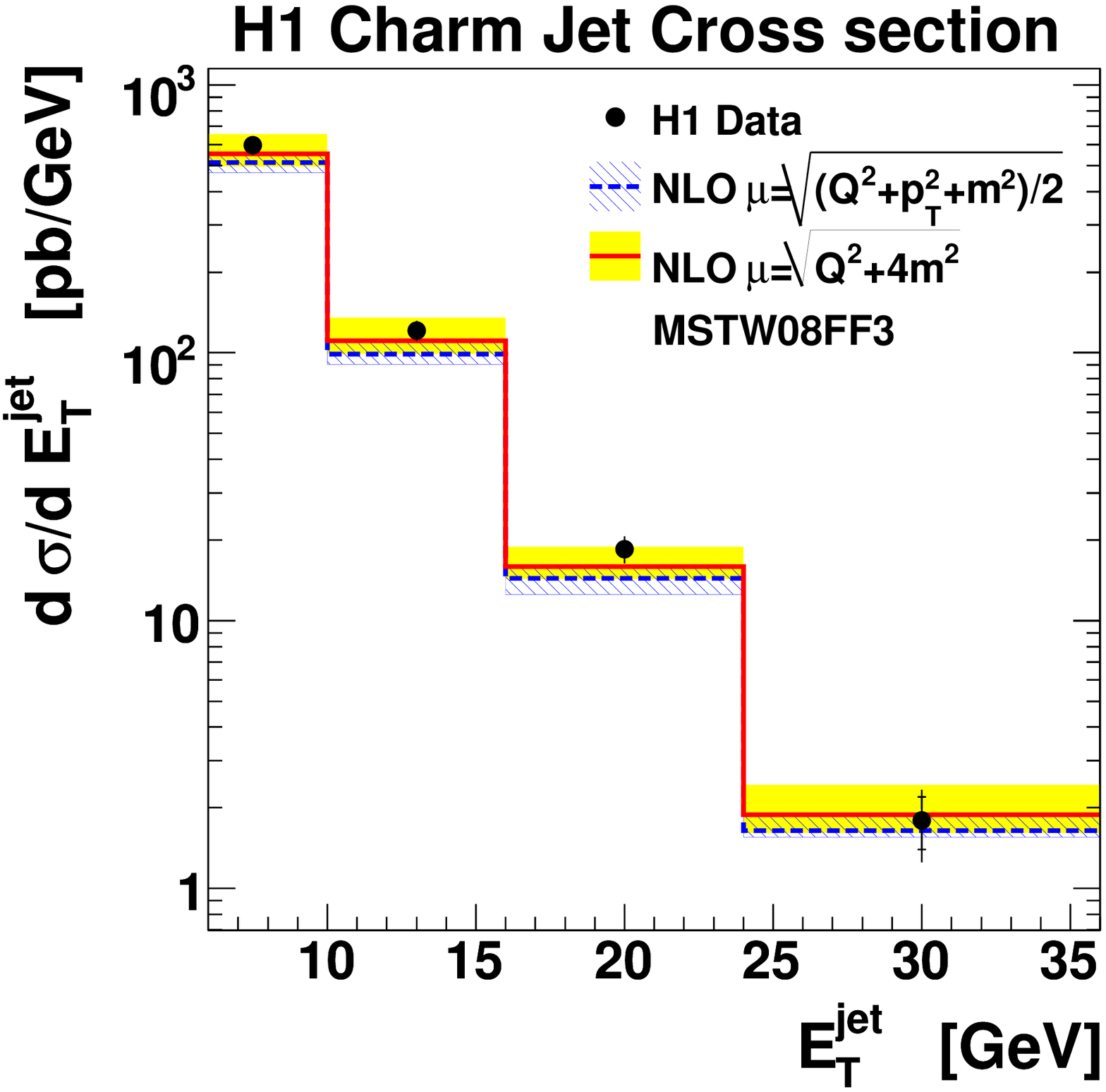}
  \includegraphics[width=0.49\textwidth]{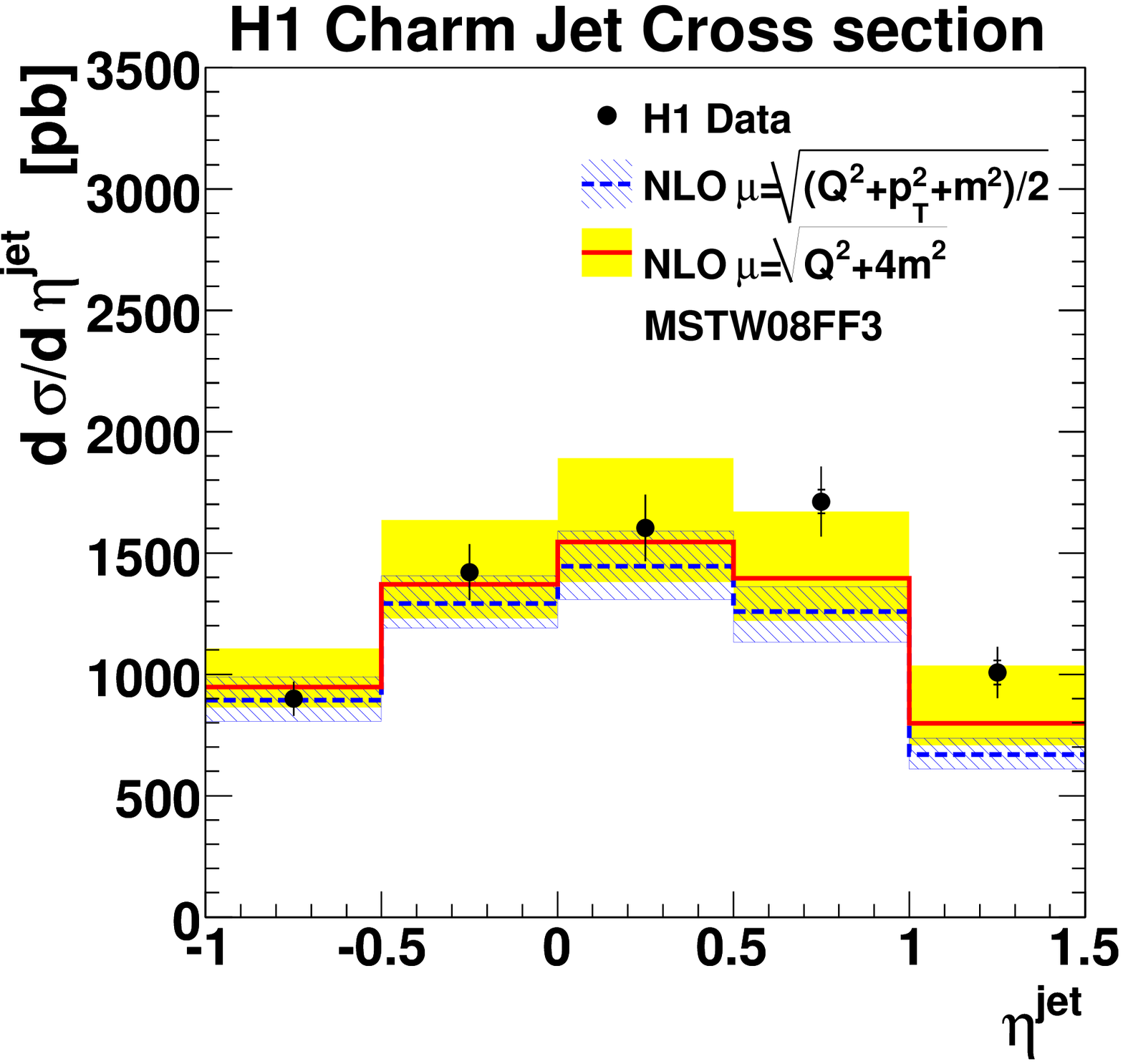} 
  \includegraphics[width=0.49\textwidth]{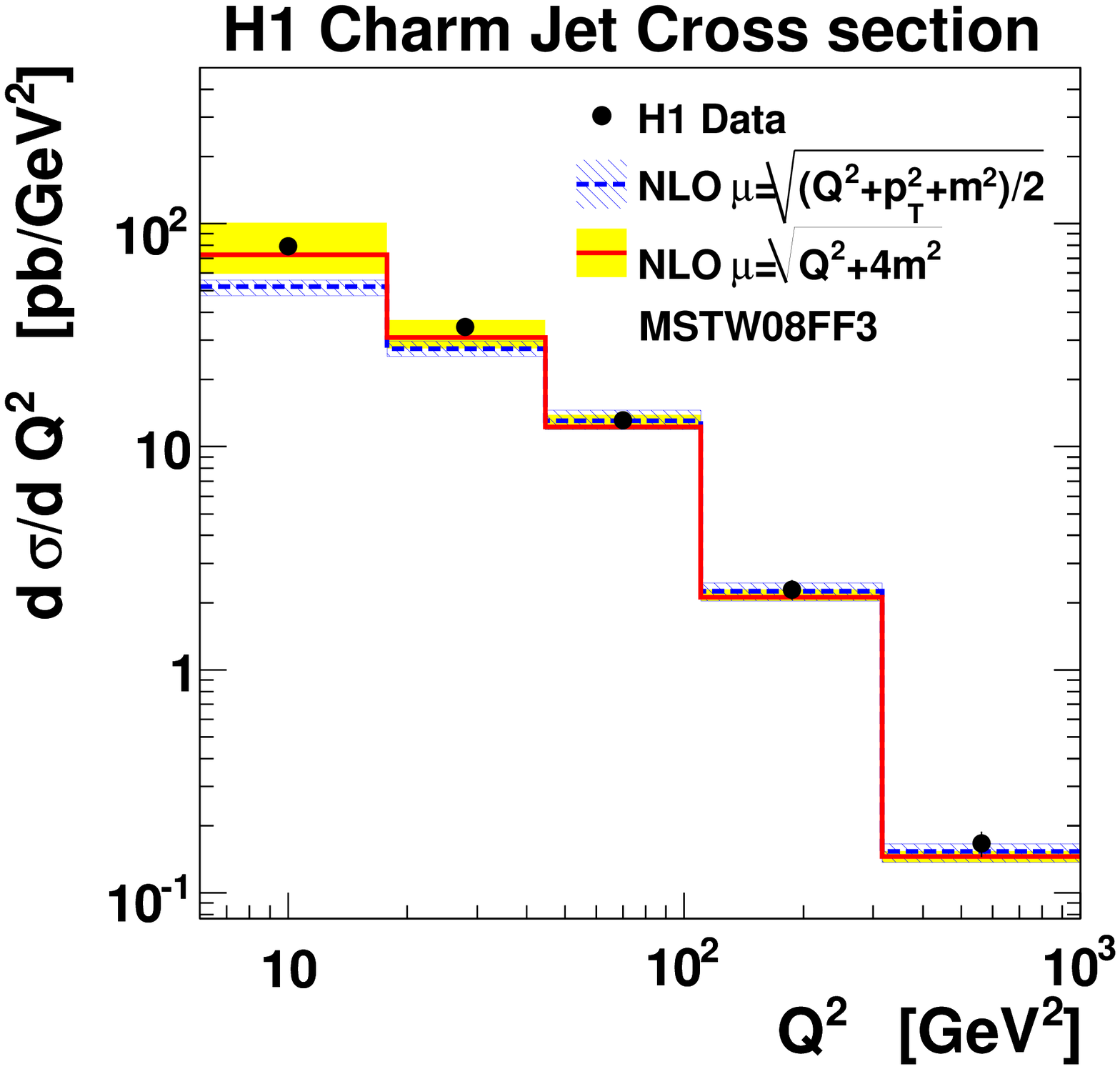}
  \includegraphics[width=0.49\textwidth]{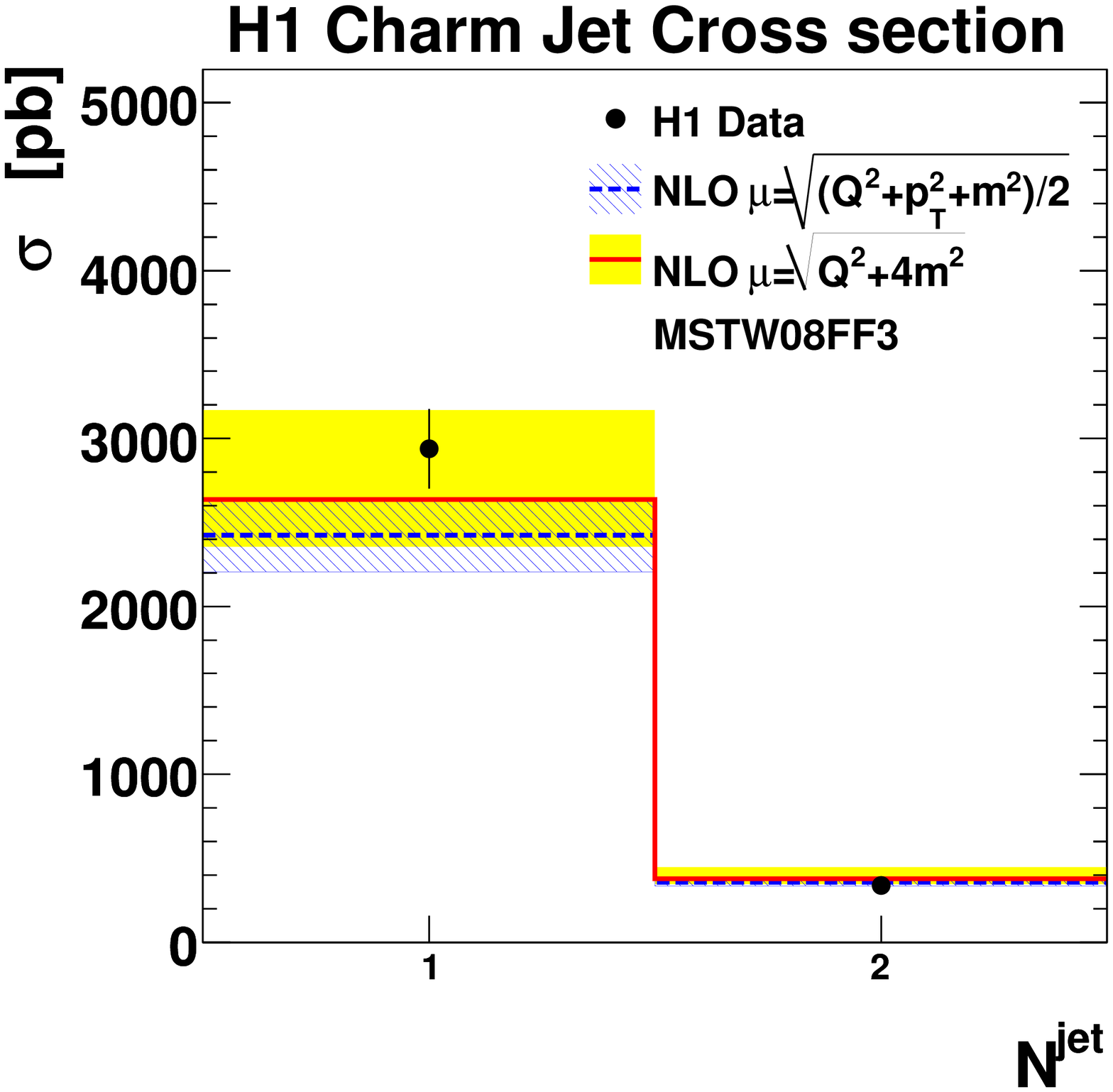} 
\caption{
The differential cross sections for
the highest transverse energy charm jet in the laboratory frame 
as a function of $E_T^{\rm jet}$, $\eta^{\rm jet}$, $Q^2$ and
the number of laboratory frame jets in the event $N^{\rm jet}$. 
The measurements are made for the kinematic range
$E_T^{\rm jet}>6~{\rm GeV}$, $-1<\eta^{\rm jet}<1.5$, $Q^2>6~{\rm GeV}^2$ and $0.07 < y <0.625$.
The inner error bars show the statistical error, the
outer error bars represent the statistical and systematic errors
added in quadrature. 
The data are compared with the predictions from NLO QCD
where the bands indicate the theoretical uncertainties.}
\label{fig:xscnlo} 
\end{center}
\end{figure}

The charm jet cross section measurements are shown in figure~\ref{fig:xscnlo}
together with the NLO
predictions of HVQDIS. In general the NLO
expectations describe the data reasonably well in all differential
distributions although the predictions with the scale $\mu =
\sqrt{(Q^2+p_T^2+m^2)/2}$ fall somewhat below the data at low $Q^2$, low
 $E_T^{\rm jet}$ and in the forward $\eta^{\rm jet}$ region.

\begin{figure}[htb]
  \begin{center} 
  \includegraphics[width=0.49\textwidth]{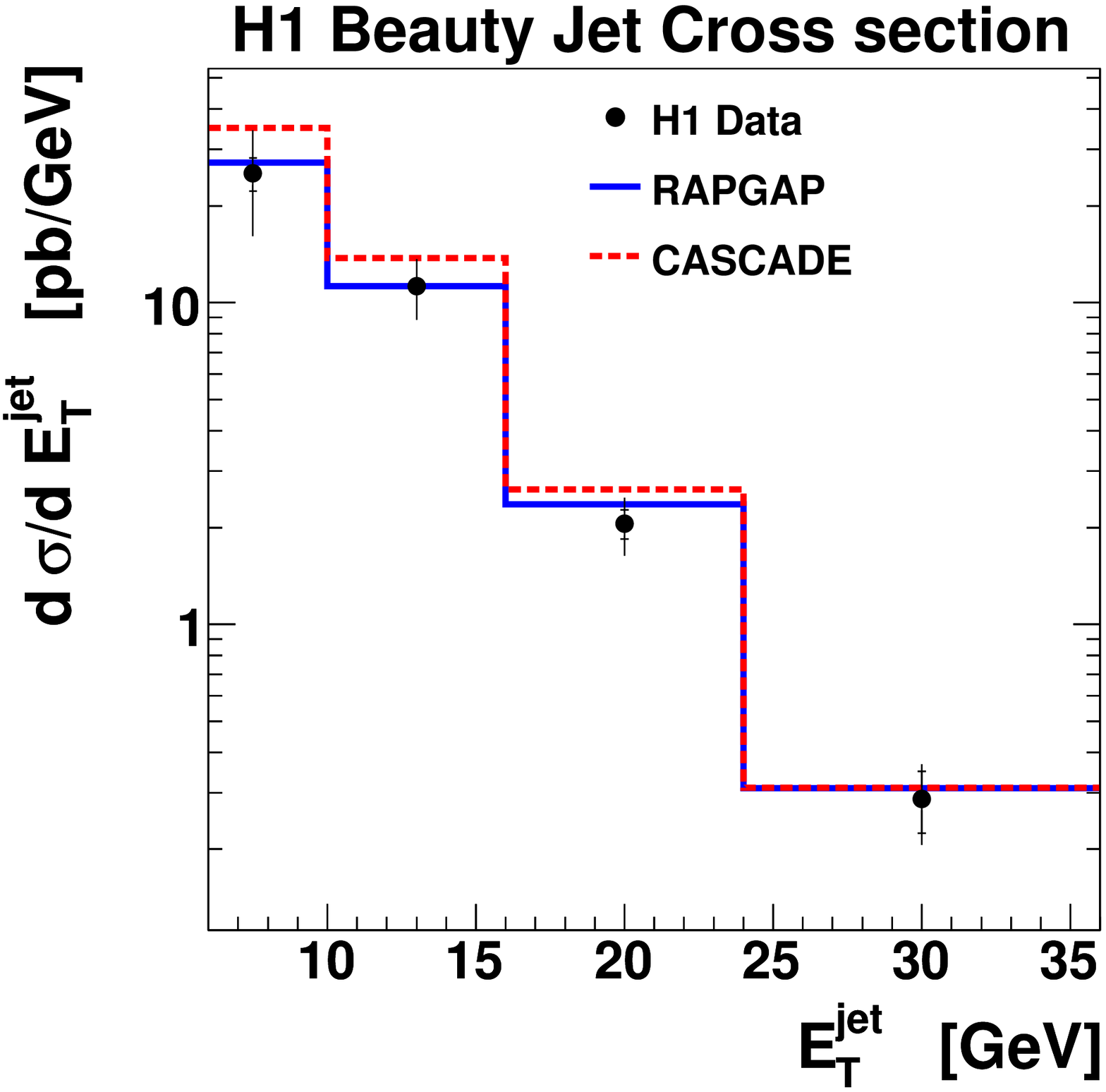}
  \includegraphics[width=0.49\textwidth]{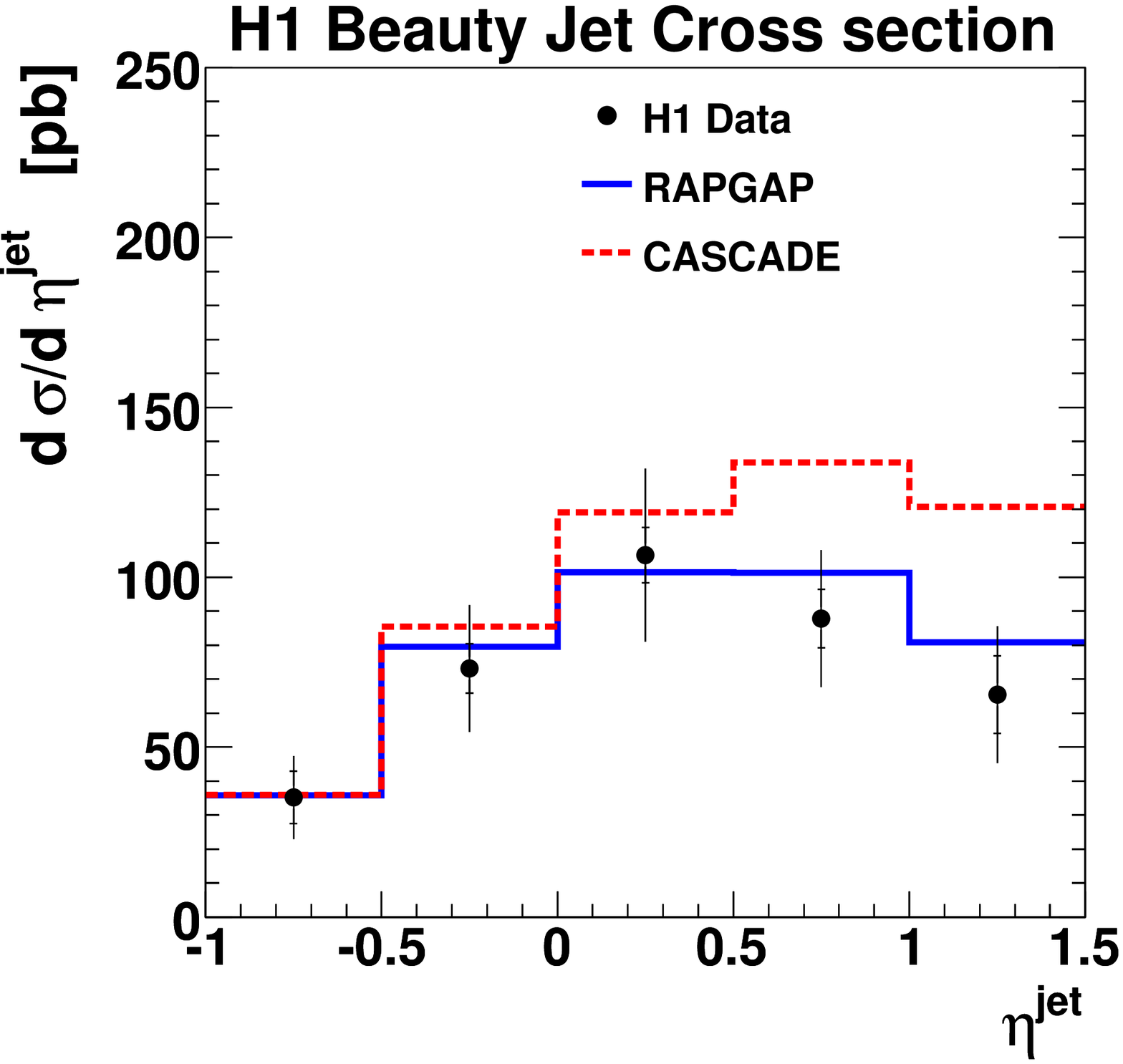}
  \includegraphics[width=0.49\textwidth]{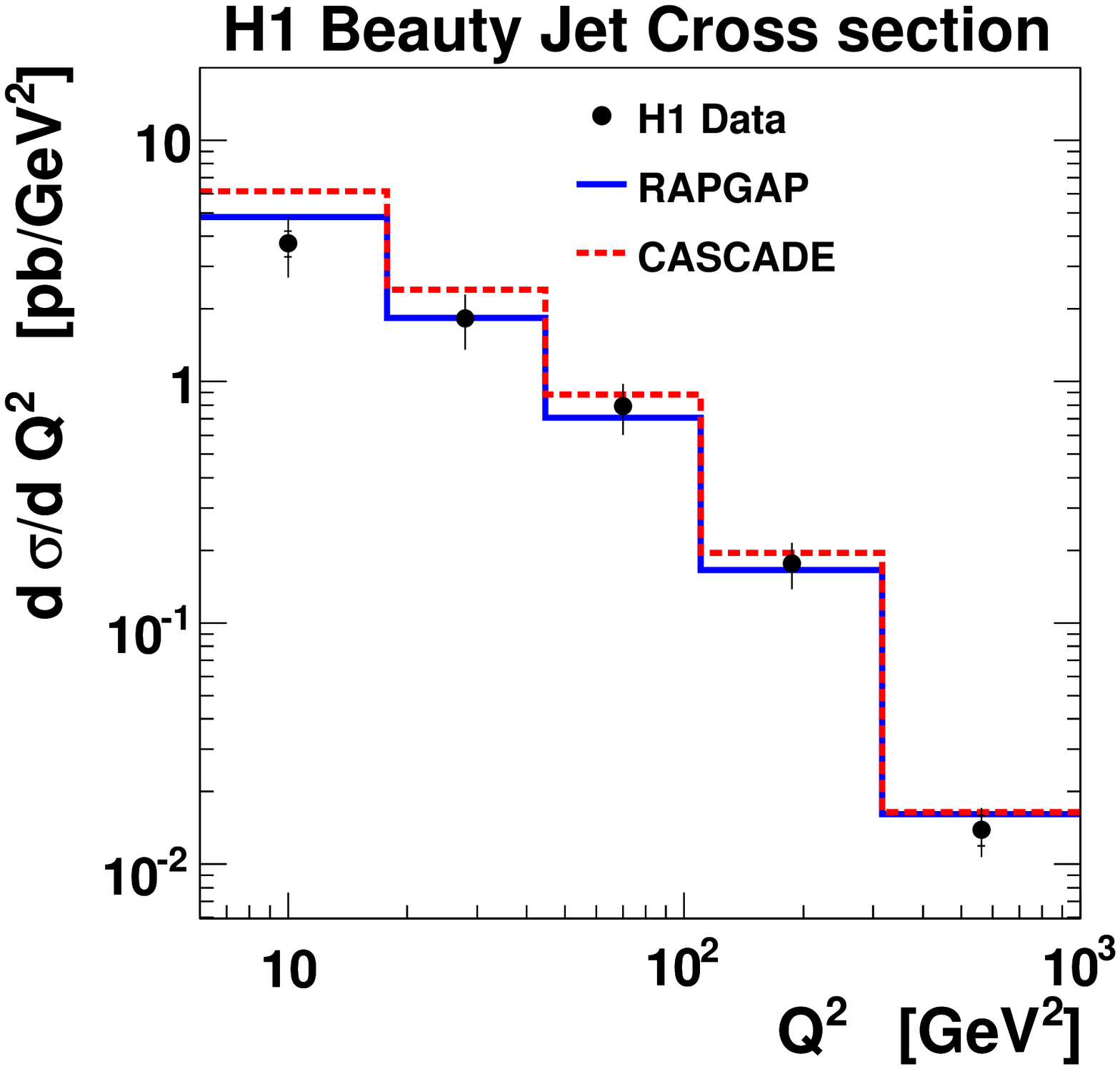}
  \includegraphics[width=0.49\textwidth]{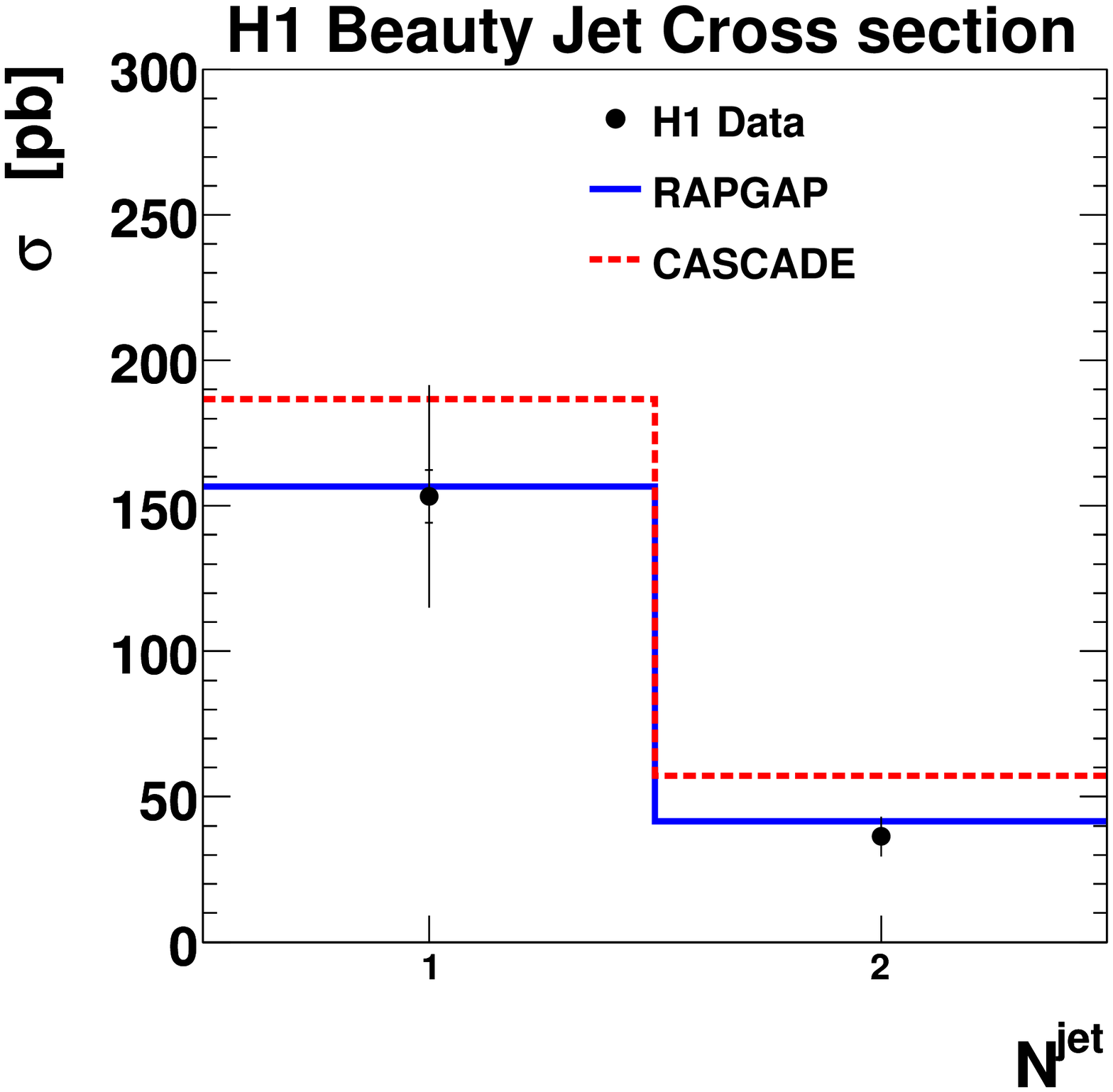}
\caption{
The differential cross sections for
the highest transverse energy beauty jet in the laboratory frame 
as a function of $E_T^{\rm jet}$, $\eta^{\rm jet}$, $Q^2$ and
the number of laboratory frame jets in the event $N^{\rm jet}$. 
The measurements are made for the kinematic range
$E_T^{\rm jet}>6~{\rm GeV}$, $-1<\eta^{\rm jet}<1.5$, $Q^2>6~{\rm GeV}^2$ and $0.07 < y <0.625$.
The inner error bars show the statistical error, the
outer error bars represent the statistical and systematic errors
added in quadrature. 
The data are compared with the predictions from
the Monte Carlo models RAPGAP and CASCADE.}
\label{fig:xsbmc} 
\end{center}
\end{figure}

\begin{figure}[htb]
  \begin{center} 
  \includegraphics[width=0.49\textwidth]{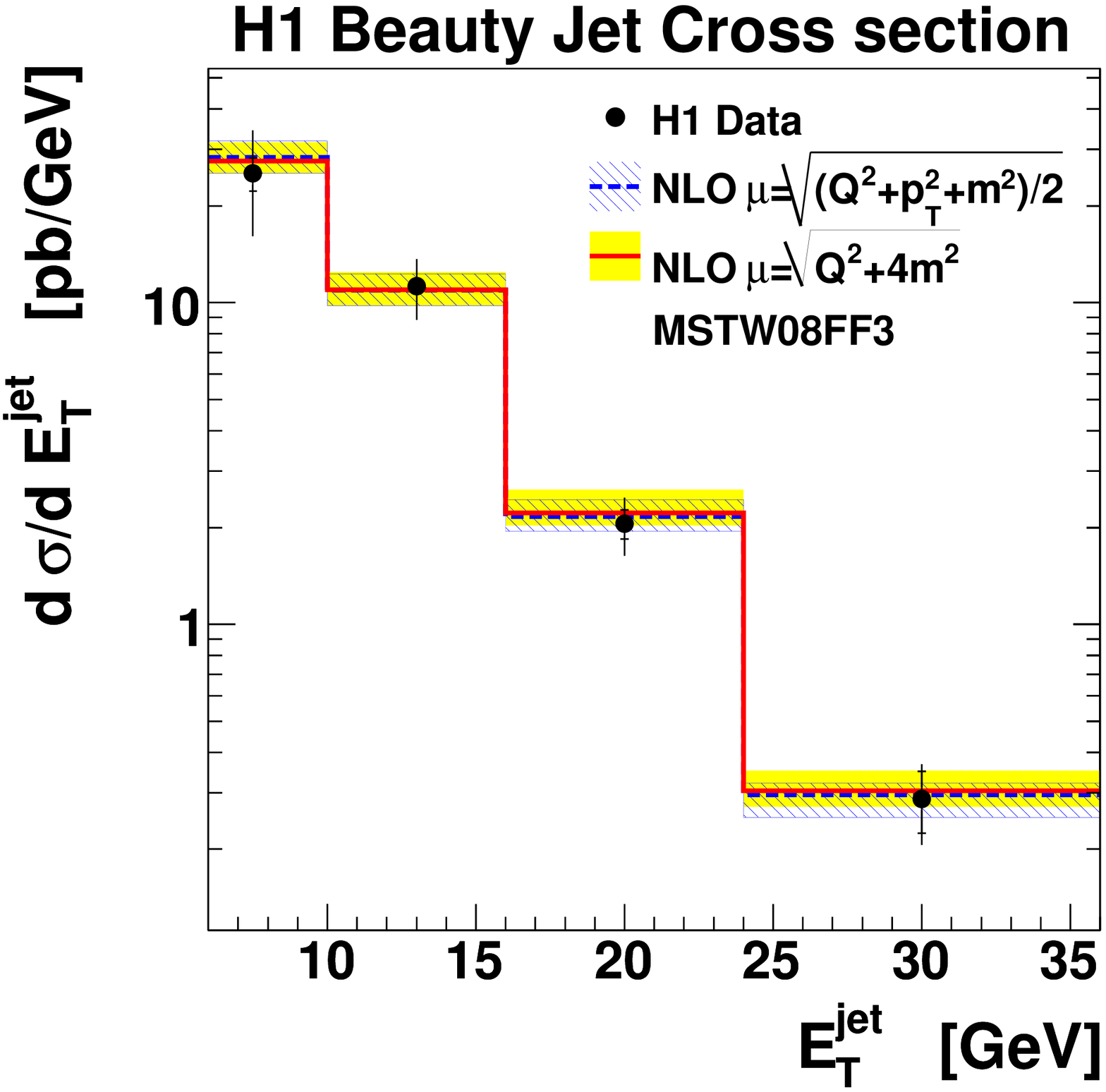}
  \includegraphics[width=0.49\textwidth]{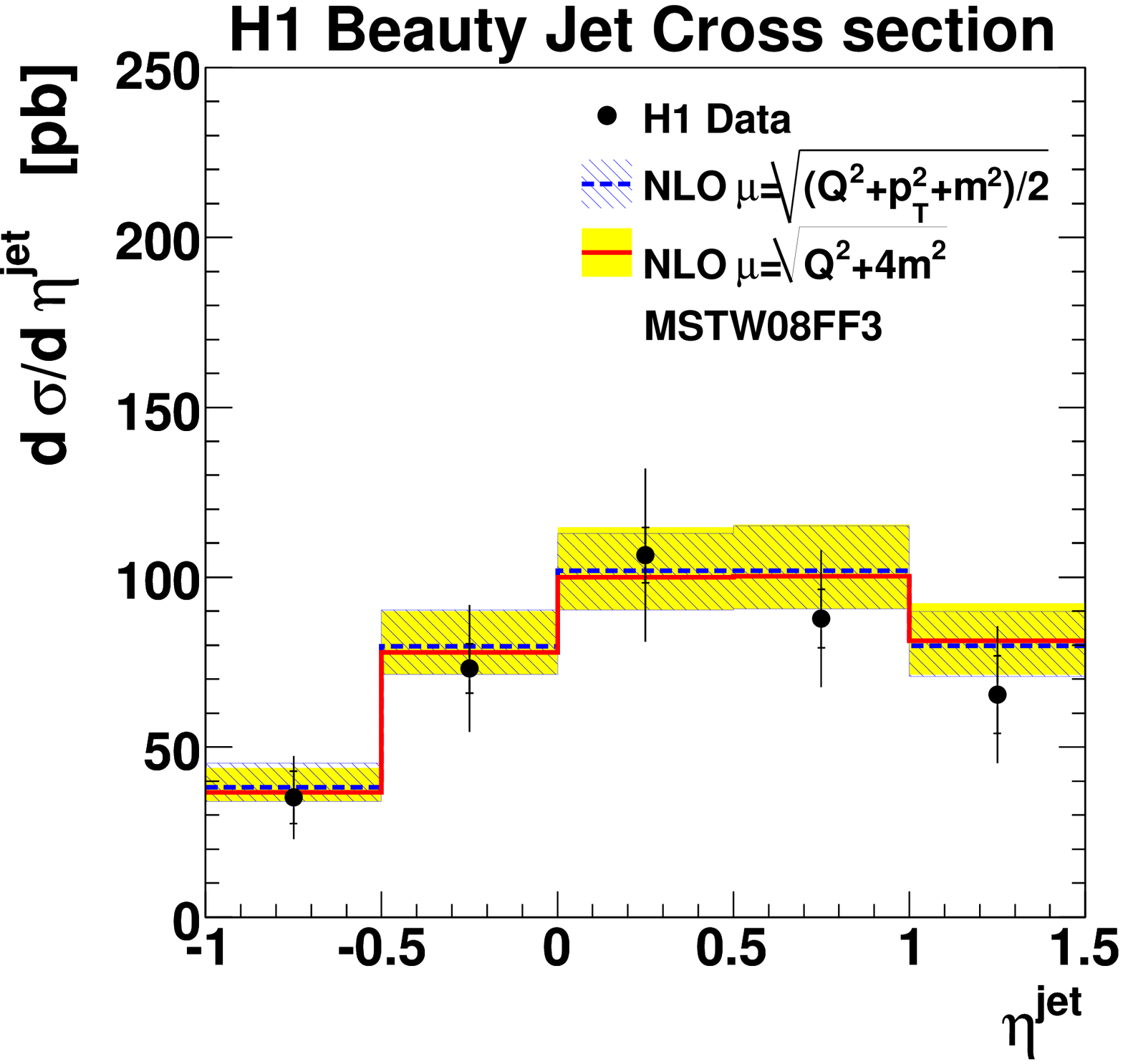} 
  \includegraphics[width=0.49\textwidth]{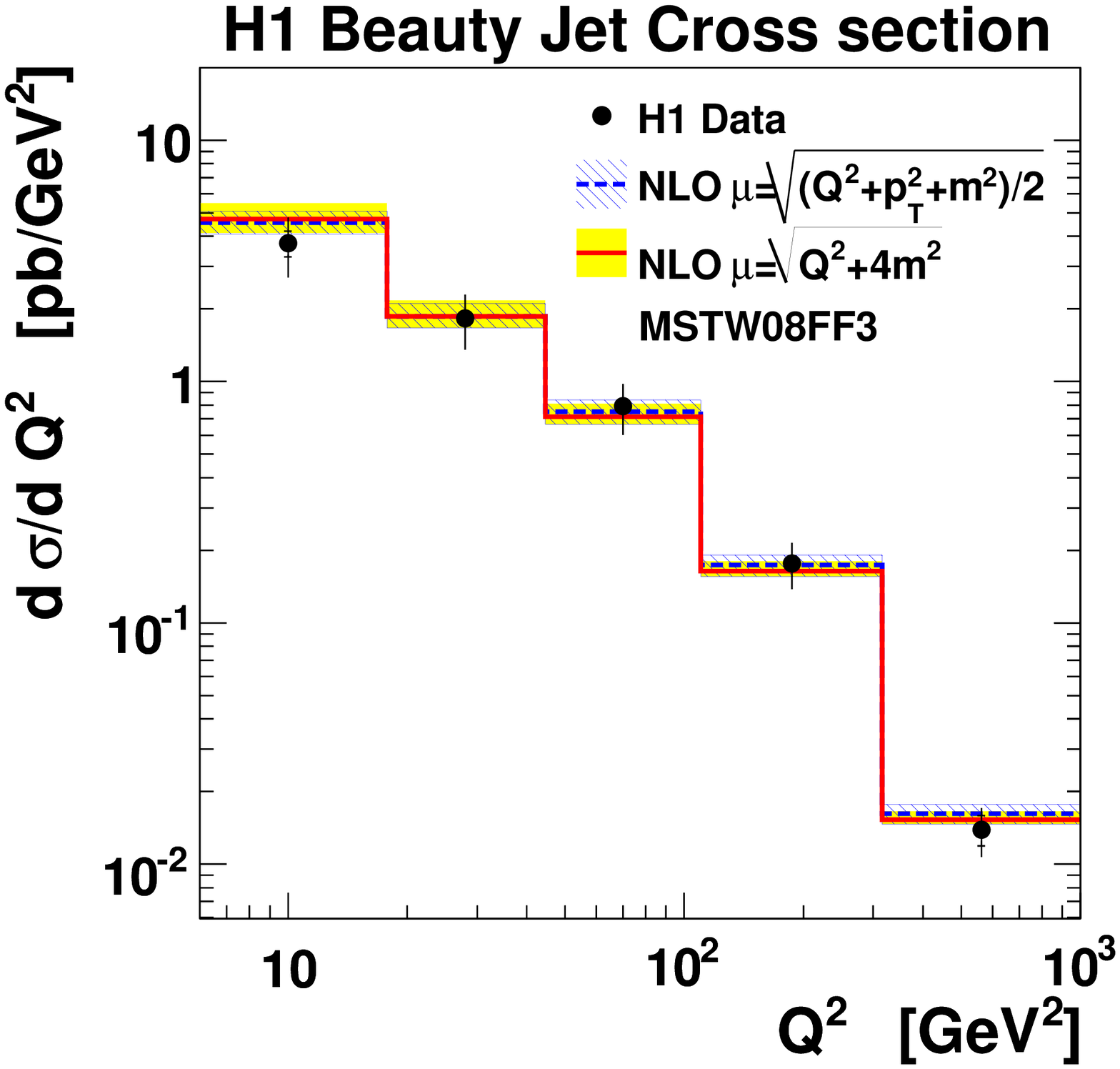}
  \includegraphics[width=0.49\textwidth]{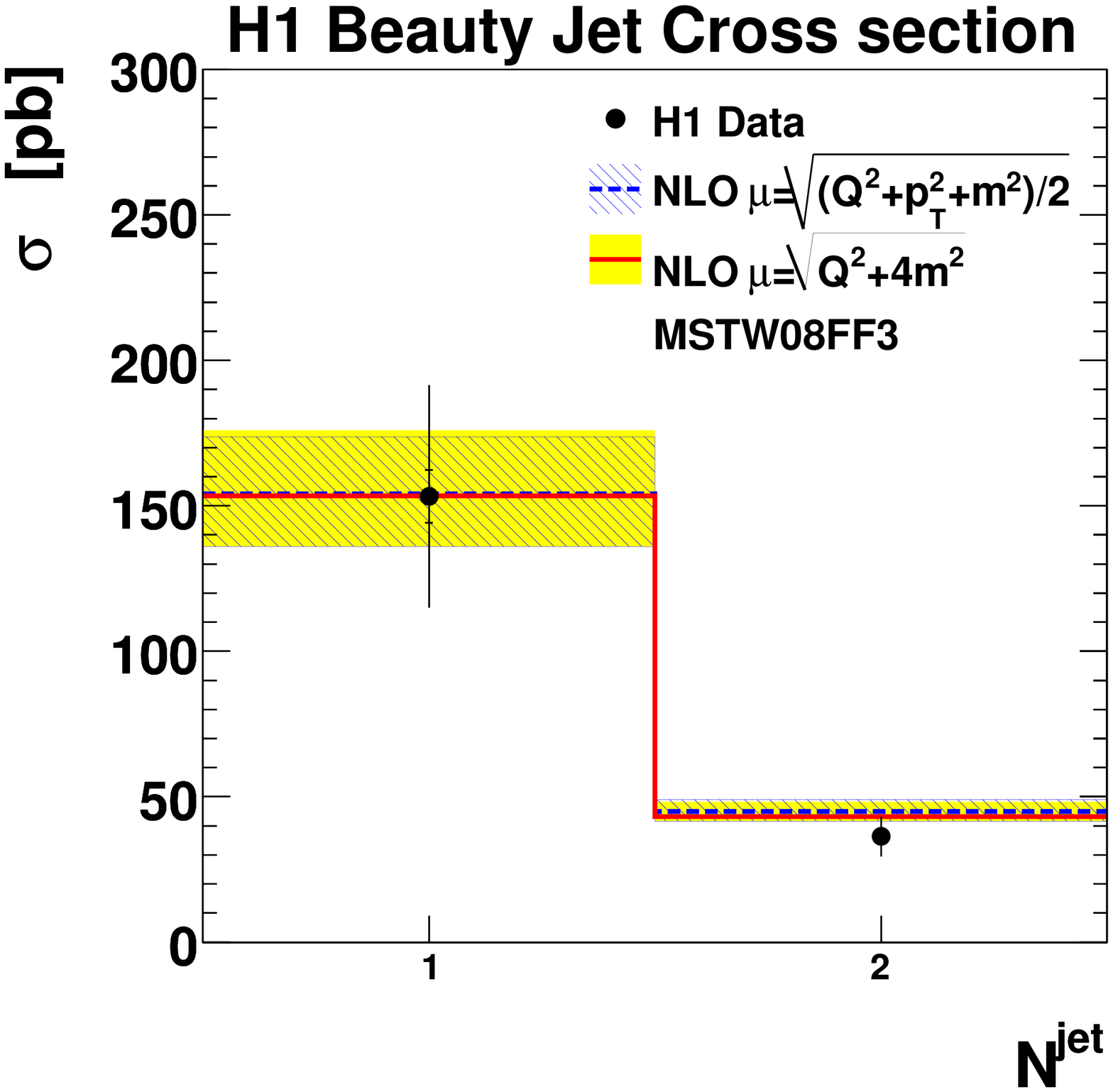} 
\caption{
The differential cross sections for
the highest transverse energy beauty jet in the laboratory frame 
as a function of $E_T^{\rm jet}$, $\eta^{\rm jet}$, $Q^2$ and
the number of laboratory frame jets in the event $N^{\rm jet}$. 
The measurements are made for the kinematic range
$E_T^{\rm jet}>6~{\rm GeV}$, $-1<\eta^{\rm jet}<1.5$, $Q^2>6~{\rm GeV}^2$ and $0.07 < y <0.625$.
The inner error bars show the statistical error, the
outer error bars represent the statistical and systematic errors
added in quadrature. 
The data are compared with the predictions from NLO QCD
where the bands indicate the theoretical uncertainties.}
\label{fig:xsbnlo} 
\end{center}
\end{figure}

In figures~\ref{fig:xsbmc} and~\ref{fig:xsbnlo} the differential $b$
cross sections are shown as a function of $E_T^{\rm jet}$, $\eta^{\rm
jet}$, $Q^2$ and $N^{\rm jet}$ in comparison to Monte Carlo and NLO
expectations, respectively. RAPGAP yields a good description of all
distributions also for beauty. CASCADE overshoots the data at small
$Q^2$, is slightly above the data at small $E_T^{\rm jet}$ and shows
an excess in the forward $\eta^{\rm jet}$ direction. These differences
are similar but less significant than for charm. HVQDIS gives a good
description of the beauty data with little dependence on the
choice of scale. 

\subsection{Jet Cross Sections in the Breit Frame}
\label{sec:results:breit}

 Differential $c$ and $b$ cross sections are also measured 
for the highest $E_T^{* {\rm jet}}$ jet in the Breit frame with
$E_T^{* {\rm jet}}>6~{\rm GeV}$ in the
kinematic range $Q^2>6~{\rm GeV}^2$, $0.07<y<0.625$  
for the heavy flavour jet with the highest $E_T^{\rm jet}$ in the laboratory 
satisfying $E_T^{\rm jet}>1.5~{\rm GeV}$ and $-1.0<\eta^{\rm jet}<1.5$.

\begin{figure}[ht]
  \begin{center} 
   \includegraphics[width=0.49\textwidth]{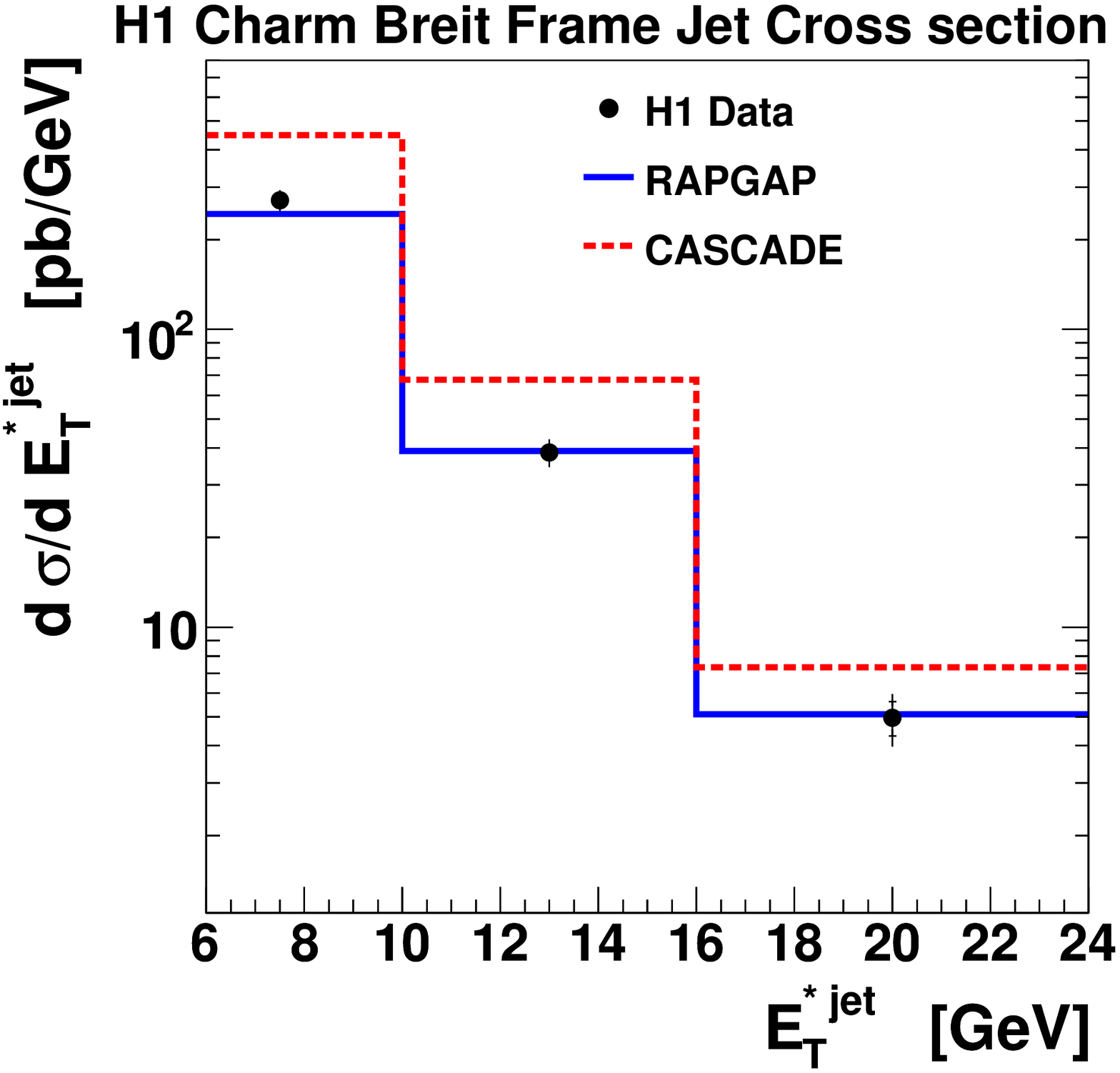}
   \includegraphics[width=0.49\textwidth]{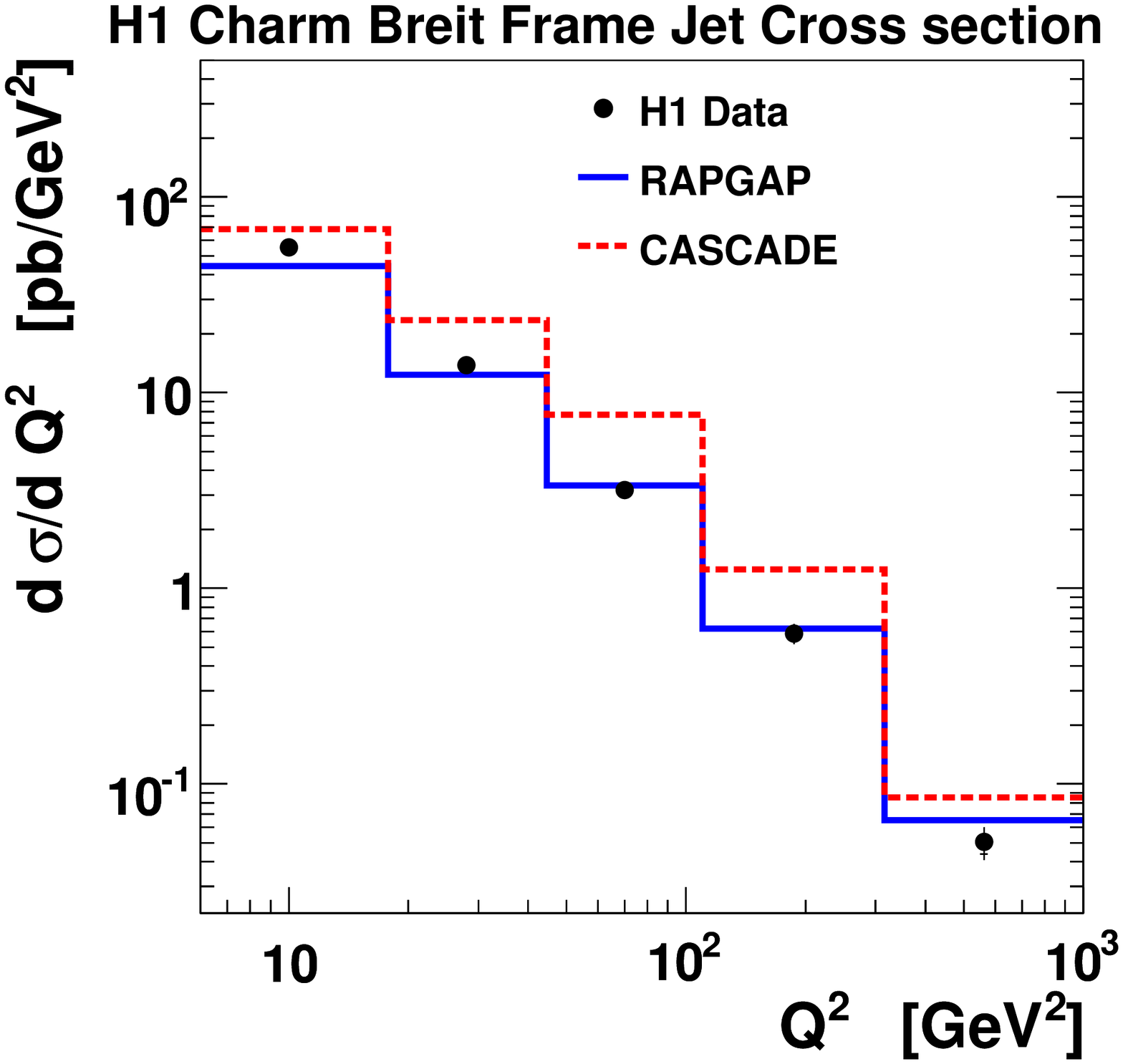}	
   \includegraphics[width=0.49\textwidth]{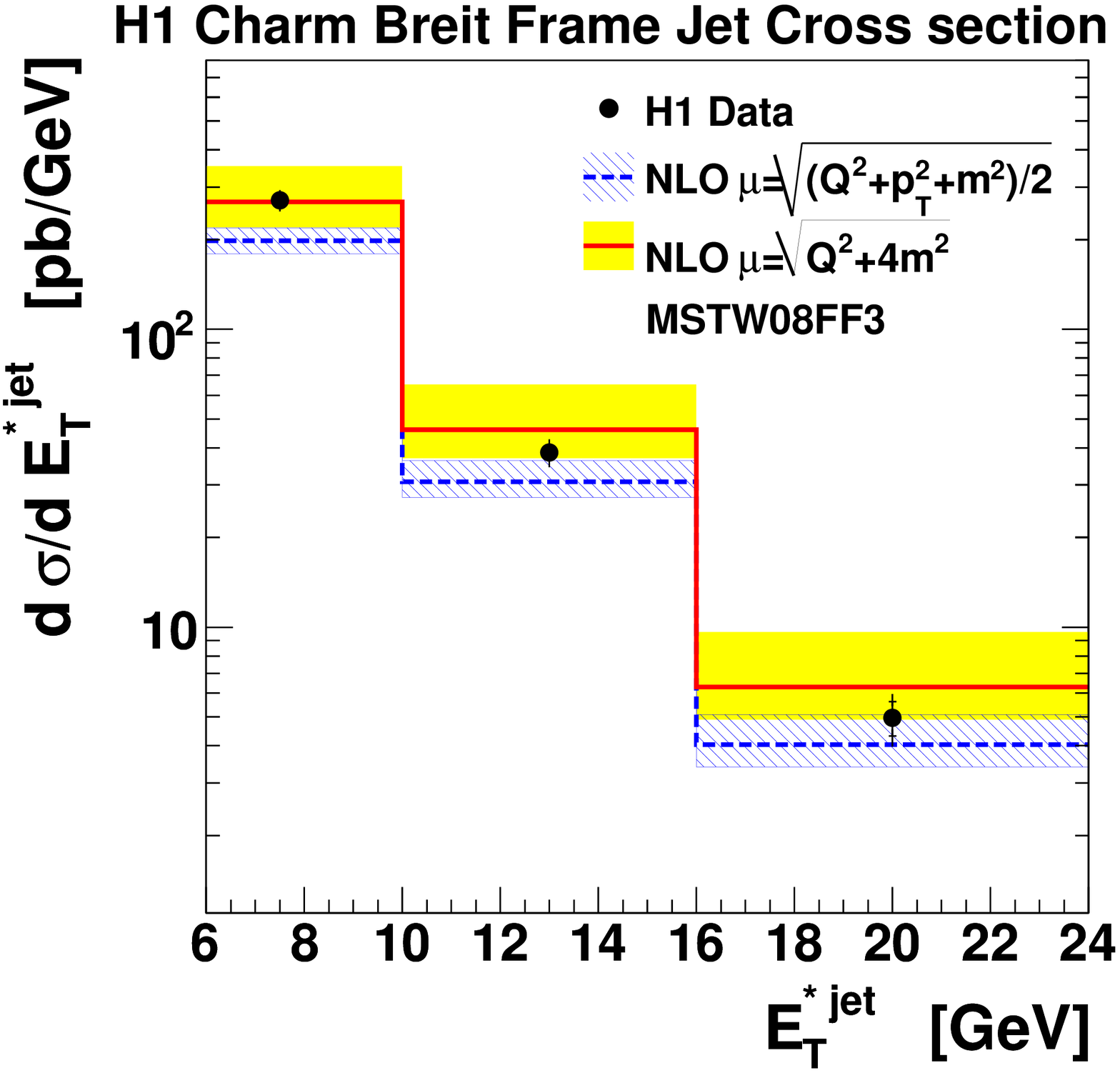}	
   \includegraphics[width=0.49\textwidth]{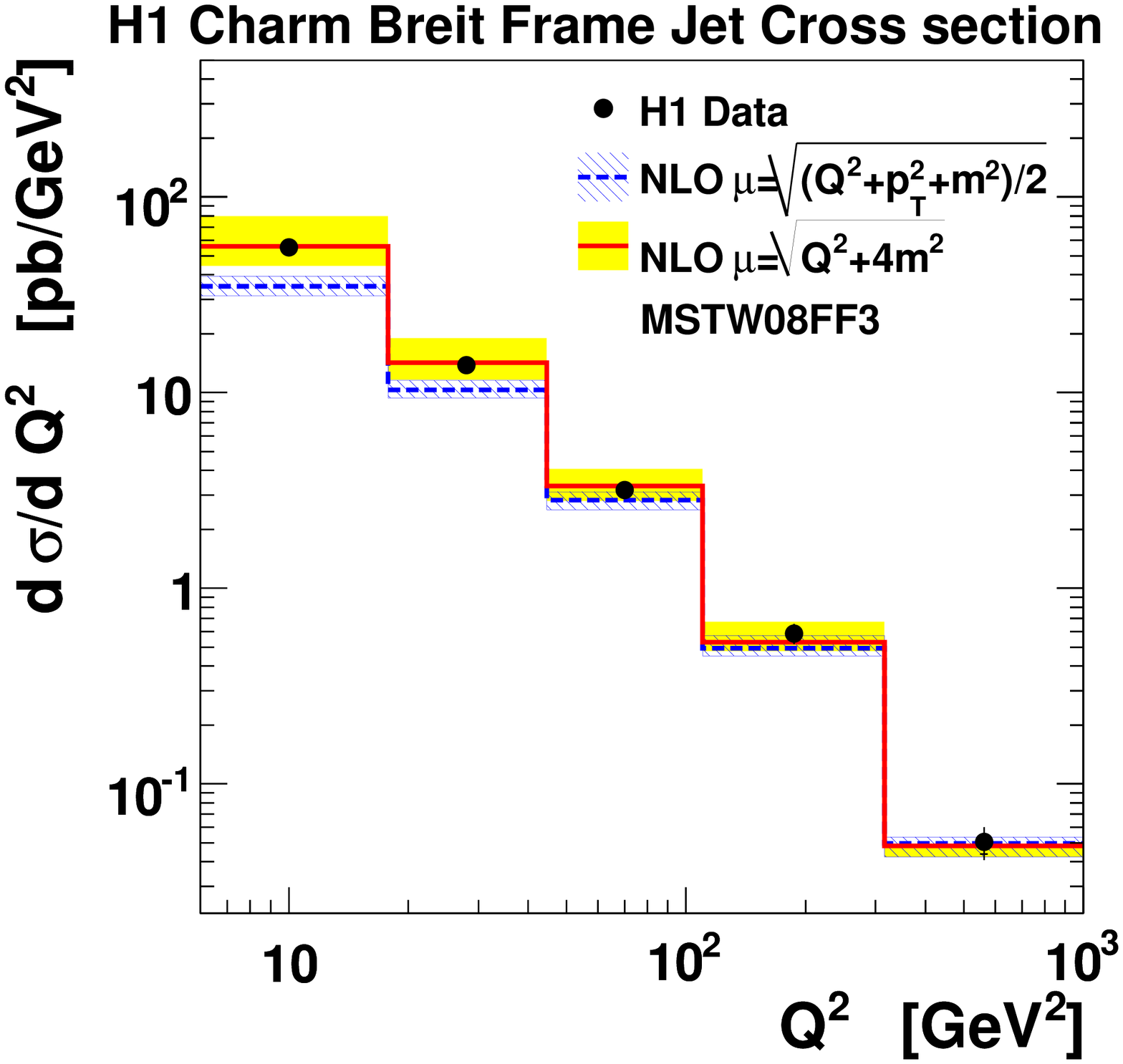}
 \caption{
  The differential cross sections ${\rm d}\sigma/{\rm d}E_T^{* {\rm jet}}$ 
  and ${\rm d}\sigma/{\rm d}Q^2$  	
  for events with a jet in the Breit frame,
  where the jet with the highest transverse energy in the laboratory
  frame satisfying $E_T^{\rm jet}>1.5~{\rm GeV}$ and $-1<\eta^{\rm jet}<1.5$
  is a charm jet.  The measurements are made for the
  kinematic range $Q^2>6~{\rm GeV}^2$ and $0.07 < y <0.625$.  
  The inner error bars show the statistical error,
  the outer error bars represent the statistical and systematic errors
  added in quadrature. The data are compared with the predictions from
  the Monte Carlo models  RAPGAP and CASCADE (upper plots) and the NLO QCD calculation 
  (lower plots), where the bands indicate the theoretical uncertainties.}
\label{fig:xscbreit} 
\end{center}
\end{figure}

The $c$ cross sections are shown as a function of $Q^2$ and $E_T^{* {\rm jet}}$
in figure~\ref{fig:xscbreit}. The data are compared to the expectations from 
RAPGAP, CASCADE and HVQDIS.
It can be seen that RAPGAP provides a good description of the data for both 
distributions.  
CASCADE overestimates the data in $Q^2$ and 
$E_T^{* {\rm jet}}$. In contrast to the observation in the laboratory frame, the deviation in $Q^2$ is found to be independent of $Q^2$ here. Nevertheless
the shapes of the predictions
are similar to those in the data.
As for the laboratory frame analysis, HVQDIS with the scale 
choice $\mu = \sqrt{Q^2+4m^2}$ reproduces the data well, while for the scale
$\mu = \sqrt{(Q^2+p_T^2+m^2)/2}$ it tends to underestimate  
the $c$ jet data at low values of $Q^2$ and  $E_T^{* {\rm jet}}$.

\begin{figure}[ht]
  \begin{center} 
  \includegraphics[width=0.49\textwidth]{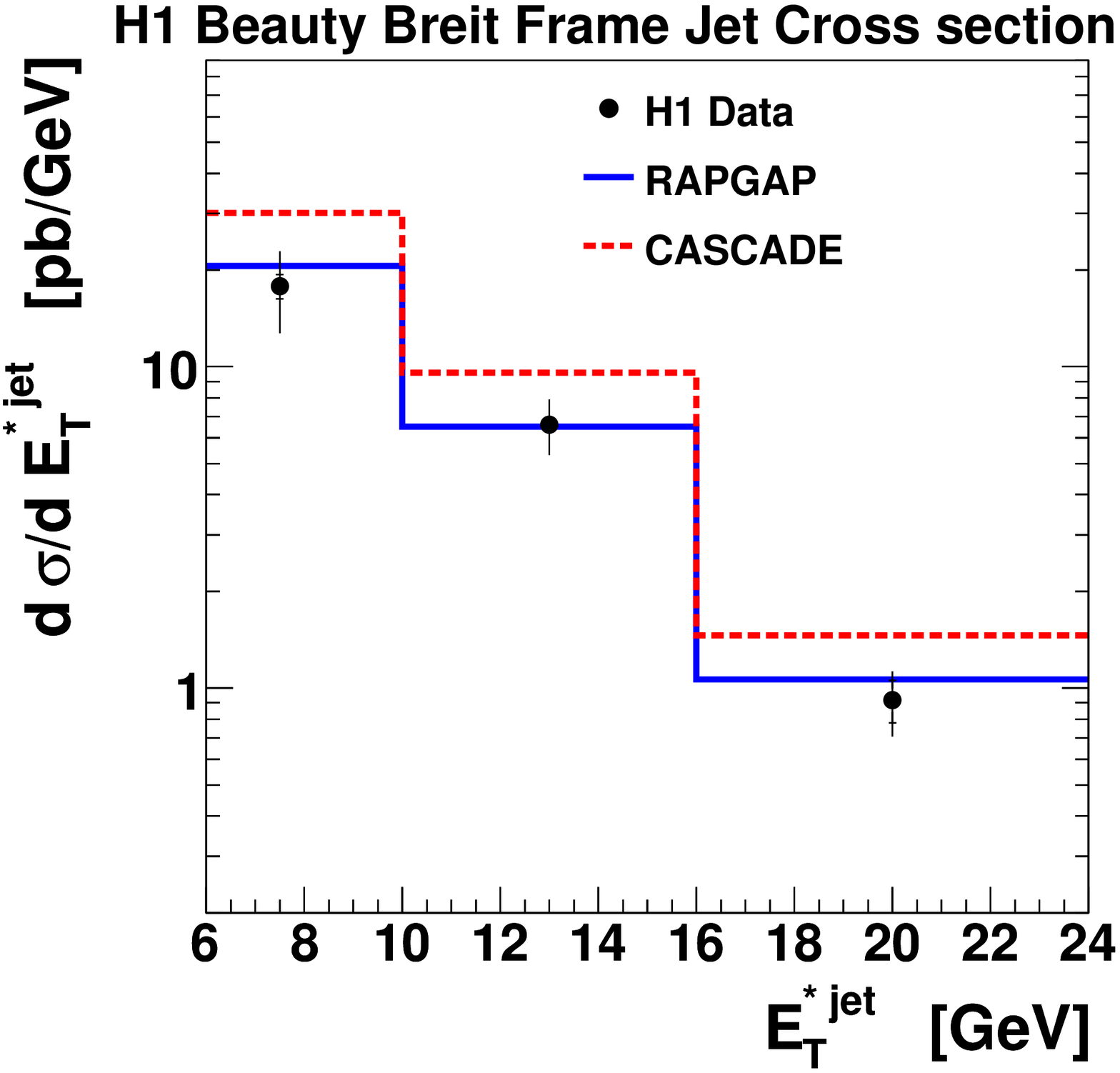}
  \includegraphics[width=0.49\textwidth]{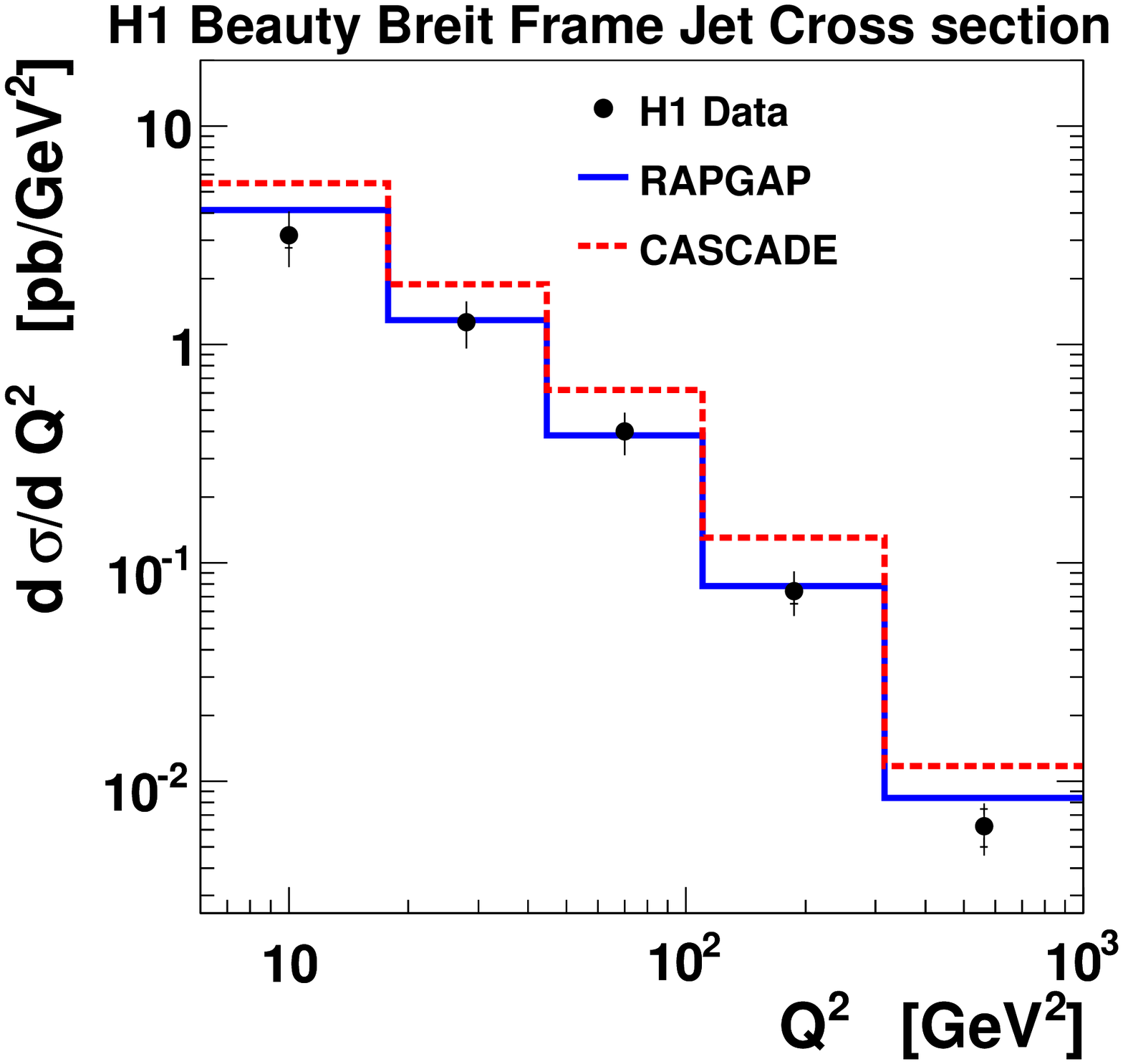}
  \includegraphics[width=0.49\textwidth]{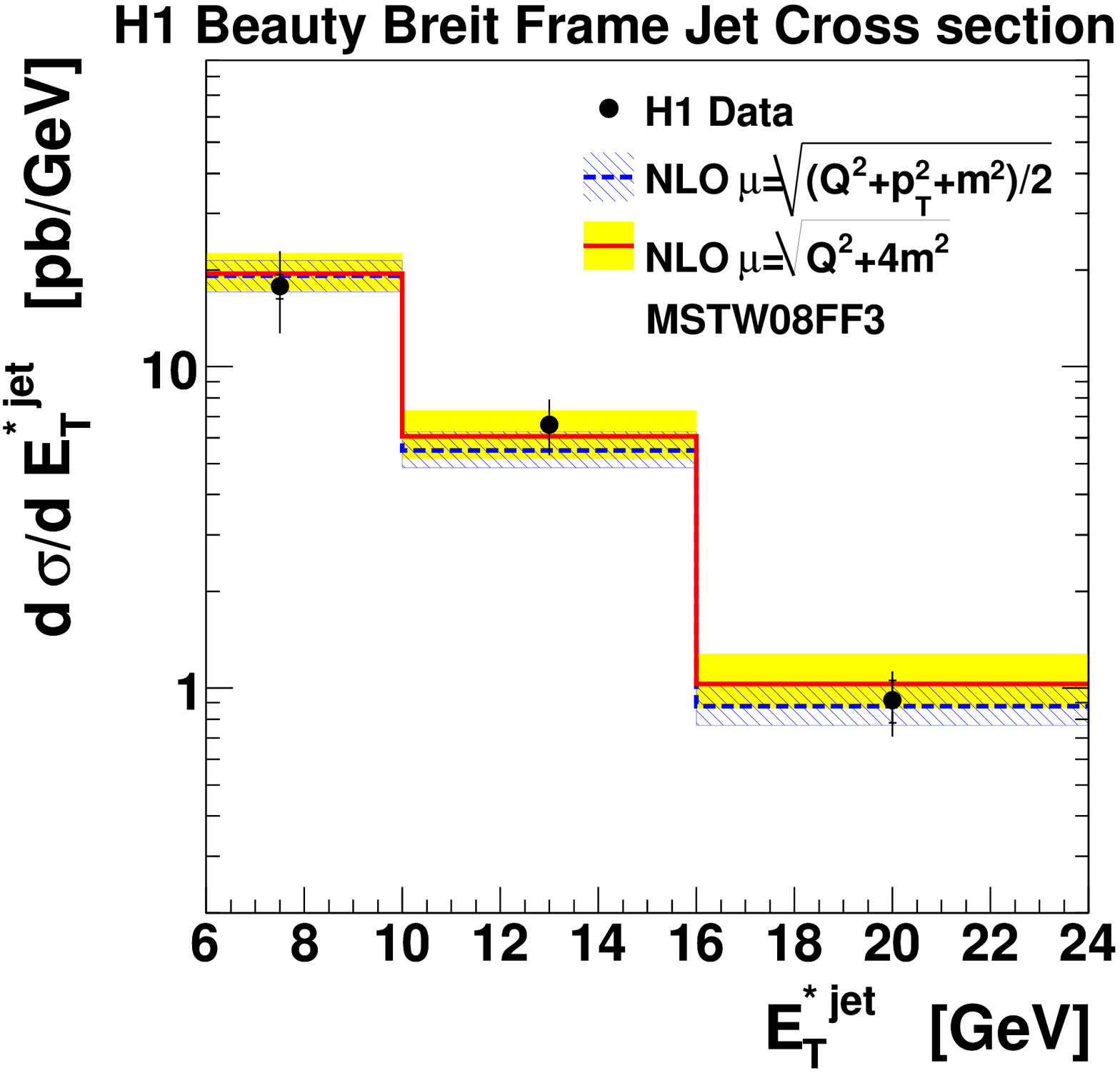}
  \includegraphics[width=0.49\textwidth]{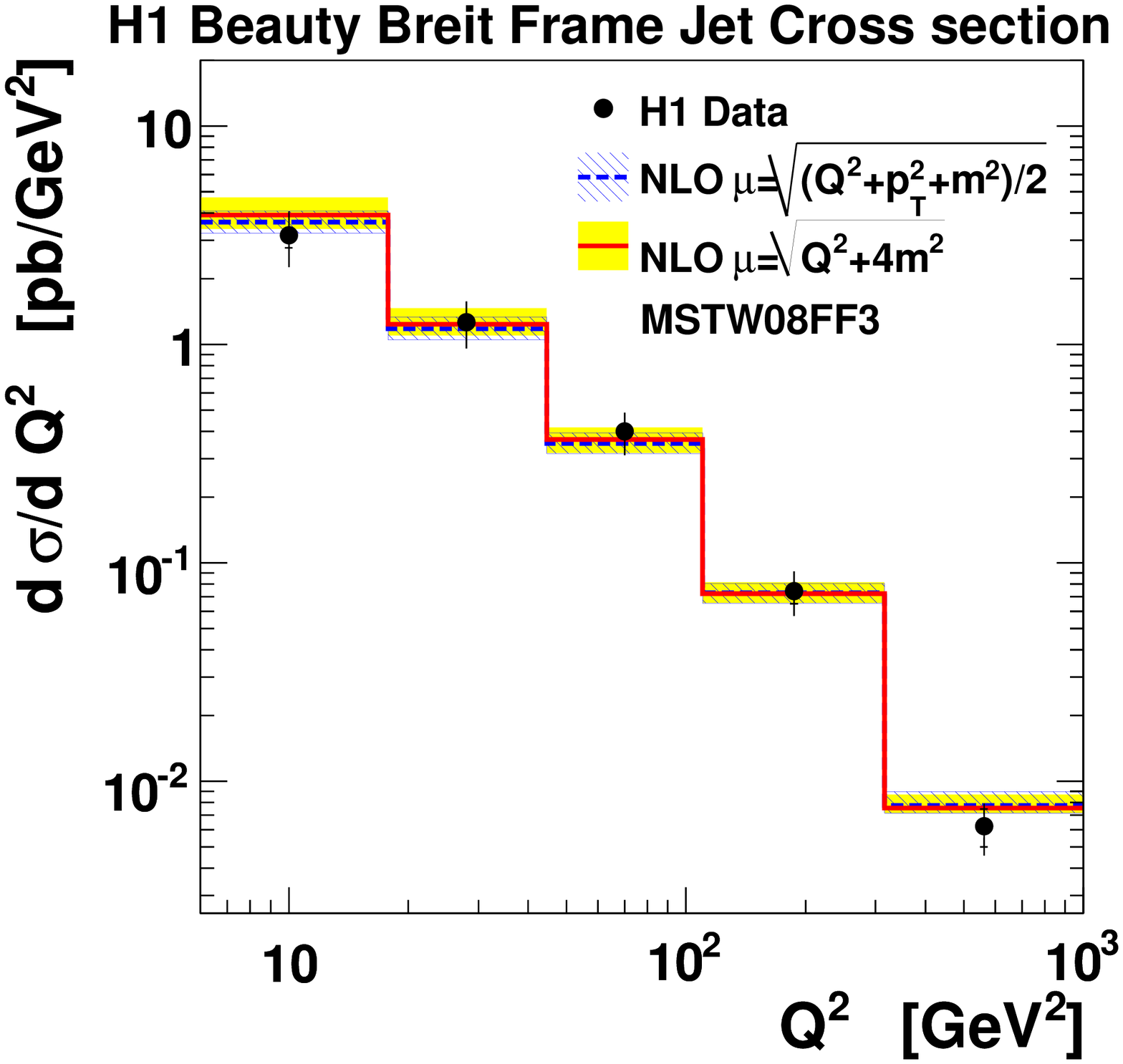} \caption{
  The differential cross sections ${\rm d}\sigma/{\rm d}E_T^{* {\rm jet}}$ 
  and ${\rm d}\sigma/{\rm d}Q^2$  	
  for events with a jet in the Breit frame,
  where the jet with the highest transverse energy in the laboratory
  frame satisfying $E_T^{\rm jet}>1.5~{\rm GeV}$ and $-1<\eta^{\rm jet}<1.5$
  is a beauty jet.  The measurements are made for the
  kinematic range $Q^2>6~{\rm GeV}^2$ and $0.07 < y <0.625$.  
  The inner error bars show the statistical error,
  the outer error bars represent the statistical and systematic errors
  added in quadrature. The data are compared with the predictions from
  the Monte Carlo models  RAPGAP and CASCADE (upper plots) 
  and the NLO QCD calculation 
  (lower plots), where the bands indicate the theoretical uncertainties.}
\label{fig:xsbbreit} 
\end{center}
\end{figure}

The differential $b$ cross sections are shown as a function of $Q^2$ and $E_T^{* {\rm jet}}$
in figure~\ref{fig:xsbbreit} together with the Monte Carlo and NLO expectations.
As for charm, RAPGAP performs well while CASCADE lies systematically above the data.
The high rate of Breit frame jets in CASCADE both for charm and beauty jet production
is related to the
transverse momentum distribution of the unintegrated gluon density used for the calculations. 
HVQDIS describes the data well showing little dependence on the choice of scales.

\subsection{Comparison with Muon Tagging Measurements }
\label{sec:results:muon}

The $b$ jet cross sections may be compared with $b$ jet measurements
obtained from muon tagging in the Breit (H1~\cite{H1muons})
and laboratory (ZEUS~\cite{ZEUSmuonslab}) frames of reference. 
The muon measurements were made requiring the presence of a muon and a jet in 
either the laboratory frame, with $E_T^{\rm jet}>5~{\rm GeV}$ 
or in the Breit frame  with $E_T^{*{\rm jet}}>6~{\rm GeV}$ 
and with a central rapidity requirement in the laboratory 
frame, similar to the present analysis. 
The measurements were also made in a similar $y$ range
but  start at lower values of $Q^2$ ($Q^2 > 2 \ {\rm GeV^2}$). 
Therefore, comparison of the cross sections with these measurements
as a function of $E_T^{\rm jet}$ or $E_T^{*{\rm jet}}$ would require 
interpolating over a large range in $Q^2$. However, the $b$ cross sections
can be compared as a function of $Q^2$ for the range where the $Q^2$
binning of the muon measurements overlaps closely
with the present analysis, namely $Q^2 > 10 \ {\rm GeV}^2$ for the laboratory
analysis and $Q^2 > 6 \ {\rm GeV}^2$ for the Breit frame analysis.

\begin{figure}[ht]
  \begin{center}
  \includegraphics[width=0.49\textwidth]{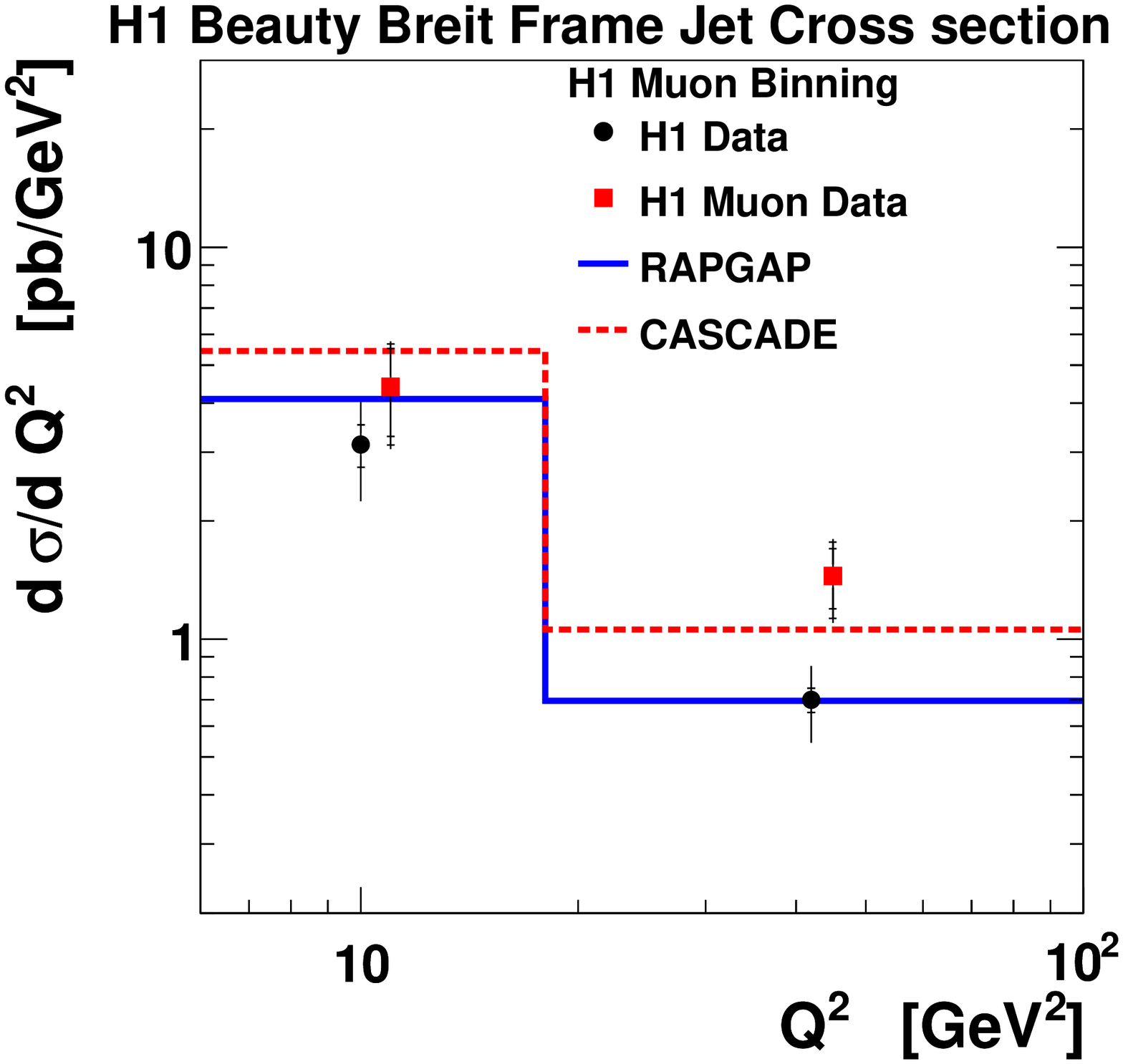}
  \includegraphics[width=0.49\textwidth]{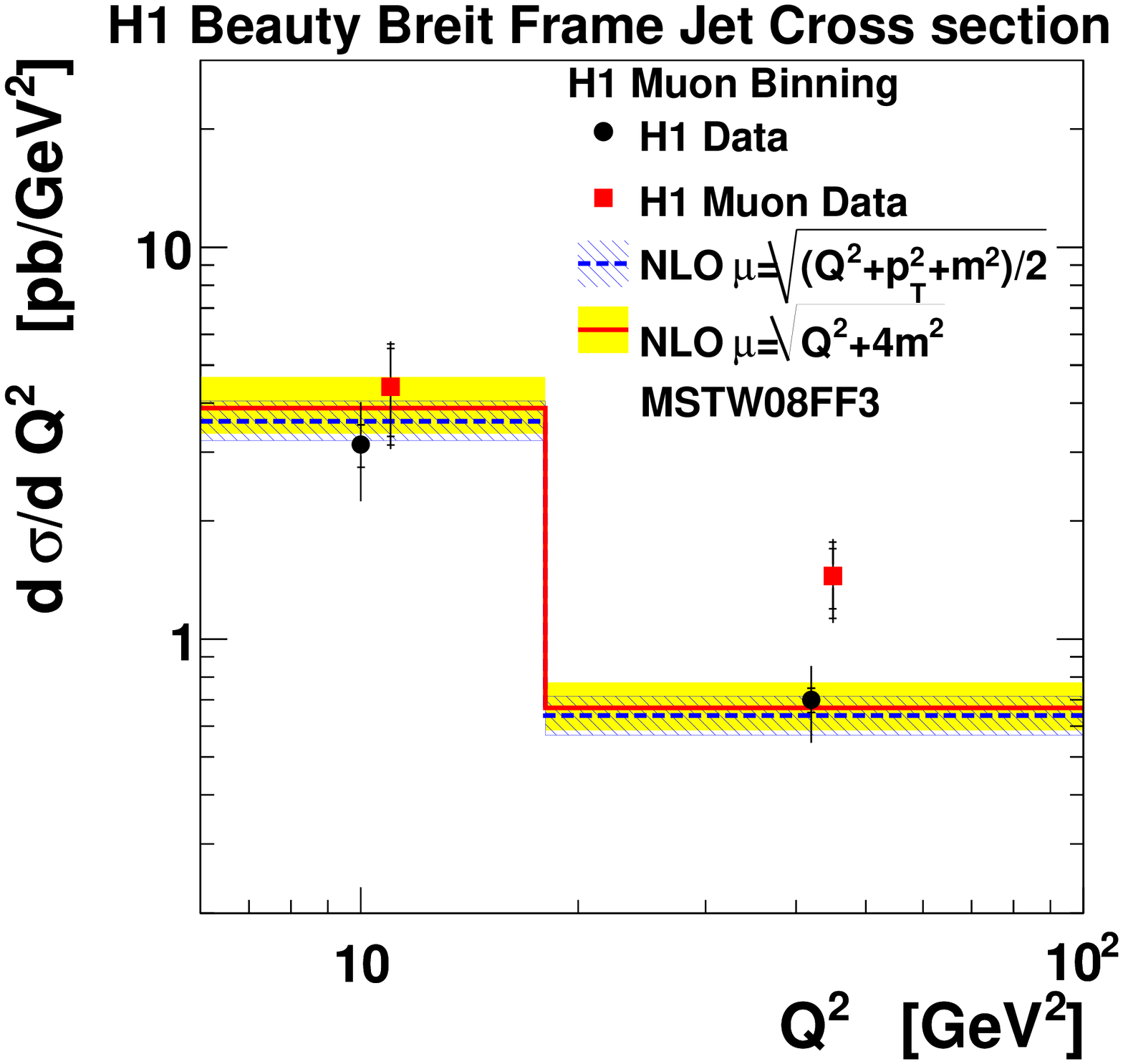}
  \includegraphics[width=0.49\textwidth]{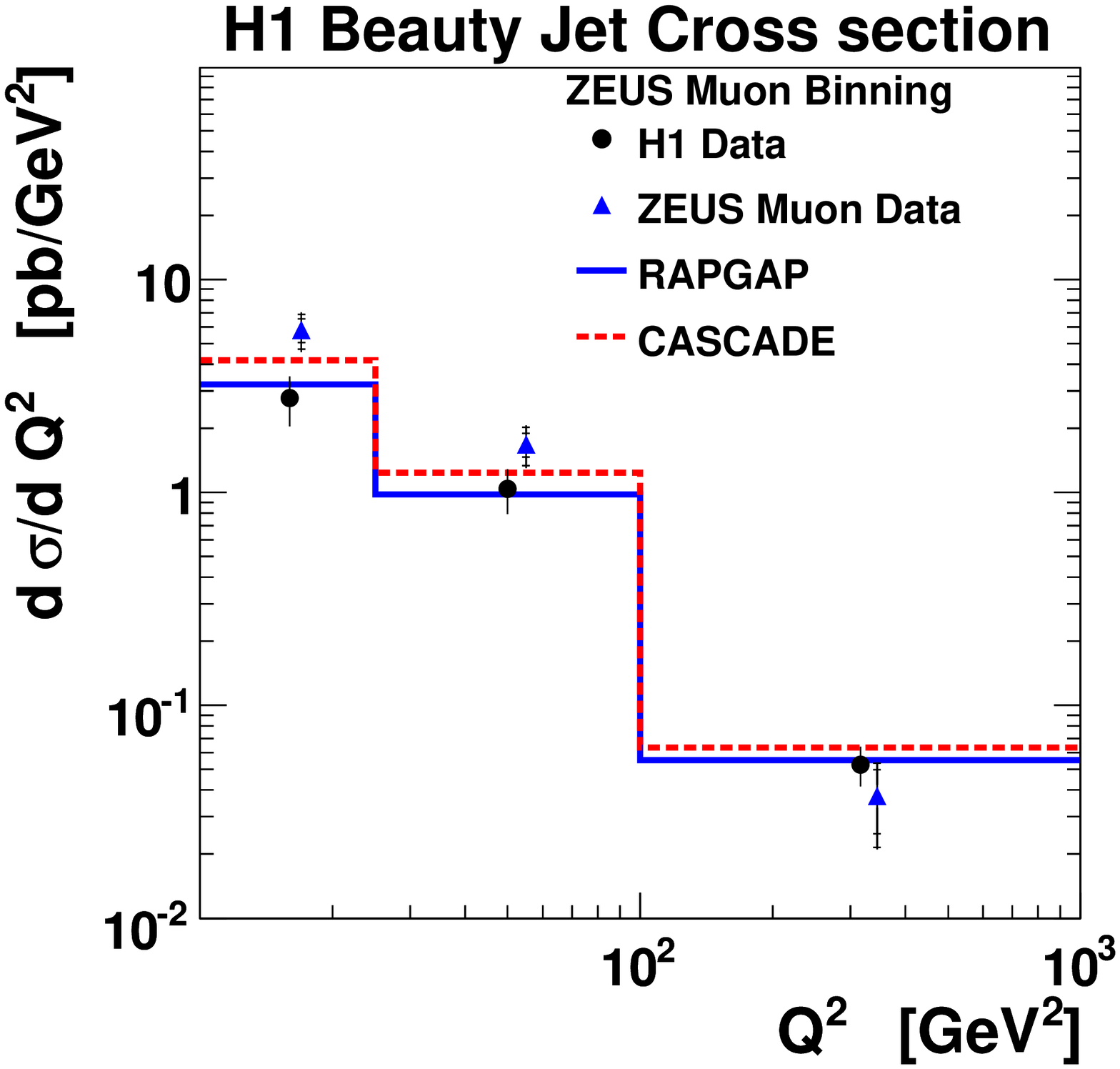}
  \includegraphics[width=0.49\textwidth]{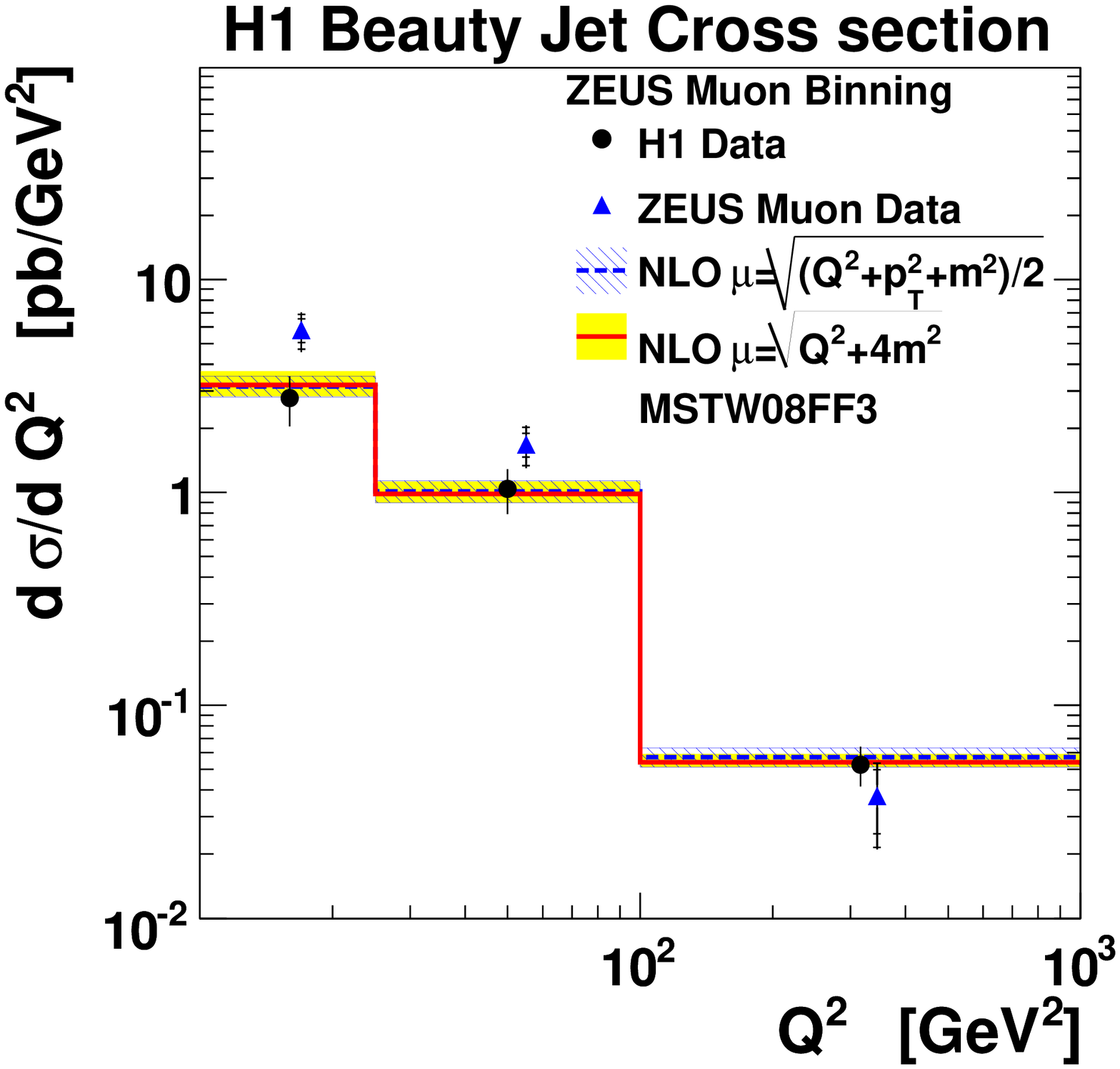}
  \caption{The upper plots show the differential cross section ${\rm
  d}\sigma/{\rm d}Q^2$ for events with a jet in the Breit frame with
  $E_T^{* {\rm jet}}>6~{\rm GeV}$, where the jet with the highest
  transverse energy in the laboratory frame satisfying $E_T^{\rm
  jet}>1.5~{\rm GeV}$ and $-1<\eta^{\rm jet}<1.5$ is a beauty jet.  
  The lower plots
  show the differential cross section ${\rm d}\sigma/{\rm d}Q^2$ for
  events with a beauty jet in the laboratory frame with $E_T^{\rm jet}>6~{\rm
  GeV}$ and $-1<\eta^{\rm jet}<1.5$.  The present measurements are
  made for the kinematic range $Q^2>6~{\rm GeV}^2$ and $0.07 < y
  <0.625$.  The inner error bars show the statistical error, the outer
  error bars represent the statistical and systematic errors added in
  quadrature. The data are compared with the measurements obtained
  using muon tagging from H1~\cite{H1muons} (upper plots) and
  ZEUS~\cite{ZEUSmuonslab} (lower plots)
  extrapolated to the present phase space and
  shifted in $Q^2$ for visual clarity.  For the muon data the outer
  error bars show the statistical, systematic and extrapolation
  uncertainties added in quadrature. The data are also compared with
  the predictions from the Monte Carlo models  RAPGAP and CASCADE (left) 
  and the NLO QCD
  calculation (right), where the bands indicate the theoretical
  uncertainties.}
\label{fig:xsq2breith1zeus} 
\end{center}
\end{figure}

The present analysis is repeated with two different sets of $Q^2$ bins
chosen to match the H1 and ZEUS muon measurements as closely as
possible.  The cross sections are shown as a function of $Q^2$ for the
two sets of bins in figure~\ref{fig:xsq2breith1zeus}.  The H1 muon
data are corrected by factors of about $15$ which are obtained
using the RAPGAP Monte Carlo. The dominant corrections account for the
$b\rightarrow \mu$ branching fraction and for the extrapolation from
the phase space of the muon measurement, which had restrictions on
$p_T^{\mu}$ and $\eta^{\mu}$, to the phase space of the present analysis. 
The ZEUS muon data are
corrected to the present phase space by factors of around $6$. These
corrections are smaller than in the case of the H1 data because
the ZEUS data have a wider $\eta^{\mu}$ and $p_T^{\mu}$
coverage.  The corrections also include smaller effects due to the
difference in the $E_T$ range of the jets for the ZEUS laboratory
frame analysis, differences in the $\eta$ ranges of the jets, the
difference in the $y$ ranges, the difference in the jet cross section
definitions and jet finding algorithms and the fact that the lower
edge of the lowest $Q^2$ bin is $Q^2 \ge 5 \ {\rm GeV^2}$ for the H1
muon measurement.  An additional uncertainty of around $10\%$ is added
to the corrected muon measurements to account for theoretical
uncertainties on the extrapolation factors coming from uncertainties
on the perturbative scales and fragmentation model used. The central
values of the present data in the Breit frame are found to lie below
the adjusted H1 muon data at high $Q^2$.  The present data in the
laboratory frame are found to lie significantly below the ZEUS muon
data at low $Q^2$, where the difference is a factor $2.1$.
The comparison suggests a systematic difference between the H1 inclusive 
lifetime tagged data and the muon tagged data, particularly in 
comparison with the ZEUS muon tagged data at low $Q^2$.

\section{Conclusion}

The  cross sections for events with charm and beauty jets have been measured 
in deep inelastic scattering  at the HERA
electron--proton collider. Measurements are made in the laboratory 
frame for $E_T^{\rm jet}>6$ GeV and $-1.0<\eta<1.5$ for the 
kinematic region of photon virtuality $ Q^2 >6~{\rm GeV}^2$ and 
inelasticity variable $0.07 < y < 0.625$. 
Measurements are also made in the Breit frame of reference.
The analysis uses the precise spatial information from the
H1 vertex detector to distinguish those jets that contain
$c$ and $b$ flavoured hadrons from jets containing only light flavoured
hadrons. 

The laboratory frame jet data are compared with the Monte Carlo models 
RAPGAP and CASCADE.
RAPGAP is generally found to give a good description of the
data. CASCADE is found to lie above the charm data, especially at low $Q^2$ 
and high $\eta^{\rm jet}$. After accounting for the
difference in normalisation CASCADE generally gives a good description of 
the shape of the differential cross section measurements. 
CASCADE provides a better prediction of the beauty cross section normalisation
than it does for charm but still tends to overestimate the data at low $Q^2$ 
and high $\eta^{\rm jet}$.
The data are also compared
with NLO QCD calculations made using the HVQDIS program. The beauty
data are well described by the calculation. 
The charm expectations are
found to depend strongly on the choice of renormalisation and factorisation
scale. The differential cross sections are described within the
experimental and theoretical uncertainties with a scale choice of 
$\mu = \mu_r = \mu_f =\sqrt{Q^2+4m^2}$. The predictions tend to lie below the 
data at low $Q^2$ and high $\eta$ with a choice of 
$\mu=\sqrt{(Q^2+p_T^2+m^2)/2}$.

For the measurements of the cross section requiring a jet in the Breit frame 
with $E_T^{* {\rm jet}}>6~{\rm GeV}$ RAPGAP is found to
give a good description of the data while CASCADE again overestimates the cross 
sections. The NLO QCD predictions for charm jets 
with a scale choice of $\mu=\sqrt{Q^2+4m^2}$ are compatible with the data while
the predictions with the choice of scale $\mu=\sqrt{(Q^2+p_T^2+m^2)/2}$
fail to describe the data at low $Q^2$.  
The $b$ jet data are described by NLO QCD
for all choices of scale.  

The $b$ jet data are compared with H1 and ZEUS data obtained from muon 
tagging by adjusting that data 
mainly for the extrapolation of the measured to the full muon phase space
and for the $b \rightarrow \mu$ branching fraction.  
The $b$ jet data from the present analysis 
are found to lie systematically below those obtained from ZEUS at low $Q^2$
and below the H1 muon tagged data at high $Q^2$.

The present measurements show that charm and beauty production in deep
inelastic scattering, 
adequately described by NLO QCD in the inclusive case, is also described in the
presence of an additional hard scale provided by a jet.

\section*{Acknowledgements}

We are grateful to the HERA machine group whose outstanding efforts
have made this experiment possible.  We thank the engineers and
technicians for their work in constructing and maintaining the H1
detector, our funding agencies for financial support, the DESY
technical staff for continual assistance and the DESY directorate for
support and for the hospitality which they extend to the non-DESY
members of the collaboration.  


\newpage
\begin{sidewaystable}[h]
\footnotesize
\begin{center}
 \begin{tabular}{|l|q|e|x|c|c|c|c|c|c|c|c|}  \hline  
 bin & \multicolumn{1}{c|}{$Q^2$ range} &  \multicolumn{1}{c|}{$E^{(*)\rm jet}_T$ range} &  \multicolumn{1}{c|}{$\eta^{\rm jet}$ range} & \multicolumn{1}{c|}{$N^{\rm jet}$} & $\rho_l$ &  $\rho_c$ & $\rho_b$ & $\chi^2/{\rm n.d.f.}$ & $C_{lc}$  & $C_{lb}$ &  $C_{bc}$ \bigstrut[t] \\  
  &  \multicolumn{1}{c|}{(GeV$^2$)} &  \multicolumn{1}{c|}{(GeV)} & & & & & & & & &  \\ \hline
     1   &  Q^2>     6    &  E^{ \rm jet}_T >      6 &   -1.0 < \eta^{\rm jet}<   1.5   & $\ge 1$ & $   1.178  \pm    0.004  $ & $    1.040 \pm    0.015 $ & $     0.95 \pm    0.05 $ & $  46.3 /  49 $ & $   -0.95 $ & $    0.52 $ & $    -0.66 $ \\ 
\hline 
     2 & Q^2>     6 &      6 <E^{ \rm jet}_T <    10 &   -1.0 < \eta^{\rm jet}<   1.5  & $\ge 1$ & $   1.203  \pm    0.006  $ & $    1.028 \pm    0.020 $ & $     0.93 \pm    0.11 $ & $  54.8 /  46 $ & $   -0.95 $ & $    0.61 $ & $    -0.77 $ \\ 
     3 & &     10 <E^{ \rm jet}_T <    16 &   &  & $   1.142  \pm    0.009  $ & $    1.078 \pm    0.030 $ & $     1.00 \pm    0.06 $ & $  34.8 /  45 $ & $   -0.95 $ & $    0.54 $ & $    -0.67 $ \\ 
     4 & &     16 <E^{ \rm jet}_T <    24 &     & & $   1.060  \pm    0.016  $ & $    1.092 \pm    0.066 $ & $     0.87 \pm    0.09 $ & $  41.0 /  40 $ & $   -0.95 $ & $    0.51 $ & $    -0.64 $ \\ 
     5 & &     24 <E^{ \rm jet}_T <    36 &     &  & $   1.080  \pm    0.030  $ & $    0.713 \pm    0.158 $ & $     0.93 \pm    0.20 $ & $  16.0 /  30 $ & $   -0.96 $ & $    0.50 $ & $    -0.62 $ \\ 
\hline 
     6   &  Q^2>     6    &  E^{ \rm jet}_T >      6 &   -1.0 < \eta^{\rm jet}<  -0.5   & $\ge 1$ & $   1.112  \pm    0.012  $ & $    0.883 \pm    0.035 $ & $     0.98 \pm    0.21 $ & $  36.9 /  40 $ & $   -0.95 $ & $    0.57 $ & $    -0.71 $ \\ 
     7   &     &  &   -0.5 < \eta^{\rm jet}<   0.0   &  & $   1.127  \pm    0.009  $ & $    0.985 \pm    0.027 $ & $     0.92 \pm    0.09 $ & $  47.5 /  44 $ & $   -0.95 $ & $    0.52 $ & $    -0.66 $ \\ 
     8   &     &   &    0.0 < \eta^{\rm jet}<   0.5   &  & $   1.199  \pm    0.008  $ & $    1.020 \pm    0.029 $ & $     1.05 \pm    0.08 $ & $  46.5 /  45 $ & $   -0.94 $ & $    0.51 $ & $    -0.64 $ \\ 
     9   &      &  &    0.5 < \eta^{\rm jet}<   1.0   &  & $   1.213  \pm    0.009  $ & $    1.172 \pm    0.033 $ & $     0.87 \pm    0.08 $ & $  37.9 /  43 $ & $   -0.95 $ & $    0.50 $ & $    -0.64 $ \\ 
    10   &      &  &    1.0 < \eta^{\rm jet}<   1.5   &  & $   1.188  \pm    0.014  $ & $    1.209 \pm    0.060 $ & $     0.81 \pm    0.14 $ & $  39.8 /  41 $ & $   -0.97 $ & $    0.60 $ & $    -0.73 $ \\ 
\hline 
    11   &     6 < Q^2<    18    &  E^{ \rm jet}_T >      6 &   -1.0 < \eta^{\rm jet}<   1.5  & $\ge 1$  & $   1.387  \pm    0.011  $ & $    1.174 \pm    0.033 $ & $     0.78 \pm    0.09 $ & $  40.4 /  43 $ & $   -0.95 $ & $    0.53 $ & $    -0.67 $ \\ 
    12   &    18 < Q^2<    45    &  &     &   & $   1.153  \pm    0.008  $ & $    1.109 \pm    0.027 $ & $     1.00 \pm    0.10 $ & $  44.2 /  43 $ & $   -0.95 $ & $    0.54 $ & $    -0.68 $ \\ 
    13   &    45 < Q^2<   110    &  &     &   & $   1.091  \pm    0.007  $ & $    0.986 \pm    0.028 $ & $     1.11 \pm    0.10 $ & $  44.0 /  45 $ & $   -0.95 $ & $    0.53 $ & $    -0.67 $ \\ 
    14   &   110 < Q^2<   316    &   &     &   & $   1.177  \pm    0.011  $ & $    0.917 \pm    0.039 $ & $     1.06 \pm    0.10 $ & $  41.2 /  42 $ & $   -0.95 $ & $    0.53 $ & $    -0.67 $ \\ 
    15   &   316 < Q^2<  1000    &  &     &   & $   1.084  \pm    0.016  $ & $    0.866 \pm    0.065 $ & $     0.86 \pm    0.12 $ & $  39.0 /  39 $ & $   -0.95 $ & $    0.50 $ & $    -0.63 $ \\ 
\hline 
    16   & Q^2>     6    &  E^{ \rm jet}_T >      6 &   -1.0 < \eta^{\rm jet}<   1.5  & =  1  & $   1.208  \pm    0.005  $ & $    1.034 \pm    0.016 $ & $     0.98 \pm    0.06 $ & $  47.4 /  49 $ & $   -0.95 $ & $    0.53 $ & $    -0.67 $ \\ 
    17   &     &   &    & =  2  & $   0.918  \pm    0.013  $ & $    1.074 \pm    0.045 $ & $     0.88 \pm    0.07 $ & $  42.7 /  40 $ & $   -0.95 $ & $    0.53 $ & $    -0.67 $ \\ 
\hline \hline 
    18 & Q^2>     6 &      6 <E^{* \rm jet}_T <    10 &   -1.0 < \eta^{\rm jet}<   1.5  & $\ge 1$ & $   1.068  \pm    0.009  $ & $    1.111 \pm    0.029 $ & $     0.86 \pm    0.08 $ & $  44.7 /  44 $ & $   -0.95 $ & $    0.56 $ & $    -0.72 $ \\ 
    19 & &     10 <E^{* \rm jet}_T <    16 &    &  & $   0.948  \pm    0.016  $ & $    0.987 \pm    0.055 $ & $     1.02 \pm    0.07 $ & $  41.9 /  41 $ & $   -0.95 $ & $    0.53 $ & $    -0.67 $ \\ 
    20 &  &     16 <E^{* \rm jet}_T <    24 &     & & $   0.803  \pm    0.033  $ & $    0.973 \pm    0.129 $ & $     0.86 \pm    0.13 $ & $  33.1 /  33 $ & $   -0.96 $ & $    0.50 $ & $    -0.64 $ \\ 
\hline 
    21   &     6 < Q^2<    18    &  E^{* \rm jet}_T >      6 &   -1.0 < \eta^{\rm jet}<   1.5  & $\ge 1$  & $   1.276  \pm    0.016  $ & $    1.248 \pm    0.047 $ & $     0.76 \pm    0.10 $ & $  35.6 /  40 $ & $   -0.95 $ & $    0.54 $ & $    -0.69 $ \\ 
    22   &    18 < Q^2<    45    &  &   &   & $   0.999  \pm    0.014  $ & $    1.120 \pm    0.045 $ & $     0.98 \pm    0.09 $ & $  29.7 /  40 $ & $   -0.95 $ & $    0.54 $ & $    -0.69 $ \\ 
    23   &    45 < Q^2<   110    &   &    &   & $   0.864  \pm    0.014  $ & $    0.949 \pm    0.049 $ & $     1.04 \pm    0.11 $ & $  42.6 /  39 $ & $   -0.95 $ & $    0.53 $ & $    -0.68 $ \\ 
    24   &   110 < Q^2<   316    &   &     &   & $   0.927  \pm    0.018  $ & $    0.945 \pm    0.068 $ & $     0.95 \pm    0.12 $ & $  38.5 /  37 $ & $   -0.95 $ & $    0.52 $ & $    -0.67 $ \\ 
    25   &   316 < Q^2<  1000    &   &    &   & $   0.865  \pm    0.024  $ & $    0.775 \pm    0.102 $ & $     0.74 \pm    0.15 $ & $  16.2 /  31 $ & $   -0.95 $ & $    0.48 $ & $    -0.62 $ \\ 
\hline \hline
    26   &     6 < Q^2<    18    &  E^{* \rm jet}_T >      6 &   -1.0 < \eta^{\rm jet}<   1.5  & $\ge 1$  & $   1.272  \pm    0.016  $ & $    1.250 \pm    0.047 $ & $     0.77 \pm    0.10 $ & $  36.5 /  40 $ & $   -0.95 $ & $    0.54 $ & $    -0.68 $ \\ 
    27   &    18 < Q^2<   100    & &    &   & $   0.944  \pm    0.010  $ & $    1.055 \pm    0.034 $ & $     1.01 \pm    0.07 $ & $  31.9 /  43 $ & $   -0.95 $ & $    0.54 $ & $    -0.69 $ \\ 
\hline 
    28   &    10 < Q^2<    25    &  E^{* \rm jet}_T >      6 &   -1.0 < \eta^{\rm jet}<   1.5  & $\ge 1$  & $   1.277  \pm    0.010  $ & $    1.127 \pm    0.029 $ & $     0.86 \pm    0.09 $ & $  42.9 /  43 $ & $   -0.95 $ & $    0.53 $ & $    -0.67 $ \\ 
    29   &    25 < Q^2<   100    &  &    &   & $   1.109  \pm    0.006  $ & $    1.044 \pm    0.022 $ & $     1.06 \pm    0.08 $ & $  45.9 /  46 $ & $   -0.95 $ & $    0.53 $ & $    -0.68 $ \\ 
    30   &   100 < Q^2<  1000    &   &     &  & $   1.131  \pm    0.009  $ & $    0.919 \pm    0.031 $ & $     0.96 \pm    0.08 $ & $  37.7 /  44 $ & $   -0.95 $ & $    0.51 $ & $    -0.65 $ \\ 
\hline 
 \end{tabular} 

\end{center}
\normalsize
    \caption{The fit parameters $\rho_l$, $\rho_c$ and $\rho_b$ along
    with their errors, the $\chi^2$ per degree of freedom and the
    correlation coefficients. The first row lists the results of the
    fit used to evaluate the integrated cross sections (bin 1). The
    remaining rows lists the fits used to evaluate the differential
    cross sections for jets in the laboratory frame (bins 2--17 and
    28-30) and those requiring at least one jet in the Breit frame
    (bins 18--27).}
\label{tab:rhotable}
\end{sidewaystable}

\begin{table}[h]
\begin{center}
\renewcommand{\arraystretch}{1.2}
 \begin{tabular}{|c|c|c|c|c|}  \hline  
 &\multicolumn{2}{l|}{ } & charm jet & beauty jet \bigstrut[tb] \\

 &\multicolumn{2}{l|}{ }&  $\sigma~{\rm [pb]}$ &  $\sigma~{\rm [pb]}$  \\[2pt] \hline  
& \multicolumn{2}{l|}{ }  & & \\[-12pt]
{\bf H1 Data} &\multicolumn{2}{l|}{ }   &\boldmath $  3290 \pm  50 \pm 260 $\unboldmath &\boldmath $   189 \pm    9 \pm   42$\unboldmath \\[5pt] 
\hline\hline
Model& $\mu$ & PDF &&\\ \hline
 RAPGAP & $ Q^2 $ &  MRST2004F3LO  & $  3170$ & $  199 $  \bigstrut[tb] \\ 
CASCADE & $ \sqrt{Q^{2}+p_{T}^{2}+4m^{2}} $ &  A0  & $  3900$ & $   248 $  \bigstrut[tb] \\  \hline

NLO HVQDIS & $ \sqrt{(Q^{2}+p_{T}^{2}+m^{2})/2} $ &  MSTW08FF3  & $  2780 ^{+ 230}_{- 230}$ & $   199 ^{+  23}_{-  22}$  \bigstrut[tb] \\ 
 & $ \sqrt{Q^{2}+4m^{2}}            $ &         & $  3020 ^{+ 600}_{- 320}$ & $   197 ^{+  28}_{-  22}$  \bigstrut[tb] \\ \cline{2-5}
 & $ \sqrt{(Q^{2}+p_{T}^{2}+m^{2})/2} $ &  CTEQ6.6  & $  2780 ^{+ 240}_{- 240}$ & $   196 ^{+  24}_{-  21}$  \bigstrut[tb] \\ 
 & $ \sqrt{Q^{2}+4m^{2}}            $ &         & $  3000 ^{+ 600}_{- 310}$ & $   194 ^{+  27}_{-  22}$  \bigstrut[tb] \\ \cline{2-5}
 & $ \sqrt{(Q^{2}+p_{T}^{2}+m^{2})/2} $ &  CTEQ5F3  & $  2550 ^{+ 210}_{- 230}$ & $   180 ^{+  21}_{-  19}$  \bigstrut[tb] \\ 
 & $ \sqrt{Q^{2}+4m^{2}}            $ &         & $  2800 ^{+ 550}_{- 320}$ & $   180 ^{+  24}_{-  21}$  \bigstrut[tb] \\ \hline
 \end{tabular} 

\end{center}
\normalsize
    \caption{ The cross sections for events with $c$ and $b$ jets for
    the kinematic range $Q^2>6~{\rm GeV}^2$, $0.07<y<0.625$, $E_T^{\rm
    jet}>6~{\rm GeV}$ and $-1.0<\eta^{\rm jet}<1.5$.  The measured
    data cross sections are shown with their statistical and
    systematic uncertainties. The data are compared with the
    predictions from the Monte Carlos RAPGAP and CASCADE and with NLO
    QCD, calculated using HVQDIS.  The NLO QCD predictions are shown
    for three sets of parton distribution functions and two choices of
    renormalisation and factorisation scales. The errors are obtained
    by changing the scales by factors of $0.5$ and $2$, by varying the
    quark masses and using a different model for the fragmentation of
    the quarks.}
\label{tab:sigtheory}
\end{table}
\renewcommand{\arraystretch}{1.0}

\begin{table}[h]
\tiny
\begin{center}
 \begin{tabular}{|l|q|e|x|c|d|d|d|d|} \hline 
 bin & \multicolumn{1}{c|}{$Q^2$ range} &  \multicolumn{1}{c|}{$E^{(*) \rm jet}_T$ range } &  \multicolumn{1}{c|}{$\eta^{\rm jet}$ range} &  \multicolumn{1}{c|}{$N^{\rm jet}$} &\multicolumn{1}{c|}{$\sigma $}  & \multicolumn{1}{c|}{$\delta_{\rm stat}$} & \multicolumn{1}{c|}{$\delta_{\rm sys}$}  & \multicolumn{1}{c|}{$C_{\rm had}$}  \\ 
  &  \multicolumn{1}{c|}{(GeV$^2$)} &  \multicolumn{1}{c|}{(GeV)} & & & \multicolumn{1}{c|}{(pb)} & \multicolumn{1}{c|}{(\%)} & \multicolumn{1}{c|}{(\%)}  &  \\ \hline
 $c$     1 & Q^2>     6 &   E^{ \rm jet}_T >     6 &   -1.0 < \eta^{\rm jet}<   1.5  & $\ge 1$  & $ 3292.3 $ & $   1.4 $   & $   7.9 $ & $  1.00 $\\ 
 $b$     1 &  & &  &   & $ 188.8 $ & $   4.8 $   & $  22.3 $  & $  1.05 $  \\ 
\hline 
 $c$     2 & Q^2>     6 &      6 <E^{ \rm jet}_T <    10 &   -1.0 < \eta^{\rm jet}<   1.5  & $\ge 1$  & $ 2386.5 $ & $   2.0 $   & $   7.8 $  & $  0.99 $ \\ 
 $b$     2 &  & &  &   & $ 100.9 $ & $  11.8 $   & $  34.3 $  & $  1.14 $  \\ 
 $c$     3 &  &     10 <E^{ \rm jet}_T <    16 &    &   & $ 727.4 $ & $   2.8 $   & $   7.9 $  & $  1.02 $ \\ 
 $b$     3 &  & &  &   & $  67.5 $ & $   6.3 $   & $  20.4 $  & $  0.96 $  \\ 
 $c$     4 &  &     16 <E^{ \rm jet}_T <    24 &  &   & $ 148.0 $ & $   6.1 $   & $   9.9 $  & $  1.06 $ \\ 
 $b$     4 &  & &  &   & $  16.5 $ & $  10.4 $   & $  17.6 $  & $  0.91 $  \\ 
 $c$     5 &  &     24 <E^{ \rm jet}_T <    36 &  &   & $  21.5 $ & $  22.1 $   & $  20.7 $  & $  1.06 $ \\ 
 $b$     5 &  & &  &   & $   3.4 $ & $  21.8 $   & $  17.8 $  & $  0.96 $  \\ 
\hline 
 $c$     6 & Q^2>     6 &   E^{ \rm jet}_T >     6 &   -1.0 < \eta^{\rm jet}<  -0.5  & $\ge 1$  & $ 449.6 $ & $   4.0 $   & $   6.9 $ & $  1.11 $\\ 
 $b$     6 &  & &  &   & $  17.6 $ & $  21.8 $   & $  27.0 $  & $  1.45 $  \\ 
 $c$     7 &  &    &   -0.5 < \eta^{\rm jet}<   0.0  & & $ 710.6 $ & $   2.7 $   & $   7.6 $ & $  1.05 $\\ 
 $b$     7 &  & &  &   & $  36.6 $ & $  10.0 $   & $  23.6 $  & $  1.04 $  \\ 
 $c$     8 &  &    &    0.0 < \eta^{\rm jet}<   0.5  &   & $ 801.2 $ & $   2.8 $   & $   8.1 $ & $  1.01 $\\ 
 $b$     8 &  & &  &   & $  53.2 $ & $   7.7 $   & $  22.7 $  & $  0.98 $  \\ 
 $c$     9 & &    &    0.5 < \eta^{\rm jet}<   1.0  &   & $ 856.0 $ & $   2.9 $   & $   7.9 $ & $  0.95 $\\ 
 $b$     9 &  & &  &   & $  43.9 $ & $   9.7 $   & $  20.8 $  & $  1.01 $  \\ 
 $c$    10 &  &    &    1.0 < \eta^{\rm jet}<   1.5  &   & $ 504.0 $ & $   4.9 $   & $   9.3 $ & $  0.84 $\\ 
 $b$    10 &  & &  &   & $  32.7 $ & $  17.5 $   & $  25.3 $  & $  1.04 $  \\ 
\hline 
 $c$    11 &     6 <Q^2<    18 &   E^{ \rm jet}_T >     6 &   -1.0 < \eta^{\rm jet}<   1.5  & $\ge 1$  & $ 934.0 $ & $   2.8 $   & $   7.3 $ & $  1.00 $ \\ 
 $b$    11 &  & &  &   & $  44.2 $ & $  12.2 $   & $  25.1 $  & $  1.07 $  \\ 
 $c$    12 &    18 <Q^2<    45 &   &    &   & $ 924.0 $ & $   2.4 $   & $   7.6 $ & $  1.00 $ \\ 
 $b$    12 &  & &  &   & $  49.1 $ & $   9.6 $   & $  23.8 $  & $  1.07 $  \\ 
 $c$    13 &    45 <Q^2<   110 &    &    &  & $ 857.3 $ & $   2.8 $   & $   8.2 $ & $  1.00 $ \\ 
 $b$    13 &  & &  &   & $  51.7 $ & $   9.2 $   & $  22.0 $  & $  1.05 $  \\ 
 $c$    14 &   110 <Q^2<   316 &    &     & & $ 471.7 $ & $   4.3 $   & $   9.2 $ & $  0.99 $ \\ 
 $b$    14 &  & &  &   & $  36.4 $ & $   9.5 $   & $  19.7 $  & $  1.01 $  \\ 
 $c$    15 &   316 <Q^2<  1000 &    &     &   & $ 113.8 $ & $   7.5 $   & $  10.3 $ & $  1.00 $ \\ 
 $b$    15 &  & &  &   & $   9.5 $ & $  14.3 $   & $  18.0 $  & $  1.00 $  \\ 
\hline 
 $c$    16 & Q^2>     6 &   E^{ \rm jet}_T >     6 &   -1.0 < \eta^{\rm jet}<   1.5  & =  1  & $ 2938.4 $ & $   1.5 $   & $   7.9 $ & $  0.99 $ \\ 
 $b$    16 &  & &  &   & $ 153.2 $ & $   5.9 $   & $  24.3 $  & $  1.02 $  \\ 
 $c$    17 & &   &  & =  2  & $ 337.3 $ & $   4.2 $   & $   7.7 $ & $  1.04 $ \\ 
 $b$    17 &  & &  &   & $  36.3 $ & $   8.2 $   & $  17.2 $  & $  1.15 $  \\ 
\hline\hline 
 $c$    18 & Q^2>     6 &      6 <E^{* \rm jet}_T <    10 &   -1.0 < \eta^{\rm jet}<   1.5  & $\ge 1$  & $ 1083.5 $ & $   2.6 $   & $   7.8 $  & $  1.00 $ \\ 
 $b$    18 &  & &  &   & $  71.3 $ & $   8.8 $   & $  27.2 $  & $  1.18 $  \\ 
 $c$    19 &&     10 <E^{* \rm jet}_T <    16 & &   & $ 231.6 $ & $   5.6 $   & $   9.1 $  & $  1.03 $ \\ 
 $b$    19 &  & &  &   & $  39.7 $ & $   7.4 $   & $  18.2 $  & $  0.95 $  \\ 
 $c$    20 & &     16 <E^{* \rm jet}_T <    24 &     &  & $  39.7 $ & $  13.2 $   & $  15.0 $  & $  1.04 $ \\ 
 $b$    20 &  & &  &   & $   7.3 $ & $  15.2 $   & $  17.4 $  & $  0.92 $  \\ 
\hline 
 $c$    21 &     6 <Q^2<    18 &   E^{* \rm jet}_T >     6 &   -1.0 < \eta^{\rm jet}<   1.5  & $\ge 1$  & $ 650.4 $ & $   3.8 $   & $   7.9 $ & $  1.01 $ \\ 
 $b$    21 &  & &  &   & $  37.2 $ & $  12.6 $   & $  25.5 $  & $  1.09 $  \\ 
 $c$    22 &    18 <Q^2<    45 &  &   &   & $ 372.1 $ & $   4.0 $   & $   8.0 $ & $  1.00 $ \\ 
 $b$    22 &  & &  &   & $  34.0 $ & $   9.6 $   & $  22.5 $  & $  1.10 $  \\ 
 $c$    23 &    45 <Q^2<   110 & &    &   & $ 207.9 $ & $   5.2 $   & $   8.4 $ & $  1.01 $ \\ 
 $b$    23 &  & &  &   & $  26.0 $ & $  10.2 $   & $  19.7 $  & $  1.09 $  \\ 
 $c$    24 &   110 <Q^2<   316 &   &    &   & $ 121.1 $ & $   7.1 $   & $   9.4 $ & $  1.02 $ \\ 
 $b$    24 &  & &  &   & $  15.3 $ & $  12.3 $   & $  19.7 $  & $  1.08 $  \\ 
 $c$    25 &   316 <Q^2<  1000 &    &    &   & $  34.5 $ & $  13.2 $   & $  13.7 $ & $  1.01 $ \\ 
 $b$    25 &  & &  &   & $   4.3 $ & $  20.0 $   & $  17.9 $  & $  1.07 $  \\ 
\hline \hline 
 $c$    26 &     6 <Q^2<    18 &   E^{* \rm jet}_T >     6 &   -1.0 < \eta^{\rm jet}<   1.5  & $\ge 1$  & $ 657.7 $ & $   3.7 $   & $   7.9 $ & $  1.01 $ \\ 
 $b$    26 &  & &  &   & $  37.6 $ & $  12.5 $   & $  25.4 $  & $  1.09 $  \\ 
 $c$    27 &    18 <Q^2<   100 &   &   &   & $ 557.6 $ & $   3.2 $   & $   8.0 $ & $  1.01 $ \\ 
 $b$    27 &  & &  &   & $  57.3 $ & $   7.1 $   & $  21.1 $  & $  1.10 $  \\ 
\hline 
 $c$    28 &    10 <Q^2<    25 &   E^{* \rm jet}_T >     6 &   -1.0 < \eta^{\rm jet}<   1.5  & $\ge 1$  & $ 811.3 $ & $   2.6 $   & $   7.1 $ & $  1.00 $ \\ 
 $b$    28 &  & &  &   & $  41.6 $ & $  10.7 $   & $  23.9 $  & $  1.07 $  \\ 
 $c$    29 &    25 <Q^2<   100 &    &    &   & $ 1400.1 $ & $   2.1 $   & $   8.0 $ & $  1.00 $ \\ 
 $b$    29 &  & &  &   & $  77.9 $ & $   7.7 $   & $  22.7 $  & $  1.06 $  \\ 
 $c$    30 &   100 <Q^2<  1000 &   &     &  & $ 664.0 $ & $   3.4 $   & $   9.1 $ & $  0.99 $ \\ 
 $b$    30 &  & &  &   & $  47.5 $ & $   7.9 $   & $  19.8 $  & $  1.01 $  \\ 
\hline 
 \end{tabular} 

\end{center}
\normalsize
    \caption{The measured charm and beauty cross sections for those
    events in which the highest $E_T^{\rm jet}$ jet is a charm or
    beauty jet. Integrated cross sections in each bin are shown.
    The first two rows (bin 1) are the integrated charm
    and beauty cross sections for the measured phase space respectively. The differential cross
    sections may be formed from the remaining rows by dividing by the
    corresponding bin width. The remaining rows list the cross
    sections for jets in the  laboratory frame (bins 2--17 and 28-30) and those
    requiring at least one jet in the Breit frame (bins 18--27).  The
    data is corrected to the hadron level.  The table also shows the
    statistical($\delta_{\rm stat}$) and systematic error
    ($\delta_{\rm sys}$), together with the hadronic correction
    $C_{\rm had}$ that is applied to the NLO theory to compare with
    the data.}
\label{tab:sig}
\end{table}

\begin{sidewaystable}
\tiny
 \begin{tabular}{|l|d|d|d|d|d|d|d|d|d|d|d|d|d|d|d|d|d|d|d|d|d|} \hline 
bin & \multicolumn{1}{r|}{$\delta_{\rm unc}$} & \multicolumn{1}{r|}{$\delta_{\rm res}$}  & \multicolumn{1}{r|}{$\delta_{\rm tr eff}$}  & \multicolumn{1}{r|}{$\delta_{\rm frag C}$} & \multicolumn{1}{r|}{$\delta_{\rm frag B}$} & \multicolumn{1}{r|}{$\delta_{uds}$} & \multicolumn{1}{r|}{$\delta^{\rm jet}_{\phi}$} & \multicolumn{1}{r|}{$\delta_{\rm had E}$}  & \multicolumn{1}{r|}{$\delta_{\rm gp}$} &   \multicolumn{1}{r|}{$\delta_{\rm E_e}$}  &   \multicolumn{1}{r|}{$\delta_{\theta_e}$} & \multicolumn{1}{r|}{$\delta_{P_T c}$}  & \multicolumn{1}{r|}{$\delta_{P_T b}$}   & \multicolumn{1}{r|}{$\delta_{\eta c}$}  & \multicolumn{1}{r|}{$\delta_{\eta b}$} & \multicolumn{1}{r|}{$\delta_{{\rm BF} D^+}$} & \multicolumn{1}{r|}{$\delta_{{\rm BF} D^0}$} &  \multicolumn{1}{r|}{$\delta_{{\rm Mult} D^+}$} & \multicolumn{1}{r|}{$\delta_{{\rm Mult} D^0}$} & \multicolumn{1}{r|}{$\delta_{{\rm Mult} D_s}$} & \multicolumn{1}{r|}{$\delta_{{\rm Mult} B}$}  \\ 
  & \multicolumn{1}{c|}{(\%)} & \multicolumn{1}{c|}{(\%)} & \multicolumn{1}{c|}{(\%)} & \multicolumn{1}{c|}{(\%)} & \multicolumn{1}{c|}{(\%)} & \multicolumn{1}{c|}{(\%)} & \multicolumn{1}{c|}{(\%)} & \multicolumn{1}{c|}{(\%)} & \multicolumn{1}{c|}{(\%)}  & \multicolumn{1}{c|}{(\%)}  & \multicolumn{1}{c|}{(\%)} & \multicolumn{1}{c|}{(\%)} & \multicolumn{1}{c|}{(\%)} & \multicolumn{1}{c|}{(\%)} & \multicolumn{1}{c|}{(\%)} & \multicolumn{1}{c|}{(\%)} & \multicolumn{1}{c|}{(\%)} & \multicolumn{1}{c|}{(\%)} & \multicolumn{1}{c|}{(\%)} & \multicolumn{1}{c|}{(\%)} & \multicolumn{1}{c|}{(\%)} \\ \hline
$c$   1  &   2.0 &   2.0 &   0.7 &  -1.2 &   0.1 &  -4.5 &   1.8 &   1.4 &   0.6 &   0.2 &   0.6 &   0.3  &   0.9 &   1.0  &  -0.1 &  -0.3  &   0.1 &  -1.2  &  -1.2 &  -1.5  &  -1.3  \\ 
$b$   1  &   2.0 &  -6.5 &   9.0 &   0.1 &  -2.1 &   2.0 &   0.9 &   7.2 &   0.8 &  -1.0 &   0.3 &  -4.1  &  -8.4 &  -0.3  &   2.1 &  -1.6  &   0.9 &  -4.4  &  -1.3 &  -1.9  &  13.1  \\ 
\hline 
$c$   2  &   2.0 &   2.4 &   0.6 &  -1.3 &   0.1 &  -4.5 &   1.4 &   1.6 &   0.8 &   0.3 &   0.7 &   0.7  &   0.2 &   1.2  &  -0.1 &  -0.1  &   0.0 &  -0.9  &  -1.2 &  -1.4  &  -0.7  \\ 
$b$   2  &   2.0 & -14.7 &  15.7 &  -1.8 &  -2.8 &   4.5 &   0.2 &  17.8 &   0.6 &  -1.2 &   0.2 &  -2.8  &  -3.7 &   0.9  &   4.1 &  -4.3  &   3.0 & -10.8  &  -1.9 &  -3.7  &  11.8  \\ 
$c$   3  &   2.0 &   2.3 &   0.3 &  -0.9 &   0.3 &  -4.2 &   2.2 &  -0.3 &   0.4 &  -0.1 &   0.4 &   1.2  &   0.3 &   1.2  &  -0.5 &  -0.2  &   0.1 &  -1.1  &  -1.1 &  -1.4  &  -2.3  \\ 
$b$   3  &   2.0 &  -4.7 &   8.0 &   0.3 &  -2.3 &   1.4 &   0.8 &   8.4 &   0.7 &  -0.7 &   0.5 &  -0.7  &  -1.1 &   0.8  &   4.1 &  -1.1  &   0.4 &  -3.0  &  -1.3 &  -1.3  &  14.0  \\ 
$c$   4  &   2.0 &   2.6 &   0.4 &  -0.6 &   0.2 &  -5.6 &   4.3 &  -0.6 &   0.1 &  -0.5 &   0.3 &   1.6  &   0.2 &   1.8  &  -0.6 &  -0.3  &   0.2 &  -1.1  &  -1.0 &  -1.3  &  -3.1  \\ 
$b$   4  &   2.0 &  -2.9 &   7.0 &   0.1 &  -1.8 &   2.0 &   0.3 &   1.6 &   1.1 &  -0.5 &   0.3 &  -0.4  &   0.4 &   0.6  &   3.6 &  -0.5  &   0.0 &  -1.9  &  -1.1 &  -1.5  &  14.2  \\ 
$c$   5  &   2.0 &   2.6 &   0.7 &  -1.1 &   0.4 & -16.1 &   9.8 &  -1.0 &  -1.0 &   0.9 &   0.0 &   1.7  &  -0.1 &   2.1  &  -1.0 &  -0.3  &   0.1 &  -1.2  &  -1.3 &  -0.6  &  -5.3  \\ 
$b$   5  &   2.0 &   0.6 &   7.5 &   0.3 &  -2.0 &   3.5 &   2.2 &   0.4 &   0.3 &  -2.1 &   0.1 &  -0.1  &   2.0 &   0.4  &   4.2 &  -0.1  &  -0.1 &  -0.8  &  -0.7 &  -1.6  &  13.8  \\ 
\hline 
$c$   6  &   2.0 &   2.3 &   0.4 &  -0.8 &   0.1 &  -3.3 &   1.2 &   1.2 &   0.3 &  -0.7 &   0.6 &   0.9  &   0.4 &  -1.0  &   0.0 &  -0.1  &   0.0 &  -1.0  &  -1.2 &  -1.5  &  -0.6  \\ 
$b$   6  &   2.0 & -10.0 &  14.2 &  -0.7 &  -1.2 &   1.6 &   2.1 &  11.0 &   1.1 &  -2.5 &  -0.3 &  -4.6  &  -4.8 &  -0.2  &  -3.1 &  -3.3  &   2.5 &  -8.8  &  -2.1 &  -2.9  &  10.6  \\ 
$c$   7  &   2.0 &   2.0 &   0.7 &  -1.0 &   0.0 &  -4.4 &   1.5 &   1.4 &   0.7 &  -0.2 &   0.6 &   0.4  &   0.7 &  -0.2  &   0.0 &  -0.3  &   0.1 &  -1.2  &  -1.1 &  -1.4  &  -1.3  \\ 
$b$   7  &   2.0 &  -6.9 &  10.6 &   0.6 &  -0.5 &   3.5 &   2.6 &   7.0 &   0.7 &  -1.1 &   0.4 &  -3.9  &  -6.9 &  -0.2  &  -1.1 &  -1.7  &   1.3 &  -4.8  &  -1.7 &  -2.8  &  14.2  \\ 
$c$   8  &   2.0 &   2.2 &   0.8 &  -1.1 &   0.0 &  -4.5 &   2.3 &   1.2 &   0.4 &  -0.1 &   0.7 &  -0.2  &   1.0 &   0.2  &   0.0 &  -0.3  &   0.1 &  -1.2  &  -1.2 &  -1.5  &  -2.0  \\ 
$b$   8  &   2.0 &  -6.0 &   8.2 &   0.3 &  -0.7 &   1.0 &  -0.5 &   5.5 &   0.9 &  -0.5 &   0.2 &  -3.1  &  -8.5 &  -0.1  &  -0.2 &  -1.2  &   0.7 &  -3.5  &  -1.1 &  -1.7  &  16.0  \\ 
$c$   9  &   2.0 &   1.8 &   1.1 &  -1.0 &   0.0 &  -4.6 &   2.1 &   1.6 &   0.8 &   0.7 &   0.6 &  -0.4  &   0.9 &   0.7  &   0.0 &  -0.3  &   0.2 &  -1.3  &  -1.2 &  -1.4  &  -1.0  \\ 
$b$   9  &   2.0 &  -4.2 &   9.0 &   0.4 &  -0.6 &   2.5 &   1.7 &   6.5 &   0.1 &  -0.9 &   0.4 &  -4.7  &  -9.6 &   0.0  &   1.0 &  -1.4  &   0.8 &  -4.2  &  -1.2 &  -1.4  &  11.0  \\ 
$c$  10  &   2.0 &   2.3 &   0.4 &  -1.6 &   0.5 &  -6.2 &   1.5 &   0.6 &   0.7 &   2.4 &   0.6 &   0.8  &   1.4 &   1.9  &   0.0 &  -0.2  &   0.2 &  -1.0  &  -1.0 &  -1.3  &  -0.9  \\ 
$b$  10  &   2.0 &  -7.6 &  10.6 &  -1.4 &  -5.8 &   3.2 &   5.0 &   9.1 &   1.2 &  -1.9 &   0.3 &  -7.5  & -11.2 &   0.0  &   2.8 &  -1.8  &   1.1 &  -4.9  &  -1.7 &  -1.8  &   8.2  \\ 
\hline 
$c$  11  &   2.0 &   2.0 &   0.6 &  -1.3 &   0.1 &  -2.9 &   1.4 &   1.2 &   1.1 &   1.3 &   0.7 &   1.0  &   0.7 &   2.2  &  -0.1 &  -0.3  &   0.0 &  -1.1  &  -1.2 &  -1.4  &  -1.0  \\ 
$b$  11  &   2.0 &  -7.6 &  11.7 &   0.3 &  -2.3 &   0.9 &   0.9 &  10.1 &   0.6 &  -0.5 &   0.5 &  -4.6  &  -7.8 &   0.0  &   4.0 &  -2.1  &   1.4 &  -6.2  &  -1.4 &  -2.4  &  12.5  \\ 
$c$  12  &   2.0 &   2.1 &   0.7 &  -1.3 &   0.1 &  -3.8 &   1.6 &   2.1 &   0.8 &   0.7 &   0.2 &   0.9  &   0.7 &   0.9  &  -0.1 &  -0.2  &   0.1 &  -1.1  &  -1.3 &  -1.5  &  -1.0  \\ 
$b$  12  &   2.0 &  -7.8 &  10.7 &  -0.2 &  -2.2 &   2.9 &   0.2 &   9.5 &   0.8 &  -0.7 &   0.2 &  -3.2  &  -7.1 &  -0.2  &   2.2 &  -2.1  &   1.3 &  -5.4  &  -1.3 &  -1.7  &  12.8  \\ 
$c$  13  &   2.0 &   1.8 &   0.7 &  -1.1 &   0.2 &  -5.3 &   1.9 &   1.0 &   0.4 &  -0.2 &   0.7 &   0.0  &   0.7 &   0.3  &  -0.1 &  -0.2  &   0.1 &  -1.2  &  -1.3 &  -1.5  &  -1.6  \\ 
$b$  13  &   2.0 &  -5.6 &  10.1 &  -0.1 &  -2.4 &   2.6 &   0.6 &   5.9 &   0.6 &  -1.2 &   0.4 &  -2.0  &  -6.4 &   0.3  &   1.4 &  -1.7  &   1.0 &  -4.5  &  -1.4 &  -2.2  &  14.2  \\ 
$c$  14  &   2.0 &   2.2 &   0.6 &  -0.6 &   0.2 &  -5.9 &   2.6 &   0.5 &   0.1 &  -1.4 &   1.3 &  -1.0  &   0.8 &   1.1  &  -0.1 &  -0.3  &   0.2 &  -1.2  &  -1.0 &  -1.4  &  -2.5  \\ 
$b$  14  &   2.0 &  -4.3 &   7.8 &  -0.1 &  -2.1 &   2.1 &   0.9 &   1.7 &   1.1 &  -0.4 &   0.0 &  -1.3  &  -6.1 &   0.1  &   2.7 &  -1.0  &   0.3 &  -3.0  &  -1.2 &  -1.2  &  14.6  \\ 
$c$  15  &   2.0 &   2.6 &   0.7 &  -0.8 &   0.1 &  -7.1 &   3.3 &   0.1 &   0.0 &  -1.4 &   0.3 &  -0.4  &   0.5 &   1.1  &  -0.1 &  -0.4  &   0.2 &  -1.3  &  -0.9 &  -1.7  &  -2.8  \\ 
$b$  15  &   2.0 &  -2.3 &   7.7 &  -0.1 &  -1.9 &   1.3 &   4.6 &   0.3 &   0.4 &  -1.6 &   0.7 &  -0.7  &  -3.3 &   0.4  &   1.9 &  -0.7  &   0.2 &  -2.0  &  -1.2 &  -1.1  &  13.6  \\ 
\hline 
$c$  16  &   2.0 &   2.0 &   0.8 &  -1.3 &   0.1 &  -4.7 &   1.7 &   1.5 &   0.6 &   0.3 &   0.6 &   0.4  &   0.9 &   1.0  &  -0.1 &  -0.3  &   0.1 &  -1.2  &  -1.2 &  -1.4  &  -1.3  \\ 
$b$  16  &   2.0 &  -7.2 &   9.8 &   0.0 &  -2.6 &   2.5 &   0.8 &   8.7 &   0.7 &  -0.9 &   0.2 &  -4.5  &  -8.6 &  -0.2  &   2.3 &  -1.8  &   1.3 &  -5.1  &  -1.4 &  -2.2  &  14.0  \\ 
$c$  17  &   2.0 &   2.6 &  -0.1 &  -0.6 &   0.1 &  -3.4 &   2.7 &  -0.1 &   0.8 &  -0.6 &   0.7 &  -0.2  &   1.0 &   1.5  &  -0.1 &  -0.2  &   0.1 &  -1.1  &  -1.0 &  -1.5  &  -2.1  \\ 
$b$  17  &   2.0 &  -4.3 &   7.7 &   0.4 &  -0.7 &   0.7 &   1.2 &   4.3 &   1.7 &  -1.0 &   0.6 &  -2.0  &  -5.7 &  -0.2  &   2.0 &  -0.7  &   0.3 &  -2.4  &  -1.2 &  -0.9  &  11.2  \\ 
\hline 
$c$  18  &   2.0 &   2.4 &   0.3 &  -1.2 &   0.2 &  -3.6 &   2.3 &   0.5 &   1.7 &   0.8 &   0.9 &  -1.7  &   0.7 &   0.7  &  -0.2 &  -0.2  &   0.0 &  -1.0  &  -1.1 &  -1.3  &  -1.4  \\ 
$b$  18  &   2.0 &  -7.3 &  10.1 &   0.5 &  -2.5 &   2.6 &   0.6 &  15.8 &   0.9 &   0.1 &   0.1 &  -3.5  &  -9.9 &   0.5  &   2.9 &  -1.8  &   1.0 &  -4.7  &  -1.5 &  -1.8  &  12.4  \\ 
$c$  19  &   2.0 &   3.6 &  -0.4 &  -0.7 &   0.3 &  -3.8 &   3.2 &  -1.2 &   0.6 &  -1.7 &   0.8 &  -0.2  &   0.9 &   1.1  &  -0.7 &  -0.2  &   0.2 &  -1.0  &  -0.9 &  -1.2  &  -3.8  \\ 
$b$  19  &   2.0 &  -3.4 &   6.0 &   0.2 &  -1.7 &   0.5 &   0.0 &   8.4 &   1.4 &   0.9 &   0.4 &  -0.9  &  -4.9 &   0.2  &   2.7 &  -0.5  &  -0.1 &  -1.4  &  -0.8 &  -0.9  &  12.3  \\ 
$c$  20  &   2.0 &   7.0 &  -0.1 &  -1.6 &   0.3 &  -6.5 &   8.1 &   0.1 &   0.2 &  -2.5 &   0.1 &   1.3  &   0.8 &   2.4  &  -0.9 &  -0.2  &   0.2 &  -0.7  &  -0.9 &  -0.9  &  -5.1  \\ 
$b$  20  &   2.0 &  -3.3 &   6.2 &   0.5 &  -1.9 &   1.7 &  -0.6 &   0.5 &   2.4 &  -0.1 &   0.6 &  -0.9  &  -2.2 &   0.1  &   3.6 &  -0.2  &  -0.2 &  -1.2  &  -0.9 &  -1.1  &  14.0  \\ 
\hline 
$c$  21  &   2.0 &   2.2 &   0.6 &  -1.0 &   0.1 &  -3.2 &   2.1 &   1.5 &   1.8 &   2.0 &   0.9 &  -1.4  &   0.8 &   1.9  &  -0.1 &  -0.3  &   0.0 &  -1.1  &  -1.1 &  -1.2  &  -1.3  \\ 
$b$  21  &   2.0 &  -6.4 &   9.5 &   0.4 &  -2.1 &   1.6 &   0.7 &  13.4 &   0.8 &  -1.7 &   1.0 &  -3.5  &  -9.5 &   0.5  &   3.9 &  -1.5  &   0.9 &  -4.6  &  -1.5 &  -2.4  &  12.5  \\ 
$c$  22  &   2.0 &   2.9 &   0.2 &  -1.3 &   0.2 &  -3.5 &   2.2 &   0.6 &   2.1 &  -0.1 &   1.0 &  -1.0  &   1.1 &   0.7  &  -0.3 &  -0.2  &   0.1 &  -1.0  &  -1.1 &  -1.3  &  -1.9  \\ 
$b$  22  &   2.0 &  -5.5 &   7.9 &   0.6 &  -2.1 &   1.8 &  -0.1 &  11.1 &   0.9 &   2.1 &  -0.6 &  -2.4  &  -9.0 &   0.2  &   2.3 &  -1.2  &   0.5 &  -3.0  &  -1.2 &  -1.0  &  12.3  \\ 
$c$  23  &   2.0 &   2.7 &   0.3 &  -1.0 &   0.3 &  -4.1 &   3.2 &   0.3 &   1.0 &  -1.2 &   0.8 &  -1.0  &   1.1 &   0.2  &  -0.4 &  -0.2  &   0.2 &  -1.1  &  -1.1 &  -1.2  &  -2.5  \\ 
$b$  23  &   2.0 &  -4.2 &   7.4 &   0.3 &  -2.2 &   1.7 &   0.2 &   7.5 &   1.7 &   0.7 &   0.6 &  -1.6  &  -7.8 &   0.4  &   1.4 &  -0.7  &   0.3 &  -2.2  &  -0.9 &  -1.3  &  12.4  \\ 
$c$  24  &   2.0 &   3.0 &   0.1 &  -0.7 &   0.2 &  -4.8 &   4.0 &  -0.3 &   0.2 &  -0.3 &   0.7 &  -1.6  &   0.9 &   1.0  &  -0.4 &  -0.2  &   0.2 &  -1.1  &  -0.9 &  -1.1  &  -3.3  \\ 
$b$  24  &   2.0 &  -3.0 &   7.5 &   0.3 &  -2.1 &   1.9 &   0.1 &   4.9 &   3.1 &   0.0 &   0.2 &  -1.1  &  -7.5 &   0.5  &   3.1 &  -0.6  &  -0.1 &  -1.9  &  -1.2 &  -0.8  &  13.7  \\ 
$c$  25  &   2.0 &   6.7 &   2.0 &  -1.3 &   0.2 &  -7.7 &   5.9 &  -2.1 &  -0.1 &  -1.1 &  -0.3 &  -1.3  &   0.6 &   0.8  &  -0.4 &  -0.3  &   0.2 &  -1.1  &  -1.0 &  -1.5  &  -3.1  \\ 
$b$  25  &   2.0 &  -5.1 &   6.7 &   0.5 &  -1.9 &   1.4 &   4.2 &   2.6 &   1.1 &  -1.4 &   0.7 &  -0.8  &  -4.5 &   0.5  &   2.3 &  -0.4  &  -0.1 &  -1.7  &  -0.9 &  -1.3  &  12.8  \\ 
\hline 
$c$  26  &   2.0 &   2.2 &   0.5 &  -1.0 &   0.1 &  -3.2 &   2.0 &   1.5 &   1.8 &   1.9 &   1.1 &  -1.4  &   0.8 &   1.9  &  -0.1 &  -0.3  &   0.0 &  -1.1  &  -1.1 &  -1.2  &  -1.3  \\ 
$b$  26  &   2.0 &  -6.4 &   9.6 &   0.4 &  -2.1 &   1.8 &   0.6 &  13.2 &   0.8 &  -1.3 &   0.4 &  -3.5  &  -9.5 &   0.5  &   3.9 &  -1.5  &   0.9 &  -4.5  &  -1.4 &  -2.3  &  12.6  \\ 
$c$  27  &   2.0 &   2.6 &   0.2 &  -1.2 &   0.2 &  -3.8 &   2.8 &   0.5 &   1.7 &  -0.3 &   0.7 &  -1.1  &   1.1 &   0.5  &  -0.3 &  -0.2  &   0.1 &  -1.1  &  -1.1 &  -1.3  &  -2.1  \\ 
$b$  27  &   2.0 &  -4.9 &   7.5 &   0.4 &  -2.1 &   1.7 &  -0.5 &   9.6 &   1.2 &   1.1 &   0.3 &  -2.0  &  -8.6 &   0.2  &   1.9 &  -1.0  &   0.4 &  -2.7  &  -1.1 &  -1.1  &  12.2  \\ 
\hline 
$c$  28  &   2.0 &   2.0 &   0.7 &  -1.4 &   0.1 &  -3.1 &   1.3 &   1.6 &   0.8 &   0.6 &   0.5 &   1.1  &   0.7 &   1.2  &  -0.1 &  -0.2  &   0.1 &  -1.1  &  -1.2 &  -1.4  &  -1.0  \\ 
$b$  28  &   2.0 &  -6.9 &  10.6 &   0.3 &  -2.4 &   1.9 &   1.1 &   9.9 &   0.4 &  -0.6 &   0.3 &  -4.0  &  -7.3 &   0.2  &   2.7 &  -2.2  &   1.5 &  -5.9  &  -1.4 &  -2.2  &  12.5  \\ 
$c$  29  &   2.0 &   1.9 &   0.7 &  -1.2 &   0.1 &  -4.8 &   1.9 &   1.6 &   0.7 &   0.3 &   0.4 &   0.3  &   0.7 &   0.7  &  -0.1 &  -0.2  &   0.1 &  -1.2  &  -1.3 &  -1.5  &  -1.2  \\ 
$b$  29  &   2.0 &  -6.7 &  10.3 &  -0.3 &  -2.3 &   3.1 &  -0.5 &   7.6 &   0.8 &  -0.9 &   0.3 &  -2.5  &  -6.8 &   0.1  &   1.8 &  -1.9  &   1.1 &  -5.0  &  -1.3 &  -1.9  &  13.3  \\ 
$c$  30  &   2.0 &   2.4 &   0.7 &  -0.7 &   0.1 &  -5.7 &   2.6 &   0.5 &   0.1 &  -1.4 &   1.0 &  -1.3  &   0.8 &   0.7  &  -0.1 &  -0.3  &   0.2 &  -1.2  &  -1.0 &  -1.5  &  -2.4  \\ 
$b$  30  &   2.0 &  -5.0 &   7.5 &   0.0 &  -2.0 &   1.2 &   1.9 &   1.6 &   0.9 &  -0.9 &   0.2 &  -1.8  &  -7.0 &   0.2  &   2.0 &  -1.0  &   0.3 &  -2.8  &  -1.3 &  -1.3  &  14.4  \\ 
\hline 
 \end{tabular} 

\normalsize
    \caption{ The contributions to the total systematic error. The bin
    numbering scheme follows that used in table~\ref{tab:sig}. The
    first column lists the uncorrelated systematic error.  The next
    $10$ columns represent a $+ 1 \sigma$ shift for the correlated
    systematic error contributions from: track impact parameter
    resolution; track efficiency; $c$ fragmentation; $b$
    fragmentation; light quark contribution; struck quark angle
    $\phi_{\rm quark}$; hadronic energy scale; photoproduction
    background; electron energy scale; electron theta; reweighting the
    jet transverse momentum distribution $P^{\rm jet}_T$ and
    pseudorapidity $\eta^{\rm jet}$ distribution for $c$ and $b$
    events; the $c$ hadron branching fractions and multiplicities; and
    the $b$ quark decay multiplicity.  Only those uncertainties where there
    is an effect of $>1\%$ in any bin are listed separately; the
    remaining uncertainties are included in the uncorrelated
    error. There is an additional contribution to the systematic error
    due to the uncertainty on the luminosity of $4\%$.}
\label{tab:sys}
\end{sidewaystable}

\clearpage
\newpage

\end{document}